\documentclass[]{aa}

\usepackage{graphics,subfigure,supertabular,array}
\input psfig

\begin{document}
\thesaurus{11(03.13.2; 
           11.05.2; 
	   13.09.1) 
           }

\title{ISOCAM observations of the Hubble Deep Field reduced with the PRETI method.\thanks{Based on observations with the Infrared Space Observatory (ISO). ISO is 
an ESA project with instruments funded by ESA Member States
(especially the PI countries : France, Germany, The Netherlands and
the United Kingdom) and with participation of ISAS and NASA.}}

\author{H. Aussel \and C.J Cesarsky \and D. Elbaz \and J.L. Starck}

\institute{Service d'Astrophysique,DSM/DAPNIA/CEA-Saclay, F-91191 Gif-sur-Yvette 
Cedex, France}

\offprints{H. Aussel}
\mail{aussel@discovery.saclay.cea.fr}

\date{Received 10 March 1998 / Accepted 25 August 1998 }

\maketitle

\titlerunning{ISOCAM observations of the Hubble Deep Field}
\authorrunning{H. Aussel et al.}

\begin{abstract}
We have developed a new ISOCAM data reduction technique based on
wavelet analysis, especially designed for the detection of faint
sources in mid-infrared surveys. This method, the Pattern REcognition
Technique for Isocam data (PRETI) has been used to reduce the
observations of the Hubble Deep Field (HDF) and flanking fields with
ISOCAM at 6.75 (LW2) and 15 $\mu$m (LW3) Rowan-Robinson et
al.~(\cite{RowanRobinson}). Simulations of ISOCAM data allow us to test
the photometric accuracy and completeness of the reduction. According
to these simulations, the PRETI source list is 95\% complete in the 15
$\mu$m band at 200 $\mu$Jy and in the 6.75 $\mu$m band at 65 $\mu$Jy,
using detection thresholds which minimize the number of false
detections. We detect 49 objects in the ISO-HDF at high confidence
secure level, 42 in the LW3 filter, 3 in the LW2 filter, and 4 in both
filters. An additional, less secure, list of 100 sources is presented,
of which 89 are detected at 15 $\mu$m only, 7 at 6.75 $\mu$m only and
4 in both filters. All ISO-HDF objects detected in the HDF itself have
optical or infrared counterparts, except for one from the additional
list. All except one of the radio sources detected in the field by
Fomalont et al.~(\cite{Fomalont}) are detected with ISOCAM. Using a
precise correction for the field of view distortion of ISOCAM allows
us to separate blended sources. This, together with the fact that
PRETI allows to correct data on the tail of cosmic rays glitches, lead
us to produce deeper source lists than previous authors. Our list of
bright sources agree with those of D\'esert et al.~(\cite{IAS}) in both
filters, and with those of Goldschmidt et al.~(\cite{Goldschmidt}) in
the LW3 filter, with systematic difference in photometry. Number
counts derived from our results show an excess by a factor of 10 with
respect to the prediction of a no evolution model
(Franceschini~\cite{Franceschini98}) in the LW3 band. On the contrary,
the number of sources in the LW2 band is compatible with the
prediction of such a model, but with greater uncertainties, given the
small number of detections.
\end{abstract}


\section{Introduction}

The observation with the Hubble Space Telescope (HST) of the Hubble
Deep Field (HDF) (Williams et al.~\cite{Williams}) has yielded
numerous new results in observational cosmology. This is due to its
depth and spatial resolution in the near-UV and optical, allowing the
computation of photometric redshifts of the observed galaxies as well
as the determination of their morphological type. With the
ground-based follow-up observations, in the near IR (Hogg et
al.~\cite{Hogg}, Connolly et al~\cite{Connolly}) for example, or at
radio wavelength (Fomalont et al.~\cite{Fomalont}), as well as
spectroscopic measurements (Steidel et al.~\cite{Steidel}, Cohen et
al.~\cite{Cohen}), this field now constitutes a very important
database for the study of galaxy formation and evolution. Among the
striking results that have been already produced (see Ellis
(\cite{Ellis}) for a review), two have strong implications on the
galaxy formation picture. First, the star formation rate (SFR) at high
redshifts has been derived (Madau et al.~\cite{Madau}), and together
with that computed from Canada France Redshifts Survey (CFRS) data at
lower redshifts (Lilly et al.~\cite{Lilly}), it seems to indicate
that the bulk of star formation arose below z $\sim$ 2. Second, an
increase of faint irregular galaxy counts around $z \approx 1$ has
been reported (Abraham et al.~\cite{Abraham}), as well as the lack of
detection of quiescent evolving early types galaxies at high
redshifts. These results tend to favor the hierarchical galaxy
formation picture.

However, the interpretation of these results is now being widely
discussed. For instance, it has already been pointed out that the
fainter objects seen in the HDF could merely be star forming regions
of galaxies and not entire galaxies, because the distance between
detected objects is small with respect to the size of a galaxy (Colley
et al.~\cite{Colley}). This idea is supported by new simulations that
show that diffuse emission from evolved stellar populations is below
the detection limit of the HDF, while star forming regions with high
UV surface brightness can be detected (Hibbard \&
Vacca~\cite{Hibbard}). For similar reasons, quiescent evolving
early-type galaxies at redshift of $z \sim 3$ are also undetectable
(Maoz~\cite{Maoz}).

Most of the data that were used to derive star formation rates are UV
and optical data from the CFRS and the HST itself, which are sensitive
to extinction. Indeed, it has been shown that star formation at high
redshift is obscured, typically by $\sim$ 2 to 3 mag at
$\lambda=1620$\AA (Meurer et al.~\cite{Meurer}). The SFR inferred
might therefore change if the reprocessing of UV by dust into IR is
taken into account. Knowledge of the dust content of galaxies is
therefore relevant to assess the results obtained so far, both at
$z<1$ with the CFRS and at higher redshifts. Additional information on
this problem can be obtained by observing the HDF field in the
infrared.

Many models have been built to compute the infrared spectra of various
types of galaxies (AGN, starbursts, early types, {\it etc}) and make
predictions on the number counts for surveys in the infrared (see, for
example, Franceschini et al.~(\cite{Franceschini91}). Moreover,
various groups have derived new IRAS 60 $\mu$m counts, going to the
extreme limits of the sensitivity of this satellite. Their results
show an excess of sources with respect to the predictions of a
scenario with only passive luminosity evolution of galaxies (Bertin et
al.~\cite{IRAS_Bertin}, Gregorich et
al.~\cite{IRAS_Gregorich}). However, the exact amount of evolution
needed is still a matter of debate, because the number counts derived
by the various teams are slightly different. This issue needs to be
revisited on deeper samples. From this point of view, the HDF field is
a great opportunity to test these models, first because their number
count predictions can be checked against observations, and second
because all the information at other wavelengths is of great help to
test the hypotheses of the various models.

Since the HDF field is rather small ($\sim$ 5 arcmin$^{2}$) and very
deep, infrared observations require a good spatial resolution as well
as a good sensitivity. This can be achieved by the mid-IR imager
ISOCAM of the Infrared Space Observatory (ISO) (Kessler et
al.~\cite{ISO}), which has proven to be 1000 times more sensitive than
IRAS, with a 50 times finer spatial resolution (Cesarsky et
al.~\cite{ISOCAM}). Part of the Director's Time was dedicated to
complete these observations, at $6.75 \mu$m and $15
\mu$m, for a total amount of 12.5 hours, with M. Rowan-Robinson as
Principal Investigator (Serjeant et al.~cite{Serjeant}, Goldschmidt et
al.~\cite{Goldschmidt}, Oliver et al.~\cite{Oliver}, Mann et
al.~\cite{Mann}, Rowan-Robinson et al.~\cite{RowanRobinson}. The data
of the ISO-HDF are now in the public domain, and we present here the
results of our processing.

The ISOCAM observation of the HDF, hereafter ISO-HDF, is at the moment
the deepest survey obtained in the 15 $\mu$m band. It is also deep in
the 6.75 $\mu$m band, although a deeper integration on the Lockman
Hole was obtained by Taniguchi et al. (\cite{Taniguchi}) with this
filter.  We have developed a new technique for ISOCAM data reduction,
based on wavelet analysis (Starck et al.~\cite{PRETI_moriond})~: the
Pattern REcognition Technique for ISOCAM data (PRETI). Compared to
data that we have at hand from less deep surveys in the Lockman Hole,
from the Guaranteed Time program by Cesarsky et al. (\cite{Catherine}),
and in the shallow survey of the ELAIS program, this set has a much
higher redundancy, and thus is useful for finding the ultimate
capacities of our data reduction method. In addition, since the
ISO-HDF has been analyzed by other teams such as Rowan-Robinson et
al. (\cite{RowanRobinson}) and D\'esert et al. (\cite{IAS}), this
observation allows us to compare the results from our method to those
of more classical ones. All these methods aim at extracting very faint
sources in ISOCAM data, and independent reduction techniques improve
the quality of the produced catalogs. In section 2, the
characteristics of the emission expected from galaxies in the MIR are
outlined, and compared to the bandpasses of the ISOCAM filters that
were used for the HDF observation. In section 3 we present the data
acquisition and stress the critical steps of the reduction
process. The way in which these difficulties are handled by the PRETI
method is described in section 4. The efficiency of the method for
source detection and its photometric accuracy was also tested on
simulated data, which are presented in a section 5. Finally, we give
our source catalog and number counts, compare them with those of other
groups, and discuss the results. A more detailed interpretation of the
results, source by source, will be presented in a forthcoming paper
(Elbaz et al., {\it in prep.}).

\section{MIR emission in ISOCAM filters}

\begin{figure*}
\resizebox{\hsize}{!}{\psfig{file=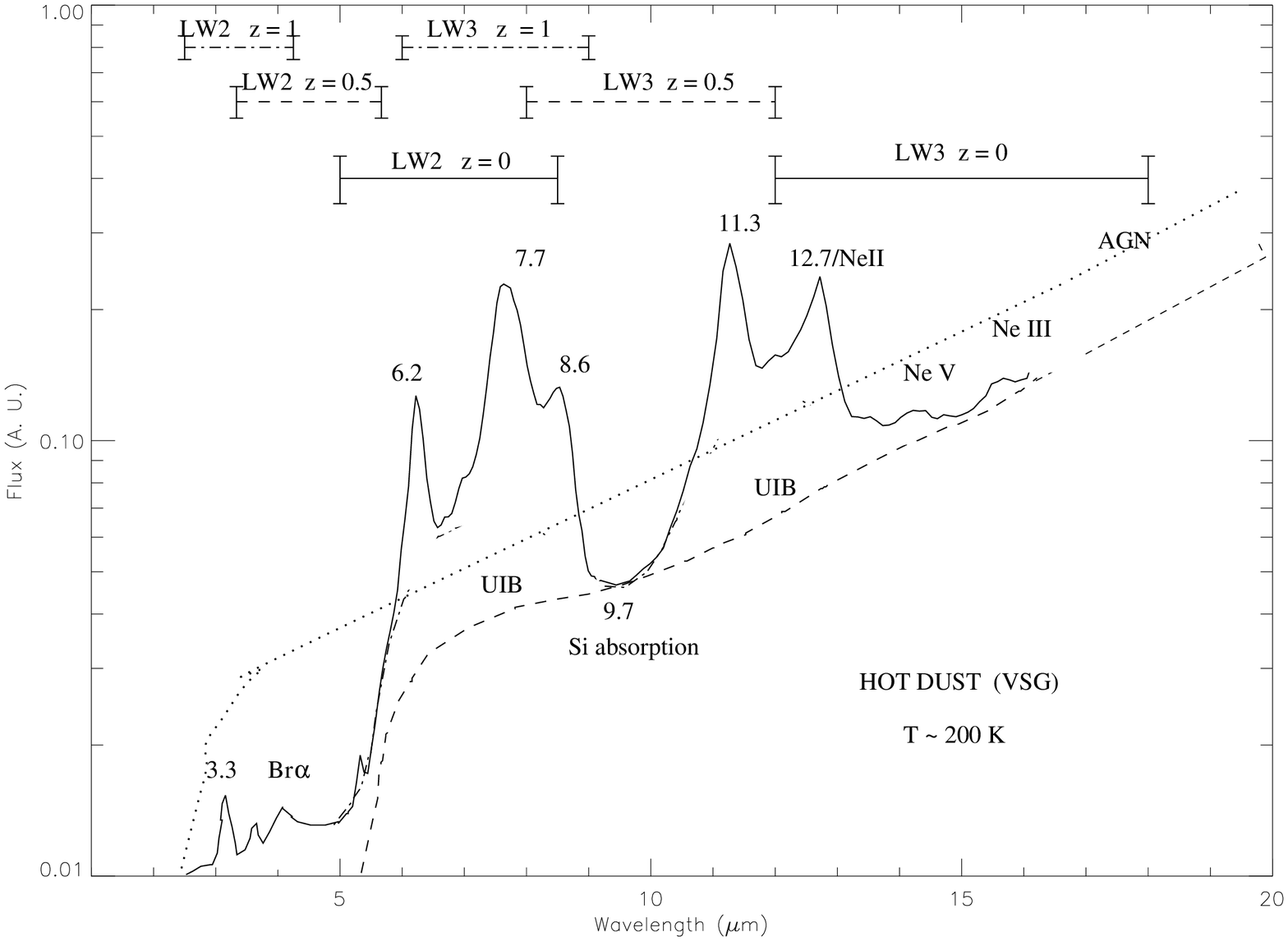,width=1.0\textwidth,angle=0}}
\caption[]{Sketch of the mid IR emission of a starburst galaxy ({\it after an original sketch of A. Jones}) and of an AGN, compared with ISOCAM LW2 and LW3 filters bandpasses at various redshifts.

Solid line: total MIR emission of a starburst galaxy. The $5 \mu$m -
$18 \mu$m part is a an ISOCAM CVF observation showing all the typical
UIB bands and ionization lines ({\it spectra kindly provided by
D. Tran}). The $2 \mu$m - $5 \mu$m part was taken from a SWS spectrum
of the Circinus galaxy (Moorwood et al.~\cite{Moorwood}), and scaled
to the CVF spectrum.

Dashed line: contribution to the MIR emission of the hot dust. This
dust is constituted of very small grains (VSG) (D\'esert et
al.~\cite{desert}) heated to a few hundred degrees.

Dotted-dashed line: Possible contribution of a continuum associated to
the carriers of UIB.

Dotted line: Hot dust (1000 K) MIR emission of an AGN central part, as
observed by Aussel et al.~(\cite{Cloverleaf}).  }
\label{sketch_starburst}
\end{figure*}

\begin{figure*}
\resizebox{\hsize}{!}{\psfig{file=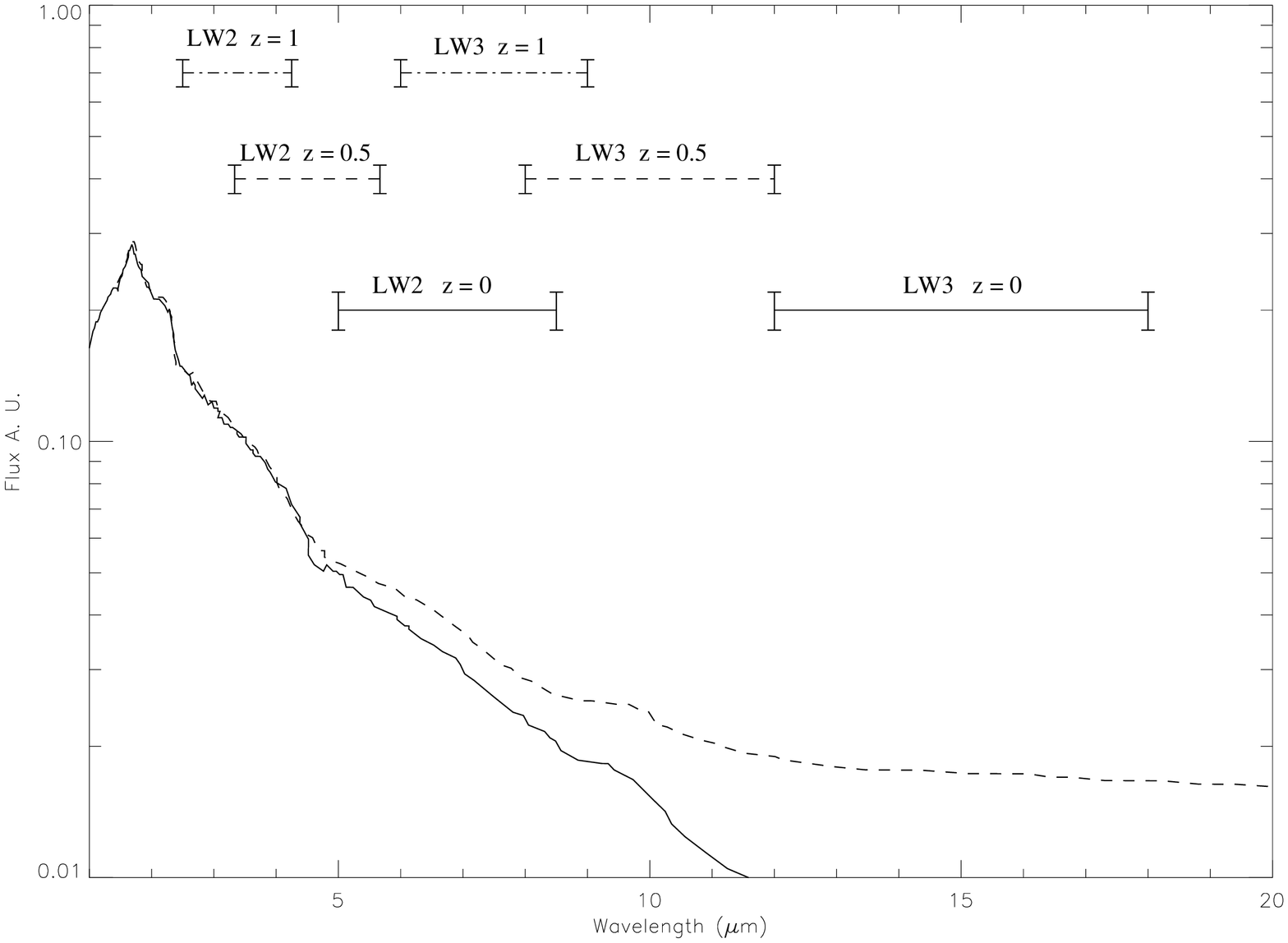,width=1.0\textwidth,angle=270}}
\caption[]{Sketch of the mid IR emission of an old stellar population, compared with ISOCAM LW2 and LW3 filters bandpasses at various redshifts.

Solid line~: model of the MIR emission of a 5 Gyr old stellar simple
population, where dust envelopes of Mira and OH/IR stars are neglected (Bressan et al. \cite{Bressan}).

Dashed line~: model of the MIR emission of a 5 Gyr old stellar simple
population, for which the dust envelopes have been taken into account
for the AGB phase, using silicate grains (Bressan et al. \cite{Bressan}).

The arbitrary units used for fluxes cannot be compared with
those of Fig.~\ref{sketch_starburst}, the relative importance of these emissions in various galaxies varies by factor of order 100. }
\label{sketch_ellipitical}
\end{figure*}

The ISO observations were performed in two broad band filters, LW2
(from 5.0 $\mu$m to 8.5 $\mu$m) and LW3 (from 12.0 $\mu$m to 18.0
$\mu$m).  Figures~\ref{sketch_starburst} and \ref{sketch_ellipitical}
present synthetic spectra of the mid IR emission of galactic
components between 1 and 20 $\mu$m. Figure~\ref{sketch_starburst} is a
spectrum of a starburst galaxy, taken from Moorwood et
al.~(\cite{Moorwood}) for the 1 to 5 $\mu$m part and from
Tran(\cite{dantran}) for the 5 to 18 $\mu$m part. Unidentified
infrared bands (UIB) and ionization lines of Neon are clearly visible,
as well as a hot dust continuum and silicate absorption. In this case,
the derived temperature for the hot dust is about 200 K, using the
model from D\'esert et al.~(\cite{desert}), and the extinction due to
dust is about 3 mag, representative of a starburst region in a
galaxy. The dotted line is a spectrum dominated by very hot dust
heated at 1000 K, that could be typical of the mid-IR emission in the
central core of an AGN, as observed by Aussel et
al.~(\cite{Cloverleaf}). Figure~\ref{sketch_ellipitical} depicts
emission dominated by old population stars as modeled by Bressan et
al.~(\cite{Bressan}), and is thus relevant for early-types galaxies.

The mid-IR (hereafter MIR) colours of galaxies are not simple, because
the observed emission of a galaxy is a mixture of all the components
shown on Fig.~\ref{sketch_starburst} and Fig.
~\ref{sketch_ellipitical}. According to the type of the object
(early-type, AGN, starburst or normal galaxy), it is expected that a
given type of emission will dominate ({\it e.g.} stellar, very hot
dust or UIB). This is not always the case, as shown by Madden et al.~
(\cite{Madden}) on a sample of elliptical galaxies~: most exhibit
stellar colours in the MIR but some present MIR spectral energy
distributions (SED) typical of UIB emission. Thus, the relative fluxes
in figs. ~\ref{sketch_starburst} and ~\ref{sketch_ellipitical} are not
to be compared directly.

In the rest frame, the LW2 filter is dominated by UIB features
while the emission in the LW3 filter is due to the hot dust continuum,
ionization lines of neon and the 12.7 $\mu$m UIB  plus the
associated continuum. This picture changes dramatically when the
redshift rises toward z=1~: the contribution of the stellar continuum,
especially from old population stars, overtakes UIB features in the
LW2 band, which in turn are shifted to the LW3 band, together with the
silicate absorption band at 9.7 $\mu$m. The wavelength coverage for
both filters is shown on both Fig.~\ref{sketch_starburst} and
Fig.~\ref{sketch_ellipitical} for redshifts of 0, 0.5, and 1. This
clearly indicates that the mid-IR observations can only be interpreted
with the help of data at other wavelengths to disentangle the various
contributions. This is possible on the HDF field, because UV, optical
and near IR, as well as radio observations are available.

\section{Data acquisition}

The HDF field is well suited for mid-infrared deep observations. It
was selected for its low optical extinction, low HI column density
and, most important fact for ISOCAM observations, low cirrus emission
in the IRAS 100 $\mu$m maps (Williams et
al.~\cite{Williams}). Moreover, it sits at high galactic latitude, and
contains therefore only a few stars. Similar criteria were used to
select fields for other ISOCAM deep surveys such as the IDSPCO program
(Cesarsky \& Elbaz \cite{Catherine}) or the DEEPPGPQ program (Taniguchi et al.~\cite{Taniguchi}).

\begin{table}[h]
{\small
\begin{center}
\caption{ISO-HDF observation parameters. The main parameters for each raster and for the whole observation are listed}\label{obsparam}
\begin{tabular}{|m{3cm}cc|}
\hline
Parameter & LW2 (6.75 $\mu$m) & LW3 (15 $\mu$m) \\
\hline
\hline
\multicolumn{3}{|c|}{Raster} \\
\hline
\hline
Integration time  & 10 s & 5 s \\
Detector Gain     & 2  & 2 \\
Number of exposures & 9 or 10 & 19 or 20 \\ 
Pixel field of view & 3 '' & 6 '' \\
Number of pointings & 8x8 & 8x8 \\
Step between two pointings & 5'' & 9 ''\\
\hline
\hline
\multicolumn{3}{|c|}{Observation}\\
\hline
Number of rasters & 3 & 3 \\
Total area covered & 10.45'$^{2}$ & 24.3'$^{2}$  \\
\hline
Area of maximum sensitivity & 0.98'$^{2}$  & 7'$^{2}$ \\
Sensitivity (Jy/$''^{2}$) & $\sigma < 0.2$& $\sigma < 0.34 $ \\
Exposure time & $> 1.94$ hrs    & $> 2.25$ hrs \\
\hline
Area of mean sensitivity & 7.25'$^{2}$ & 21.5'$^{2}$ \\
Sensitivity (Jy/$''^{2}$) & $\sigma < 0.5 $ & $\sigma < 1.37 $  \\
Exposure time & $ > 0.31$ hrs          & $ > 0.17$ hrs \\
\hline
\end{tabular}
\end{center}
}
\end{table}

Table \ref{obsparam} describes the observation parameters for both
filters. Each consist of three 8x8 rasters, which are partly
overlapping. In both cases, the area covered is wider that the HDF
field, and even contains a large part of the flanking field for the
LW3 observation, which was done with a larger pixel field of view
(PFOV) and a greater raster step than the LW2 observation. If the
detector had been ideal, {\it i. e.} with no memory effect and insensitive
to cosmic rays, a very good sensitivity would be reached~:
\begin{eqnarray*}
\sigma & = & \sqrt{\frac{\sum_{pixels}{\sigma_{pixel}^{2}}}{N_{exp}}}
\end{eqnarray*}
where $\sigma_{pixel}$ is the noise per readout for each ISOCAM pixel
(it is therefore equal to the readout noise plus the photon noise,
which is dominated in the case of faint sources by the background
emission) , and $N_{exp}$ is the number of exposures. One can easily
compute the ideal sensitivity of the ISO-HDF if one computes
$\sigma_{pixel}$ for each pixel~: this is easily done by fitting a
Gaussian distribution on the histogram of the each pixel, after having
filtered it of all the low frequency components, because all the non
ideal behaviour of ISOCAM occurs mostly at low frequency. At a
5$\sigma$ level, one could reach 15 $\mu$Jy in the LW3 band and 5
$\mu$Jy in the LW2 band. The observation was designed to be as deep as
possible, and with a good spatial resolution in order to be able to
identify with confidence the optical counterparts of ISOCAM
sources. The observations were therefore made using the microscan mode,
that is by moving the camera by fraction of pixels between two raster
steps. With a step of 5 arcsec and a PFOV of 3 arcsec, the final
resolution of the ISOCAM map should reach 1 arcsec in LW2, and 3
arcsec in LW3.

Unfortunately, the long wavelength detector (LW) of ISOCAM is very 
sensitive to cosmic rays due to its thickness, and shows a transient 
behaviour. These two main drawbacks alter its efficiency in detecting 
faint sources. This is the reason why the PRETI method was developed. 

\section{Data reduction with the PRETI method}

The two main difficulties for the detection of faint sources with the
ISOCAM LW channel are mostly the cosmic ray effects and the transient
behaviour (Starck et al.~\cite{PRETI_moriond}). The PRETI method has
been designed to overcome these two problems. The main idea of this
method is to clean the cube of ISOCAM data as well as possible, before
reconstructing the images and performing the source detection. For the
ISO-HDF data, as for all other data, the process goes through the
following steps~:
\begin{enumerate}
\item translate the raw data (CIER and IIPH files) in a raster structure 
containing the images, the ISOCAM configuration and the pointing 
information, using Cam Interactive Analysis (CIA).
\item convert the images in normalized units of ADU/G/s and subtract 
the dark current, with CIA.
\item correct the data for short time cosmic rays and mask the 
affected readouts.
\item clean the data for long term effect of cosmic rays.
\item subtract the baseline of each pixel.
\item compute the reduced images per raster pointings.
\item derive a flat-field from the baseline and correct the reduced 
images accordingly.
\item project each raster pointing on the sky.
\item perform a spatial detection on the final image.
\end{enumerate}

The general method is described in Starck et al.~(\cite{PRETI_moriond}) and
does not need to be detailed here. However, we discuss here some of
the points above, either because they are very important, or because
they need to be specifically adapted to each observation.

\subsection{Long term effect of cosmic rays.}

\begin{figure*}
\resizebox{\hsize}{!}{\psfig{file=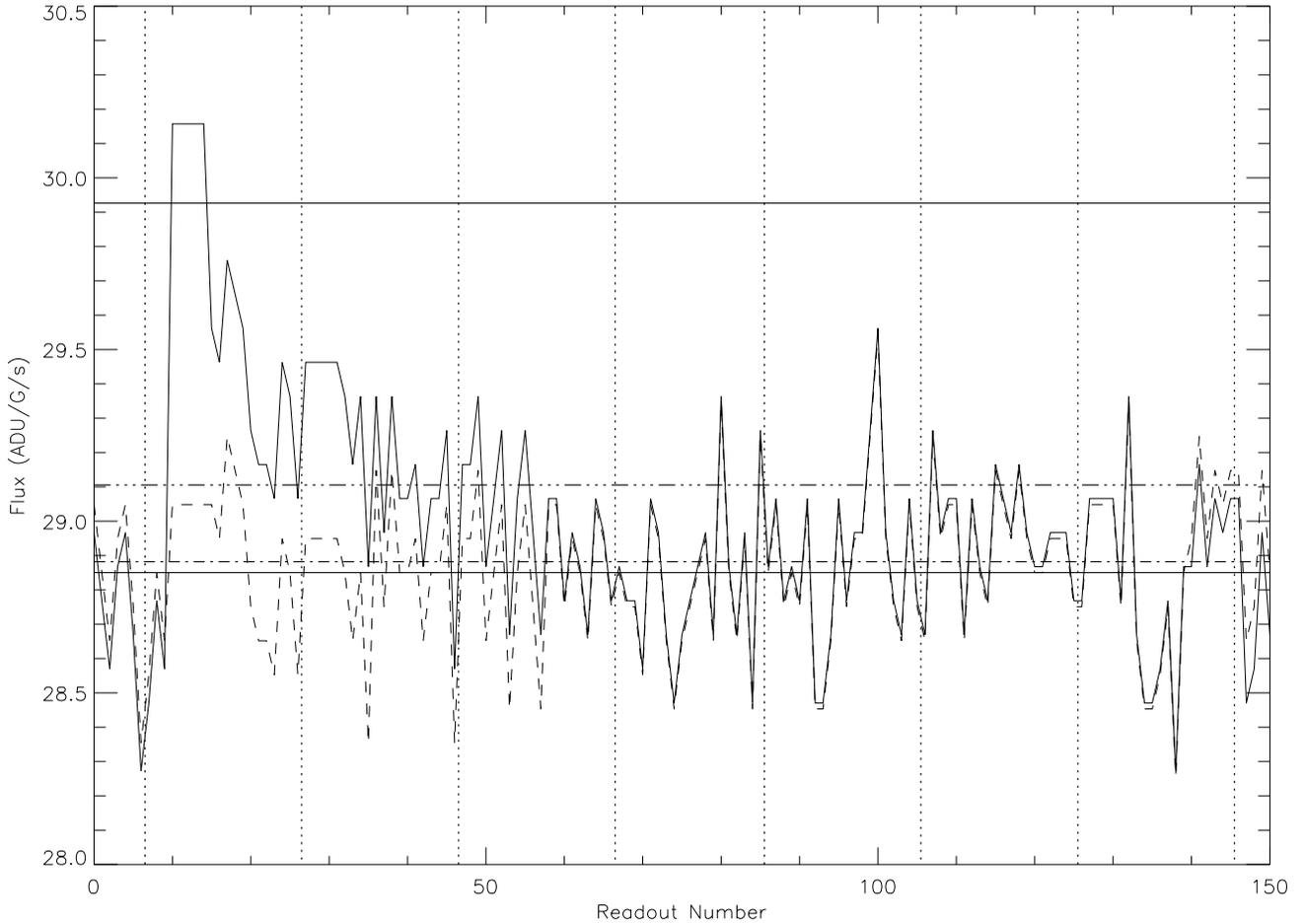,width=1.0\textwidth,angle=90}}
\caption[]{Effect of a cosmic ray impact on the ISOCAM LW detector. The `glitch' occurred on readout 9, and produced a `fader' effect. The solid horizontal line at 28.85 ADU is the average background level. The solid horizontal line at 29.92 is the 5 $\sigma$ level of the noise. The dash-3 dots horizontal line is the average 5 $\sigma$ noise level after averaging all the readouts of a pointing. The dashed-dotted horizontal line is the 5 $\sigma$ level on the final raster map. The limits of each ISOCAM pointings are represented by dotted vertical lines. This `fader' last for 3 pointings. 

solid line~: pixel response versus readout number, after correction for the short term effect of cosmic rays. Frames 9 to 15 and 26 to 30 have been masked out.

dashed line~: final pixel response after correction for the long term effect of cosmic rays.
}

\label{fig_glitch}
\end{figure*}

As explained in Starck et al.~(\cite{PRETI_moriond}), the impact of cosmic
rays, apart from the high frequency response they induce in the LW
detector, are sometimes followed by a slow decrease of the signal
(`faders') or by a depletion in the detector gain (`dippers'). This
behaviour is attributed to proton or $\alpha$ particle impacts on
the detector, while simple `glitches' are due to effects of the cosmic
ray electrons.  One has to be very careful in cleaning them, because
they will produce false detection on the final map if they are not
well removed.

This is especially true for the ISO-HDF observations, because only a
small number of readouts are done per sky position, and because of the
microscan observation mode. In this observation, one third of the `faders'
and `dippers' lasts more than 60 readouts. This means that they last
for more than 6 pointings (LW2) or 3 pointings (LW3). But each cosmic
ray hits on average 4.5 adjacent pixels, that will more or less have
the same behaviour. In a microscan, each position on the sky is
observed by two adjacent pixels during two successive
pointings. Figure~\ref{fig_glitch} illustrates such an event~: if the
`fader' effect is not corrected or masked out, it produces a
spurious source when averaging all the readouts of a same pointing,
because the flux appears to be above the 5 $\sigma$ noise level obtained on
individual pointings. This example is taken from the LW3 observation~:
the two first pointings affected by the cosmic rays are less
distant that the PSF FWHM. Since adjacent pixels of the one
in the figure present the same behaviour, a spurious source
will be `detected', because faders will be co-added together.

This effect is even more important in very deep observations where
many raster positions are coadded together~: `faders' and `dippers'
are detected in the time history of the pixel, thus against a rather
high noise level (see Fig.~\ref{fig_glitch}). On the contrary, sources
will be detected on the co-added map, with a very low noise level that
has been represented on the same figure. It may happen that the tail
of cosmic ray effects, that were undetected in the temporal history,
are coadded to themselves and therefore become visible on the
final image as spurious structure.

In order to confront these problems for the HDF survey, we put very
strong constraints on the deglitching process, that is~:

\begin{enumerate}
\item The detection of structures in the temporal history of the pixel 
is done against 4 $\sigma$ of readout and photon noise, a lower level
than for the usual PRETI method.
\item Each positive structure whose maximum is expressed at the same scale as a raster pointing is treated as a cosmic ray if~:
\subitem a) a cosmic ray has been detected at its beginning ({\it i.e.} 
the two first readouts have been masked by the short deglitching 
process).
\subitem b) more than $50 \%$ of the readouts are already masked by the short deglitching 
process
\end{enumerate}
 
This intends to avoid as much as possible false detections. There is
the possibility that we might clean out a source, but this can be
checked by the following method~: for each data set that we treat, we
produce also raster maps where all the long term effects of cosmic
rays are masked out instead of being corrected. Of course, this leads
to far less sensitive maps since 20 to 40 $\%$ of the readouts are
then eliminated. But this allows to check that the photometry of the
remaining sources is consistent between the two reductions~: this
ensures that no real source was `cleaned out' during the
reduction. Indeed, in the end, we never find discrepencies in the flux
of a given source by factors greater than $15\%$.

\subsection{Baseline subtraction.}

In order to correct for the long transient behaviour that takes place 
at the beginning of each observation (due to the change of the ISOCAM 
configuration), and to clean the data of the possible drifts of the 
detector, the PRETI method suppresses the baseline of each pixel's time 
history. This is computed at larger temporal scales of the signal, 
at least larger than the scale of one raster pointing (Starck et al.~
\cite{PRETI_moriond}). 

Because of the microscan observation, one has to extract the baseline 
at least two scales further for the ISO-HDF observation.  Since the 
step between two raster pointings is about the same size as the FWHM 
of the ISOCAM PSF (9 arcsec at 15 $\mu$m and 4.5 at 6.75 $\mu$m), each 
source is seen by the same pixel during more than a raster 
pointing.  In order to avoid to take some flux from the source into 
the baseline, it is computed two scales further as usual.

It is worth noting that the subtraction of the baseline frees the 
result of the reduction from the possible errors in the dark 
subtraction. Final images that are produced by the PRETI method are 
always on a zero background. Thus, such a method is only usable for 
observation on constant background, with no structure. This is the 
case for the zodiacal light emission, on an area as small as the HDF.

Nevertheless, it is possible to construct a background map of the
observation. This map can be used to check the photometry by comparing
the observed background with predicted values of models of zodiacal
light.

\subsection{Projection onto the sky.}
\begin{figure}
\resizebox{\hsize}{!}{\psfig{file=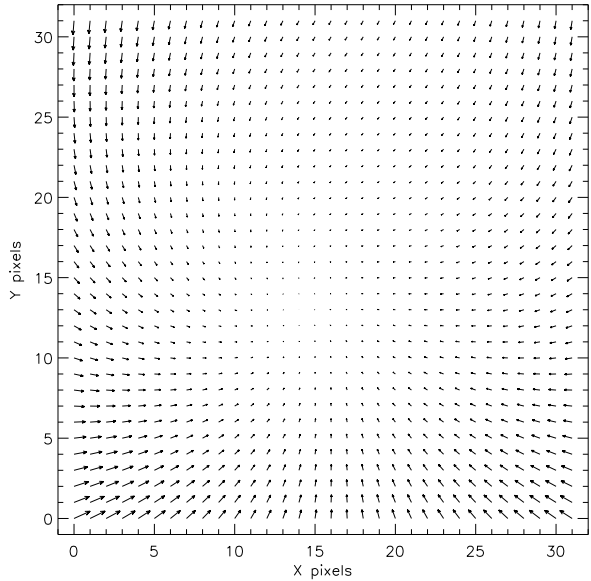,width=0.5\textwidth,angle=0}}
\caption[]{ISOCAM distortion pattern for the LW10 6'' PFOV configuration.
Each vector ends where the center of a pixel should fall if they were
no distortion and starts from its observed position.  The length of the
vectors are at the scale of the plot~: the effect is sometimes greater
than the size of one pixel.}
\label{figure_disto}
\end{figure}
 
Once each raster position has been averaged and flat-fielded, it is 
projected onto a sky map (RA, DEC, epoch 2000), whose resolution is 
computed according to the step size between raster pointings. The 
algorithm being used is derived from the classic `shift-and-add' method. 
The intersection area of each CAM pixel with the sky map is computed, 
and added to it with a weight that is equal to this area times the 
square root of the number of pixels that were coadded to give the 
raster pointing image. That is~:
\begin{eqnarray*}
R_{x,y} & = & \frac{\sum_{pointings}S_{(x,y,i,j)}\sqrt{N_{i,j}}I_{i,j}}
{\sum_{pointings}S_{(x,y,i,j)}\sqrt{N_{i,j}}}    
\end{eqnarray*}
Assuming Gaussian distributions for pixels, one has for the noise 
map~:
\begin{eqnarray*}
\sigma_{x,y} & = &\sqrt{\frac{\sum_{pointings}S^{2}_{(x,y,i,j)}N_{i,j}\sigma^{2}_{i,j}}
{\sum_{pointings}S^{2}_{(x,y,i,j)}N_{i,j}}}    
\end{eqnarray*} 
where $R_{x,y}$ is the value of the final raster map at (x,y), 
$S_{(x,y,i,j)}$ is the intercepted surface between ISOCAM pixel (i,j) 
and raster map pixel (x,y) and $N_{i,j}$ is the number of readouts that 
where averaged together to produce $I_{i,j}$, the image of the raster 
pointing. We choose to use $\sqrt{N_{i,j}}$ as weight rather than the 
R.M.S. computed during the coaddition because this is very 
poorly estimated (less than 10 readouts for the LW2 observation 
for example). Since the ideal detector should follow a Gaussian 
behaviour, its noise should vary as $\sqrt{N_{i,j}}$, which is a more  
reliable estimator.
It is straightforward from the computation of $\sigma_{x,y}$ that the 
noise map against which we do our detection is highly inhomogeneous.

\begin{figure*}
\centering
\mbox{\subfigure[Non corrected]{\psfig{file=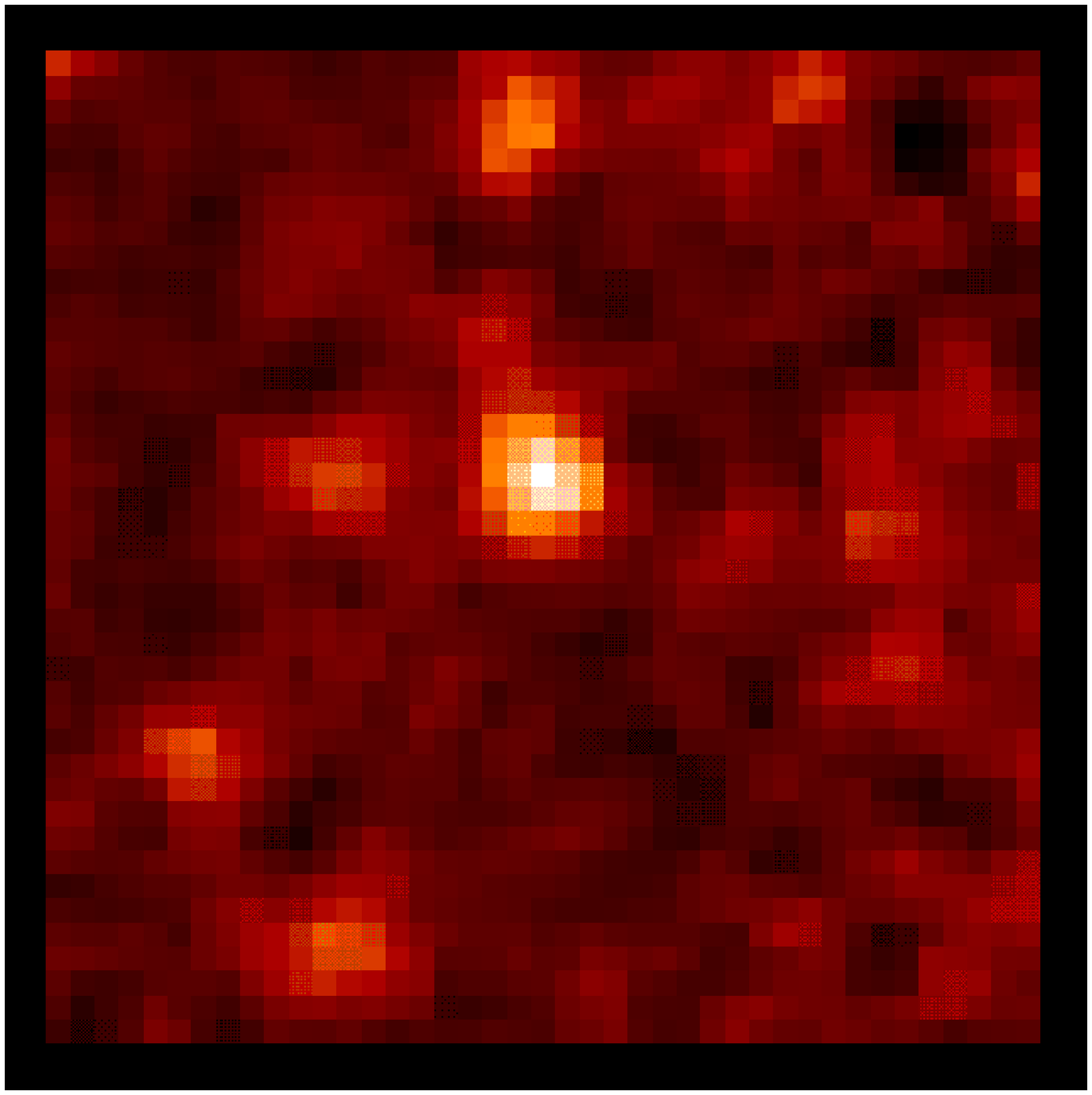,width=.45\textwidth}}\quad
      \subfigure[Corrected]{\psfig{file=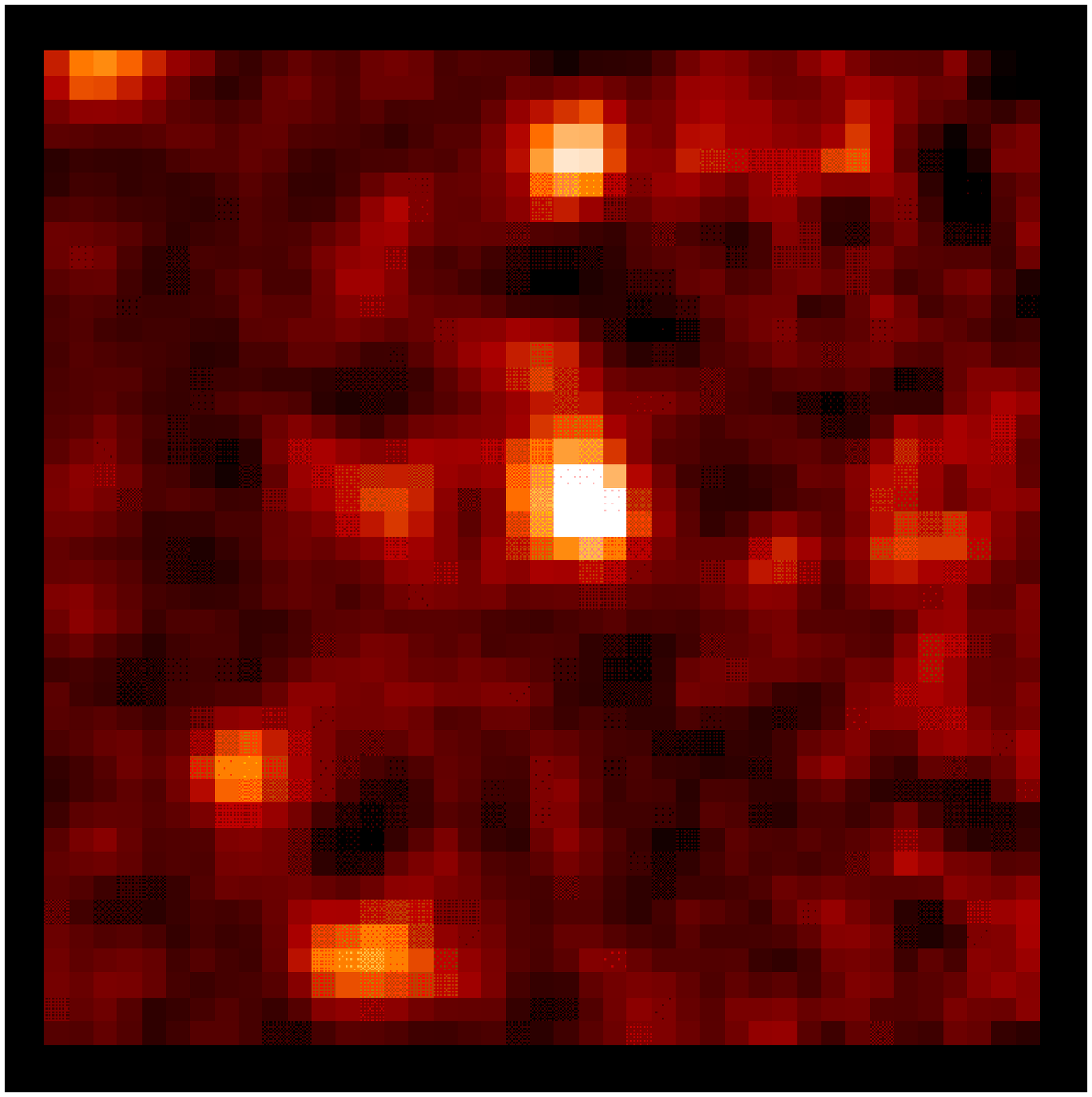,width=.45\textwidth}}}
\caption[]{Zoom on a part of the ISOCAM Hubble Deep Field observation at $15 \mu m$. The same processing was applied to the two images, except for the correction of the distortion. Cuts on images are at the same level. Pixel size is 2''.}
\label{fig_dist_smear}
\end{figure*}

The projection of each raster pointing takes into account the field
distortion of ISOCAM. It has been measured by our team
(Aussel~\cite{Distortion}), using the ISOCAM PSF model from
Okumura~(\cite{Okumura}). Following the work done on the HST WFPC by
Holtzman et al.~(\cite{holtzman}), we fit on our measurement a general
polynomial of degree 3, that allows to compute for any pixel on the
detector its position on the sky~:

\begin{eqnarray*}
x_{c} & = & a_{0}+ a_{1}x + a_{2} y + a_{3}x^{2} + a_{4}x y + a_{5}y^{2} + \\
      &   & a_{6}x^{3}+ a_{7}x^{2}y + a_{8}xy^{2}+ a_{9}y^{3}
\end{eqnarray*}
\begin{eqnarray*}
y_{c} & = & b_{0}+ b_{1}x + b_{2} y + b_{3}x^{2}+ b_{4}x y + b_{5}y^{2} + \\ 
      &   & b_{6}x^{3}+ b_{7}x^{2}y + b_{8}xy^{2}+ b_{9}y^{3} 
\end{eqnarray*}

Figure~\ref{figure_disto} shows a map of the distortion of the LW
channel of ISOCAM with the 6'' PFOV and the LW3 filter, where each
vector ends where the center of a pixel should fall if they were no
distortion, and starts from its actual position.  The length of the
vectors are at the scale of the plot~: at the lower corners of the
array (lines 0 to 5), the effect is bigger than one pixel.  If this
effect were not taken into account, the sources would be smeared when
coadding, and the final astrometry would be poor, reaching up to a 6"
error.  This is illustrated by Figure~\ref{fig_dist_smear}, where a
zoom on the outskirts of one ISO-HDF raster is displayed without and
with correction of distortion.  On the borders and corners of the
raster, the gain of signal over noise ratio is about 50 $\%$.

The measurements at various wavelengths have shown that the distortion
is a chromatic effect. We were able to measure it for both filters and
lenses of the ISO-HDF observation.

The flat fields are computed in CIA by making the assumption that the
illumination of the detector is homogenous. Because of the field
distortion, this is not truly the case since the pixel size is not
uniform on the sky (the pixels at the edges of the array cover a wider
surface). A new flat-field correction has to be applied in order to
account for it. This flat-field is of the form~:
\begin{eqnarray*}
F_{i,j} & = &\frac{S_{16,16}}{S_{i,j}}
\end{eqnarray*}

where $S_{i,j}$ is the surface on the sky of pixel i,j.  The pixel 
(16,16), which is the center pixel of ISOCAM LW array and therefore the 
least distorted, has been taken as reference.  The images should 
therefore be divided by F before reprojecting on the sky.

\subsection{Astrometric corrections} 

As the observations of the ISO-HDF have been performed in three 
separate rasters, one has to coadd all of the images in order to 
reach the maximum sensitivity.  But this can only be done if the 
pointing of each raster is well known.  This is not straightforward 
because of~:
\begin{enumerate} 
\item The absolute pointing accuracy ({\it i.e.} the 1 $\sigma$ error 
that can be made when acquiring a new target) of ISO is of 3 arcsec.  
This is much greater than the relative pointing accuracy, ({\it i.e.} 
the accuracy of pointings between two raster steps), which is less than 0.1 
arcsec.  
\item The lens wheel jitter.  In order to avoid any 
mechanical blocking, the gear wheel has been designed with a small 
play.  The position on which the lens stops is not fixed.  It has been 
shown by the Cam Instrument Dedicated Team (CIDT) and by us that there 
are only two positions that each lens can take for a commanded 
position, probably at either side of 
the play.  This can be very easily detected by a close inspection of 
the flat field derived from the data. If the leftmost column of the 
detector receives very little light, we know we are in the `left' position.  
This jitter results in an offset of about 1.2 pixels from the optical 
axis, thus $\approx$ 7" with the 6" PFOV. It modifies also the 
distortion pattern of ISOCAM that has been measured for both 
positions.  The only way to avoid such problem when coadding various 
rasters is to use the screw 129 and have them observed in a sequence
 in order to prevent any movement of the lens wheel between the 
two rasters.  This was not the case for the ISO-HDF observations.
\end{enumerate}

If all the observations suffer from the absolute pointing error, only
one raster of the LW3 observation suffers from the second problem. The
only way to coadd efficiently the rasters is to identify common
sources in each of them and to compute an offset. These sources have
to be bright enough to have a well definite position. In order to
correct the absolute pointing error, good astrometric references are
needed. The astrometry of the HST observation of the HDF has been
established very carefully (Williams et al.~\cite{Williams}), but is
not useful in our case~: since the FWHM of ISOCAM PSF is large, there
are in general several optical counterparts which could be associated
to a single ISO-HDF source.

In order to solve this problem, we used the well known MIR/radio
correlation (Helou et al.~\cite{Helou}). Since the HDF and the
flanking fields have been observed at 8.4 GHz (Fomalont et
al.~\cite{Fomalont}), we used this source list to match our brightest
objects.  We were successful in identifying at least 4 radio sources
in each LW3 raster and at least one in each LW 2 raster. These sources
are given in Table
\ref{tab_astrometrie} together with the derived offsets.

\begin{table}[h]
{\small
\begin{center}
\caption{ISO-HDF astrometric correction. The radio sources (Fomalont et al.~\cite{Fomalont}) used are listed for each field.}\label{tab_astrometrie}
\begin{tabular}{|c|c|c|c|}
\hline
Raster & 8.4 GHz  & $\alpha$ offset  & $\delta$ offset\\
\hline
LW2 Raster 2	& 3646+1404 & 4.5'' $\pm$ 2.0	& +3.0'' $\pm$ 2.0	\\
		& 3649+1313 &			&			\\
\hline
LW2 Raster 3    & 3649+1313 & 0.0'' $\pm$ 2.0	& 0.0''	$\pm$ 2.0	\\
\hline
LW2 Raster 4    & 3644+1133 & +4.5 ''$\pm$ 2.0	& +2.5'' $\pm$ 2.0		\\
\hline
		& 3634+1212 &			&			\\
		& 3634+1240 &	 		&			\\
		& 3644+1249 &			&			\\
LW3 Raster 2	& 3646+1404 &	0.0'' $\pm$ 0.5	& +4.0 '' $\pm$ 0.5 	\\
		& 3646+1448 &			&			\\
	        & 3649+1313 &		 	&			\\
\hline
	  	& 3644+1249 & 			&			\\
LW3 Raster 3	& 3646+1404 &-0.7'' $\pm 0.5$ 	& +4.0 '' $\pm$ 0.5	\\
		& 3639+1313 &  			&			\\
		& 3653+1139 &			&			\\
\hline
		& 3634+1212 &			&			\\
		& 3634+1240 &	 		&			\\
		& 3644+1249 &			&			\\
LW3 Raster 4	& 3646+1404 &	0.0'' $\pm$ 0.5	& -4.2 '' $\pm$ 0.5 	\\
		& 3646+1448 &			&			\\
	        & 3649+1313 &		 	&			\\
		& 3653+1139 &			&			\\
\hline
\end{tabular}
\end{center}
}
\end{table}

\subsection{Detection, source photometry and transients}

As explained in Starck et al.~(\cite{PRETI_moriond}), the source detection is
made on the final reprojected map, against the noise map, using a
wavelet technique. We use a b-spline wavelet transform, and detect at
each scale all pixels above a given threshold ($\tau_{w}$) of the
transformed noise map. In the ISO-HDF case, the final size of the
pixels is 3'' in LW3 (1.5'' in LW2) for a PSF that is 9" FWHM
(resp. 4.5). We therefore ask the algorithm to perform the detection
down to the $3^{rd}$ scale, that is over a box of 16 pixels, matching
the PSF size. Since the maps are not square, some sources are detected
at the edges of the field due to non continuous limit conditions. Such
sources are eliminated.

It is very important to note that the $\tau_{w}$ that is employed here 
describes the noise level of the detection map, but that this noise is 
not Gaussian. Since many treatments applied on the cube where done 
using median transform, and since the `dippers' and `faders' are 
highly non Gaussian, this means that one cannot relate a 3 $\tau_{w}$ 
detection to a 99.69 \% confidence level, as it is the case for 
Gaussian statistics. Simulations only can give confidence levels for
the value of the detection threshold.

The photometry of the objects can be measured in two ways. First, the
detection program constructs an image of each object. The total flux
in this image is computed. Second, a direct aperture photometry can be
performed on the ISOCAM map. We have used the second method because
simulations have shown us that for faint objects, the reconstruction
using wavelets is not reliable. The measured Analog to Digital Units
(ADU) are corrected to account for transients and for PSF. This correction has
been obtained by simulations (see below). The corrected ADU are then
translated to milli Jansky by using ISOCAM `cookbook' conversion
tables.

The response of the ISOCAM LW detector is affected by a transient when
its illumination changes. The response reaches instantaneously a level
corresponding to 60~\% of the final, stabilized signal, the rise for
the remaining 40~\% being a slow exponential whose time constant
depends on the past history of the pixel. First order corrections have
been produced, especially by Abergel et
al.~(\cite{Abergel}). Transients are the source of two main problems~:
\begin{enumerate}
\item an error in the photometry if the measured flux is not 
stabilized.
\item `ghost' sources appear on maps after that the detector has 
observed a bright source $(\geq 500 \mu Jy)$. 
\end{enumerate}

\begin{figure*}
\centering
\begin{tabular}{cc}
\subfigure[LW2, 5$\tau_{w}$ detection level.]{\psfig{file=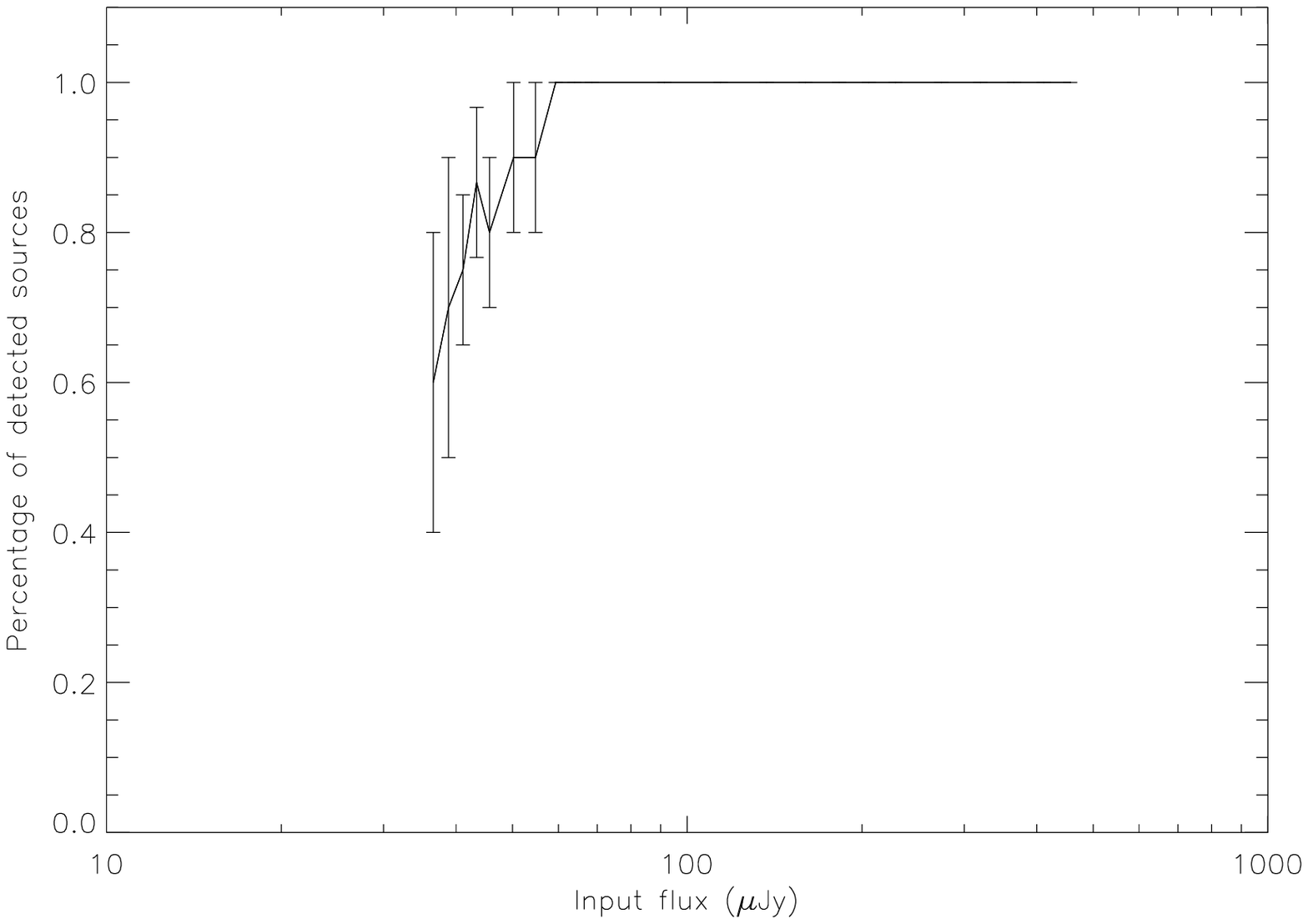,width=0.5\textwidth}} & \subfigure[LW3. Solid~:~5$\tau_{w}$ detection level. Dashed~:~7$\tau_{w}$ detection level.]{\psfig{file=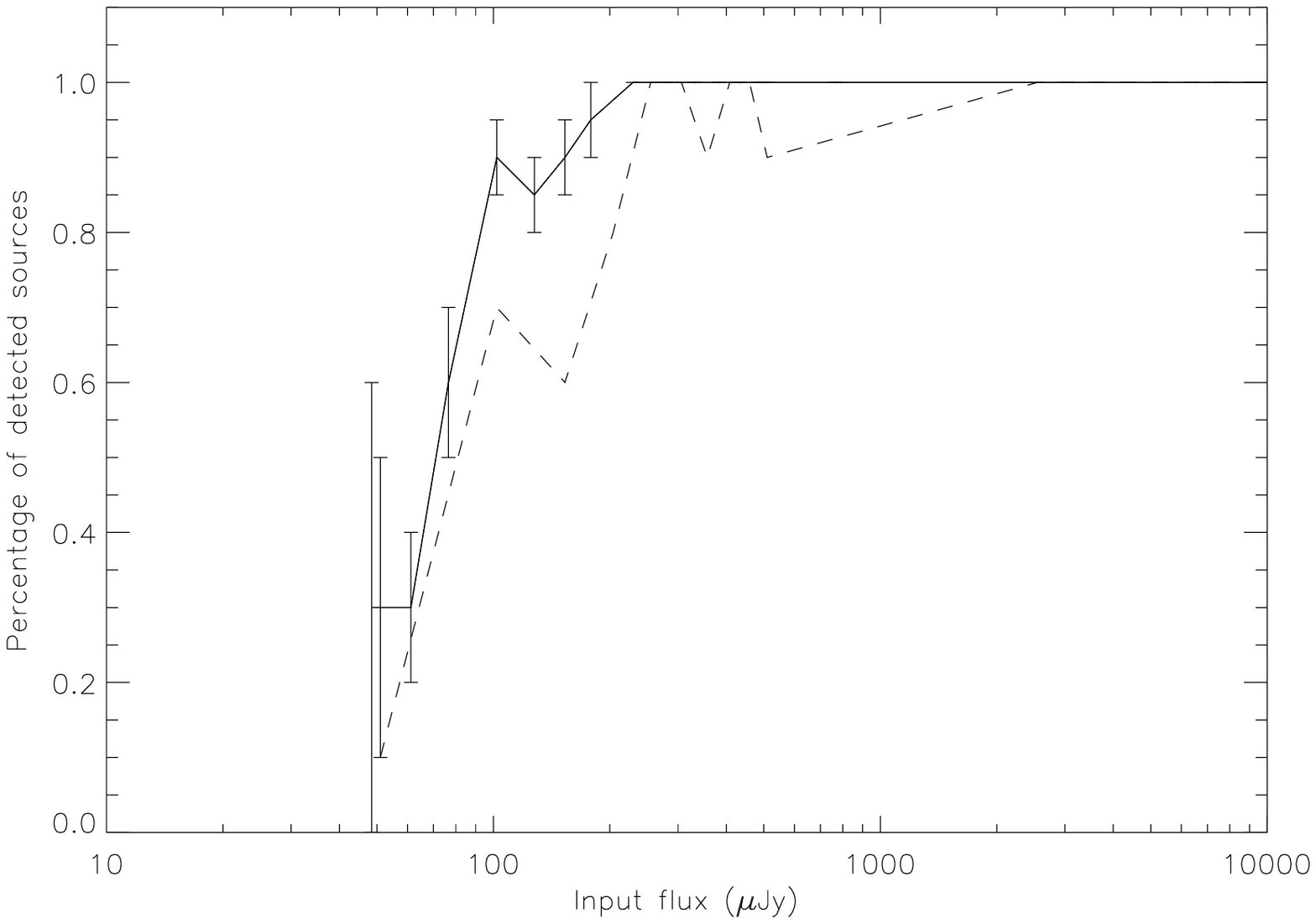,width=0.5\textwidth}}
\end{tabular}
\caption[]{Percentage of detected sources in simulations of the ISO-HDF, as a function of the input flux of the sources, for both filters}
\label{fig_sim_counts}
\end{figure*}

Since the ISO-HDF has very small raster steps and no bright source is
present in the field, the second point is not a problem in our case.
Correcting transients has an effect on the time history of each
pixel. Some glitches, with their slow decrease, mimic a transient
behaviour, and could be treated inadequately by the correction
algorithm. Since we detect these effects by criteria on their shape in
the wavelet space, we might miss some of them after the correction of
the transients. On the other hand, transient effects can be modeled,
and taken into account when converting ADU to physical units. For
these reasons, we do not perform any transient correction during the
cleaning of the data cube.

Simulations were widely used to calibrate the translation of measured 
ADUs in `real' ADUs, then in $\mu$Jy and are detailed below.

\section{Simulations}

The PRETI method has been developed to detect faint sources in all ISOCAM 
data (if they are taken in fields where the background is more or less 
constant), but has to be tuned in order to work very efficiently, as 
shown in the previous section. This fine tuning has been done 
through various tests and simulations, that have proven to be very 
useful to ascertain detections, study the completeness, the number of 
false detections and to calibrate the source photometry. 

In order to test the PRETI method, we have tried to build simulated
data. These should be as near as possible of real ISOCAM data, that
is, contain glitches, transients, and sources. A special attention has
to be payed to the `fader' and `dipper' long term effects of glitches
since they are the most limiting factor for the faint source
detection. Unfortunately, no available model is able to reproduce
accurately these behaviours of the LW detector. The only solution is
to use real ISOCAM data to build simulations.

\subsection{cosmic rays in simulated data}

Cosmic ray effects are dependent of three parameters 
\begin{enumerate}
\item the integration time because it is directly related to the mean number of cosmic rays that will hit the detector per readout.
\item the background emission, because in the long term effect a cosmic 
ray may affect the gain of the pixels (though this is not yet clearly 
established).  
\item the period of the observation since the cosmic ray density varies along the orbit of the satellite 
as it crosses the radiation belts. We have noticed variations of 40~\% 
in the number of `faders' and `dippers' between various observations, 
early part of the orbit being most affected.    
\end{enumerate}

For these reasons, data for the simulation are taken with the same
integration time, same gain, same filter and zodiacal background as
the observation that is to be simulated. Moreover, in order to avoid
any modification of the time history of pixels, these data have to be
taken on periods at least as long as the observation. For example,
data taken on a shorter period and duplicated would cause
discontinuities in the time history and produce spurious sources.

\begin{figure*}
\centering
\begin{tabular}{cc}
\subfigure[Measured flux as a function of input flux for 2 photometry methods. Solid~:~2 pixel aperture photometry (3''). Dot-dashed~:~wavelet photometry. Dashed~:~the identity curve.]{\psfig{file=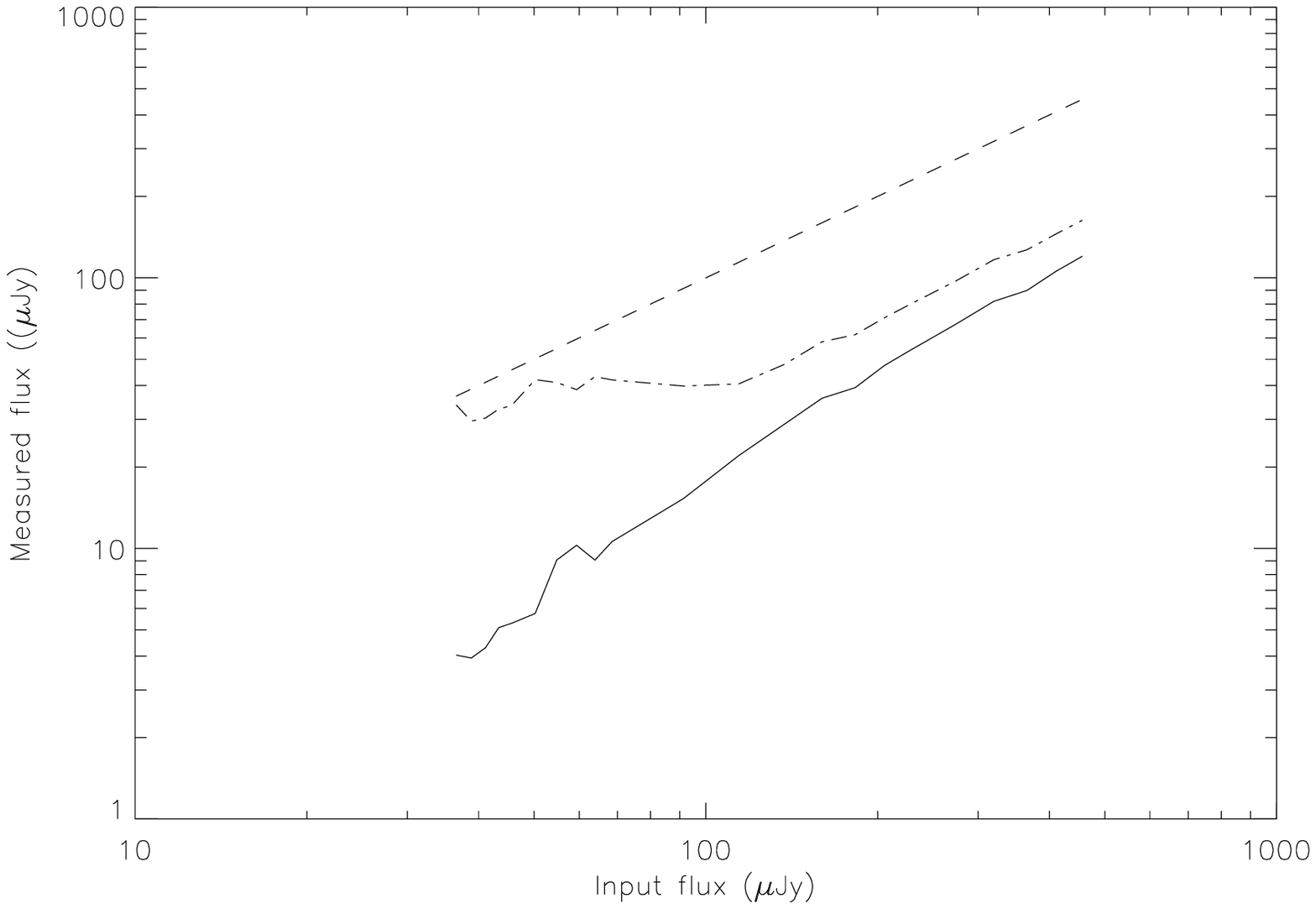,width=0.5\textwidth}} & \subfigure[Ratio of measured flux over real flux as a function of the real flux. Solid~:~2 pixel aperture photometry (3''). Dot-dashed~:~3 pixel aperture photometry (4.5''). Dashed~:~wavelet photometry.]{\psfig{file=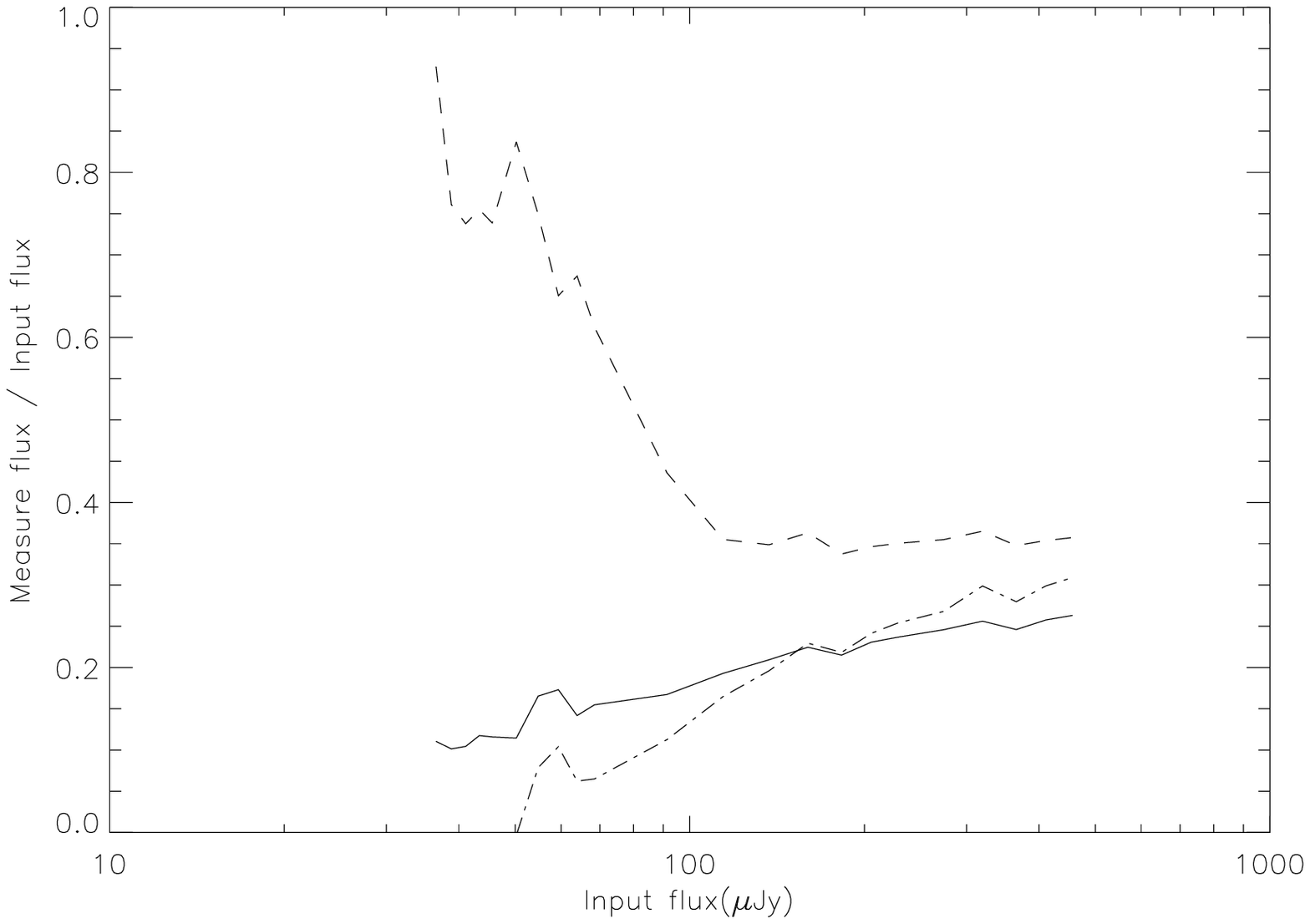,width=0.5\textwidth}} \\
\subfigure[Measured flux as a function of input flux for the wavelet method. Solid~:~measure. Dash-dotted~:~1$\sigma$ error on measure. Dotted~:~Identity.]{\psfig{file=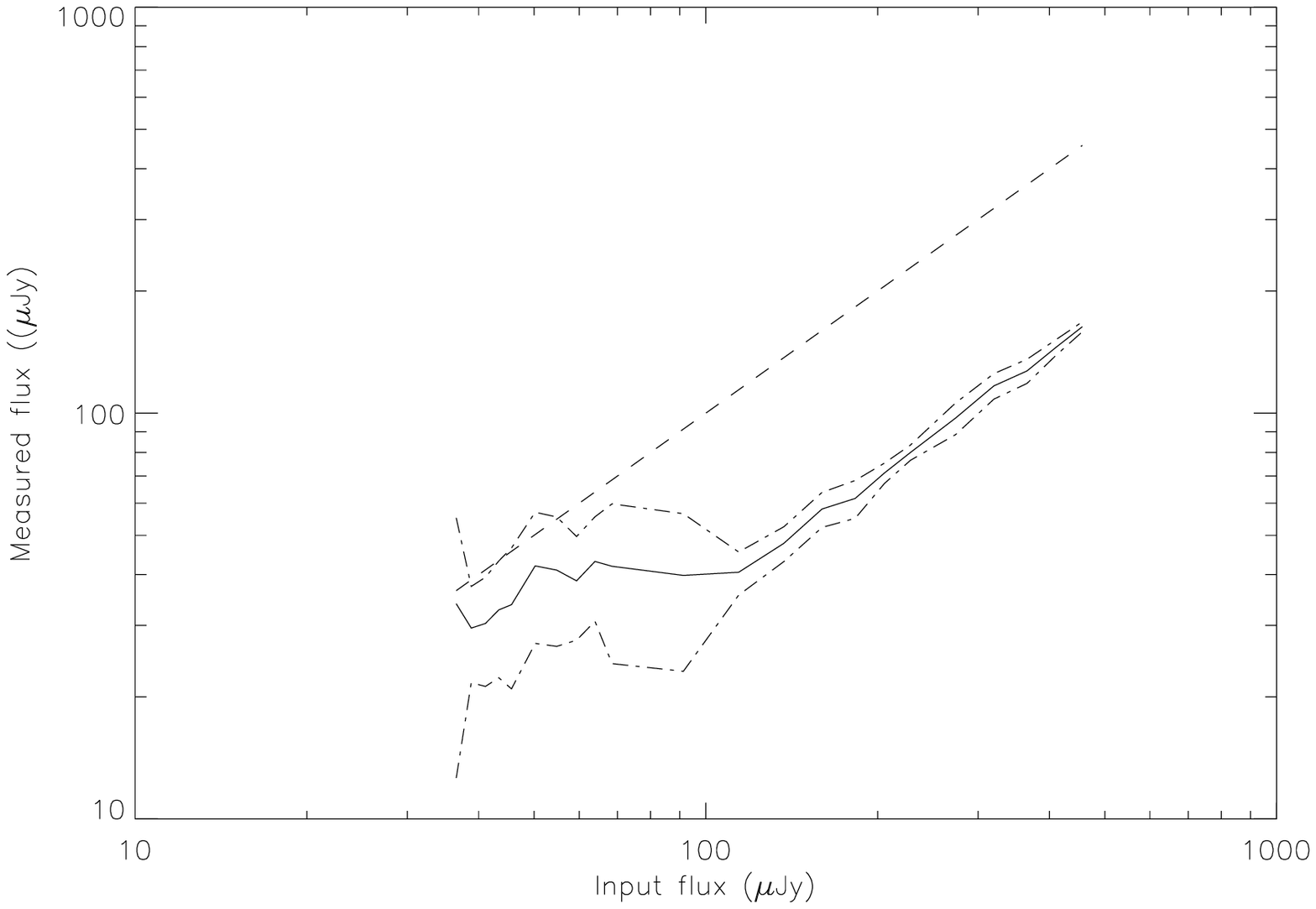,width=0.5\textwidth}} & \subfigure[Measured flux as a function of input flux for 2 pixel (3'') aperture photometry. Solid~:~measure. Dash-dotted~:~1$\sigma$ error on measure. Dotted~:~Identity.]{\psfig{file=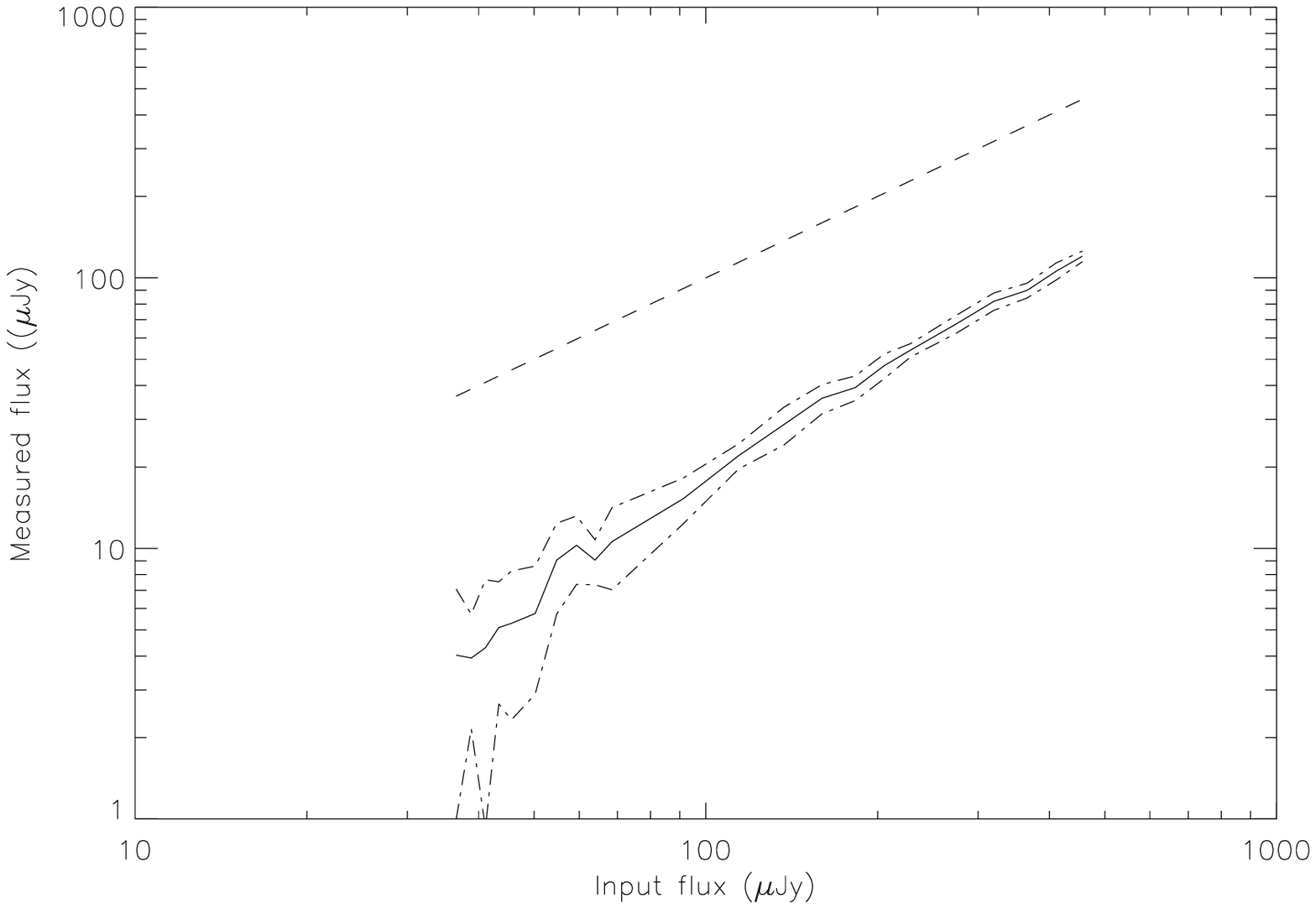,width=0.5\textwidth}} \\
\end{tabular}
\caption[]{Photometry results of LW2 simulations. Comparison of photometric methods and errors.}
\label{fig_sim_lw2}
\end{figure*}

\begin{figure*}
\centering
\begin{tabular}{cc}
\subfigure[Measured flux as a function of input flux for 2 photometry methods. Solid~:~2 pixels aperture photometry (3''). Dot-dashed~:~wavelet photometry. Dashed~:~the identity curve.]{\psfig{file=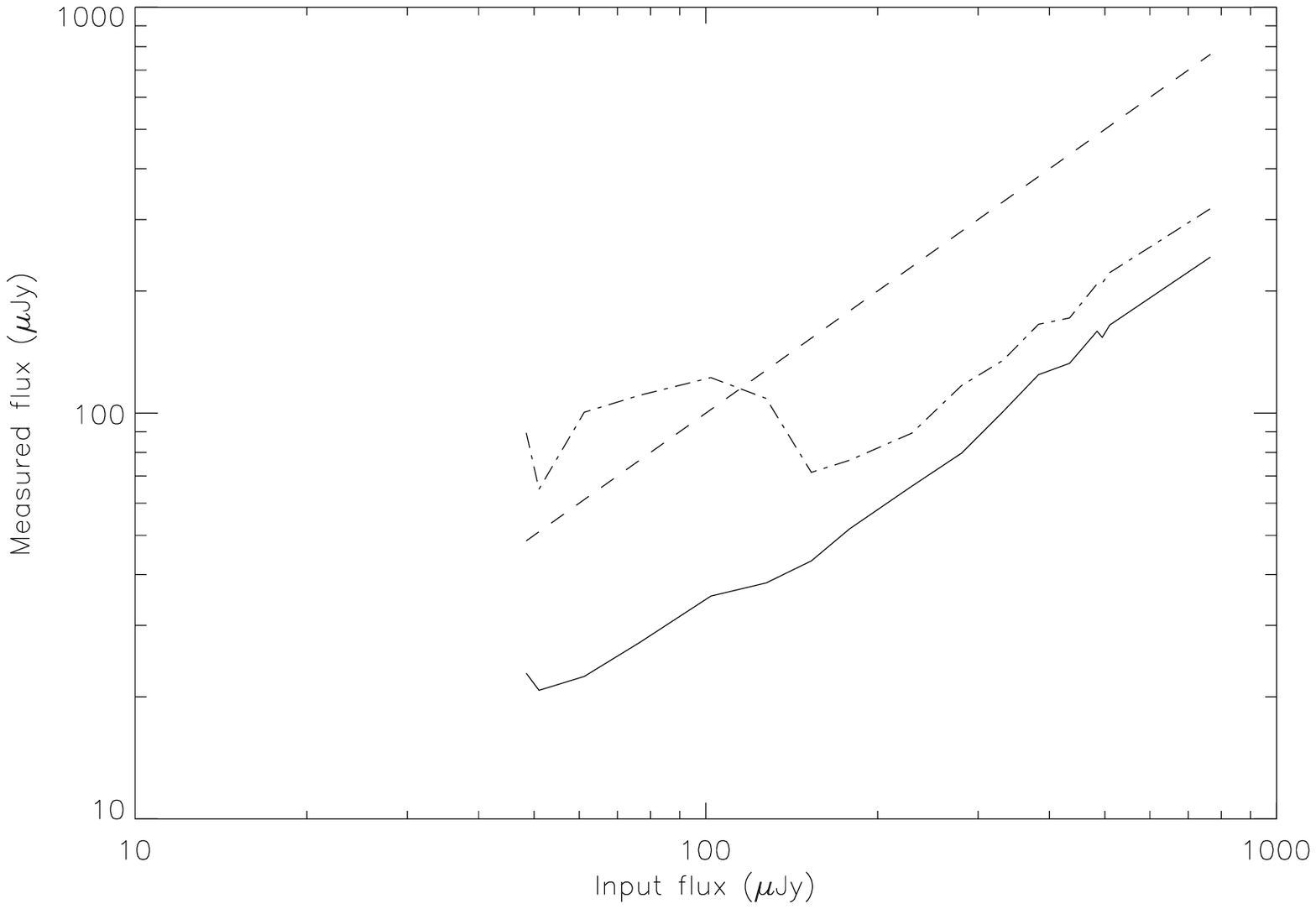,width=0.5\textwidth}} & \subfigure[Ratio of measured flux over real flux as a function of the real flux. Solid~:~2 pixel aperture photometry (3''). Dot-dashed~:~wavelet photometry.]{\psfig{file=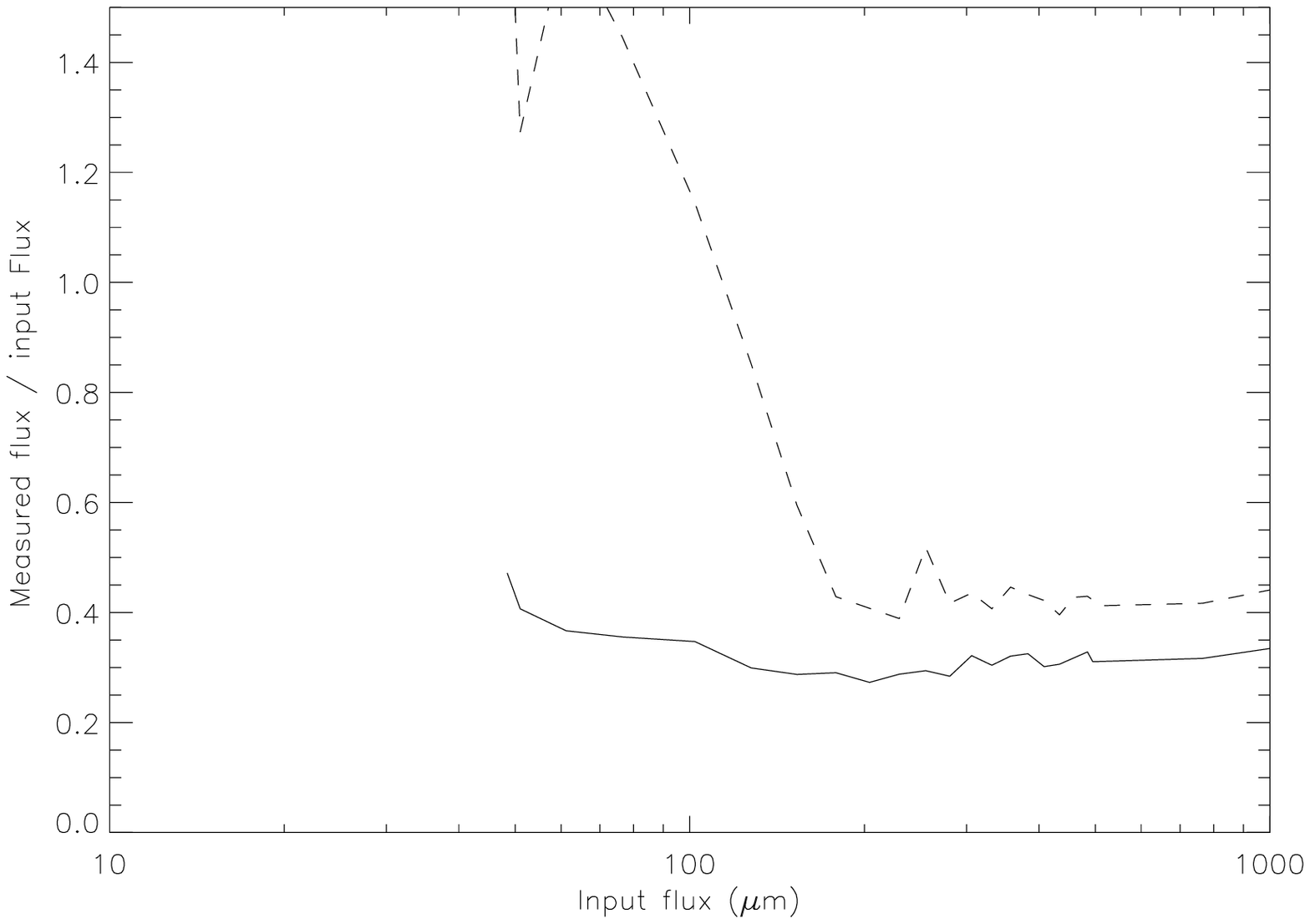,width=0.5\textwidth}} \\
\subfigure[Measured flux as a function of input flux for the wavelet method. Solid~:~measure. Dash-dotted~:~1$\sigma$ error on measure. Dotted~:~Identity.]{\psfig{file=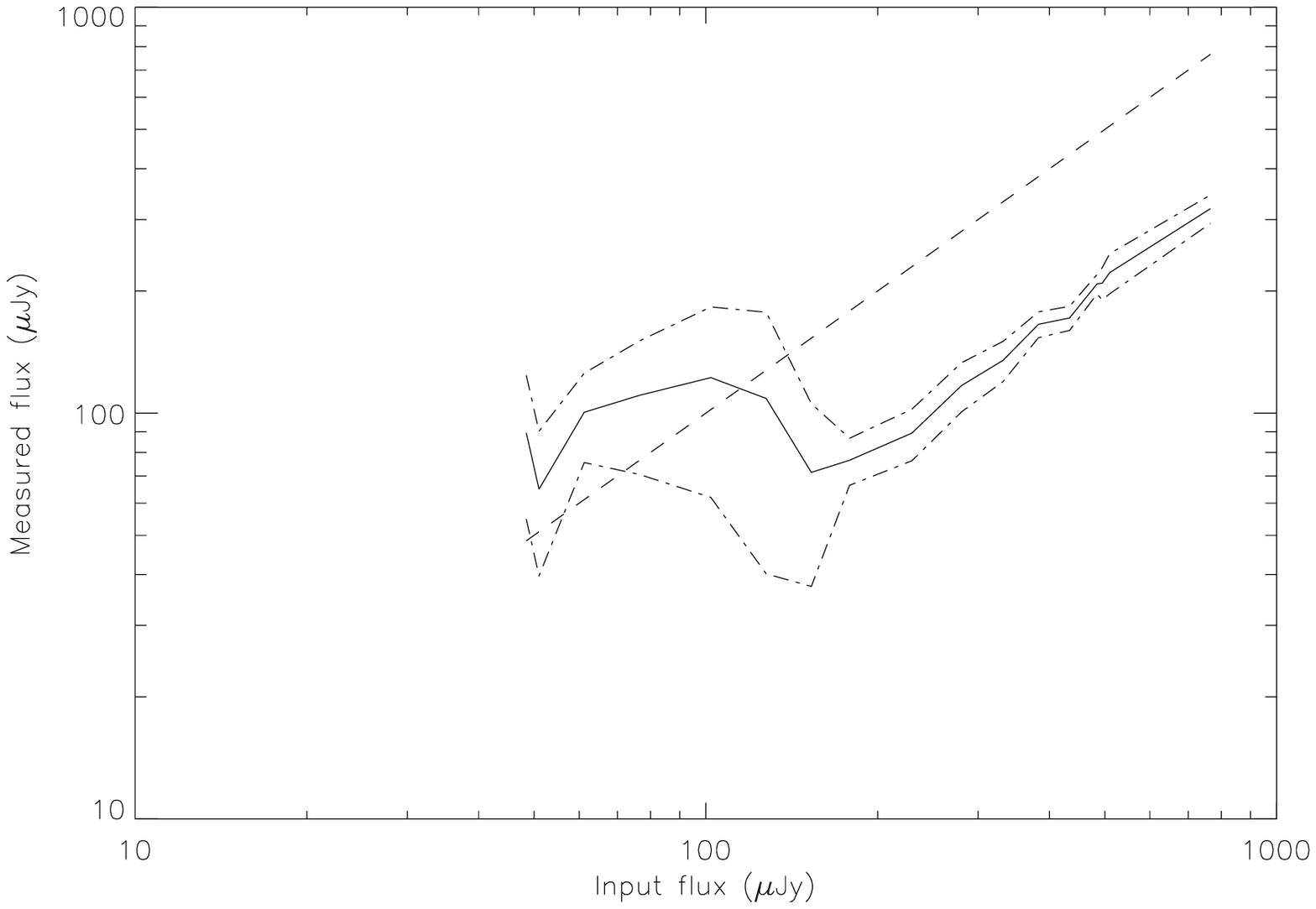,width=0.5\textwidth}} & \subfigure[Measured flux as a function of input flux for 2 pixels (3'') aperture photometry. Solid~:~measure. Dash-dotted~:~1$\sigma$ error on measure. Dotted~:~Identity.]{\psfig{file=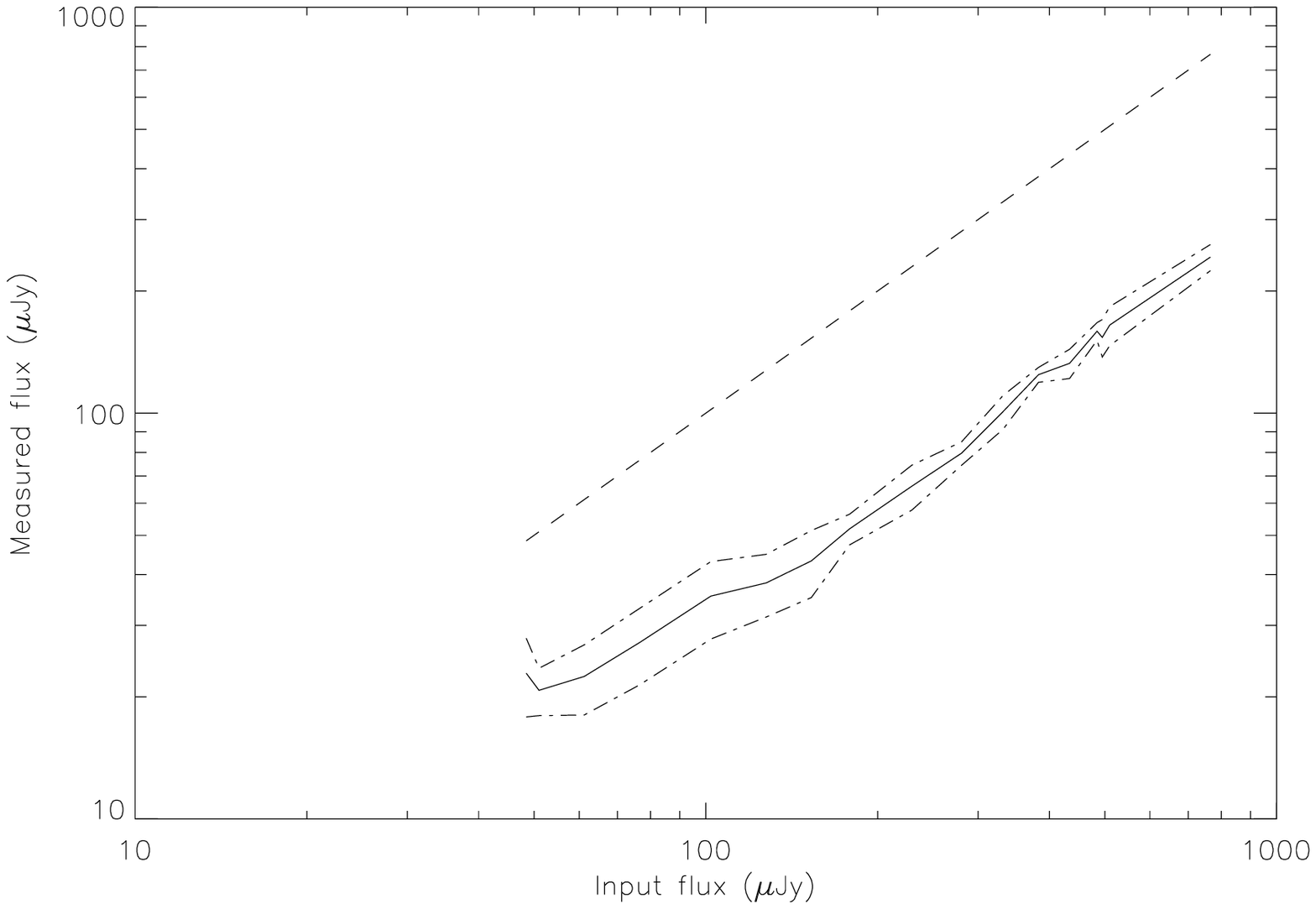,width=0.5\textwidth}} \\
\end{tabular}
\caption[]{Photometry results of LW3 simulations. Comparison of photometric methods and errors.}
\label{fig_sim_lw3}
\end{figure*}

\subsection{Suppression of real sources in the data set used for simulations.}
Since we look for faint sources, it is very likely that the
observations selected to be used for simulations will contain such
sources.
 
The ideal case is when these are `staring' observations, {\it i.e.} 
when the satellite points during a very long time at the same position. 
Since the flat-field is built directly from the observation, any 
source present in the field is cleaned during the operation.

One less ideal case is when the raster parameters of the observation
are different from the simulated data. the sources in the simulation
data set are blurred when the raster map is reconstructed with
different steps. If there are bright sources in the simulated data,
which would despite the blurring process still have a high S/N level,
these can be masked out.

\begin{figure*}
\begin{center}
\psfig{file=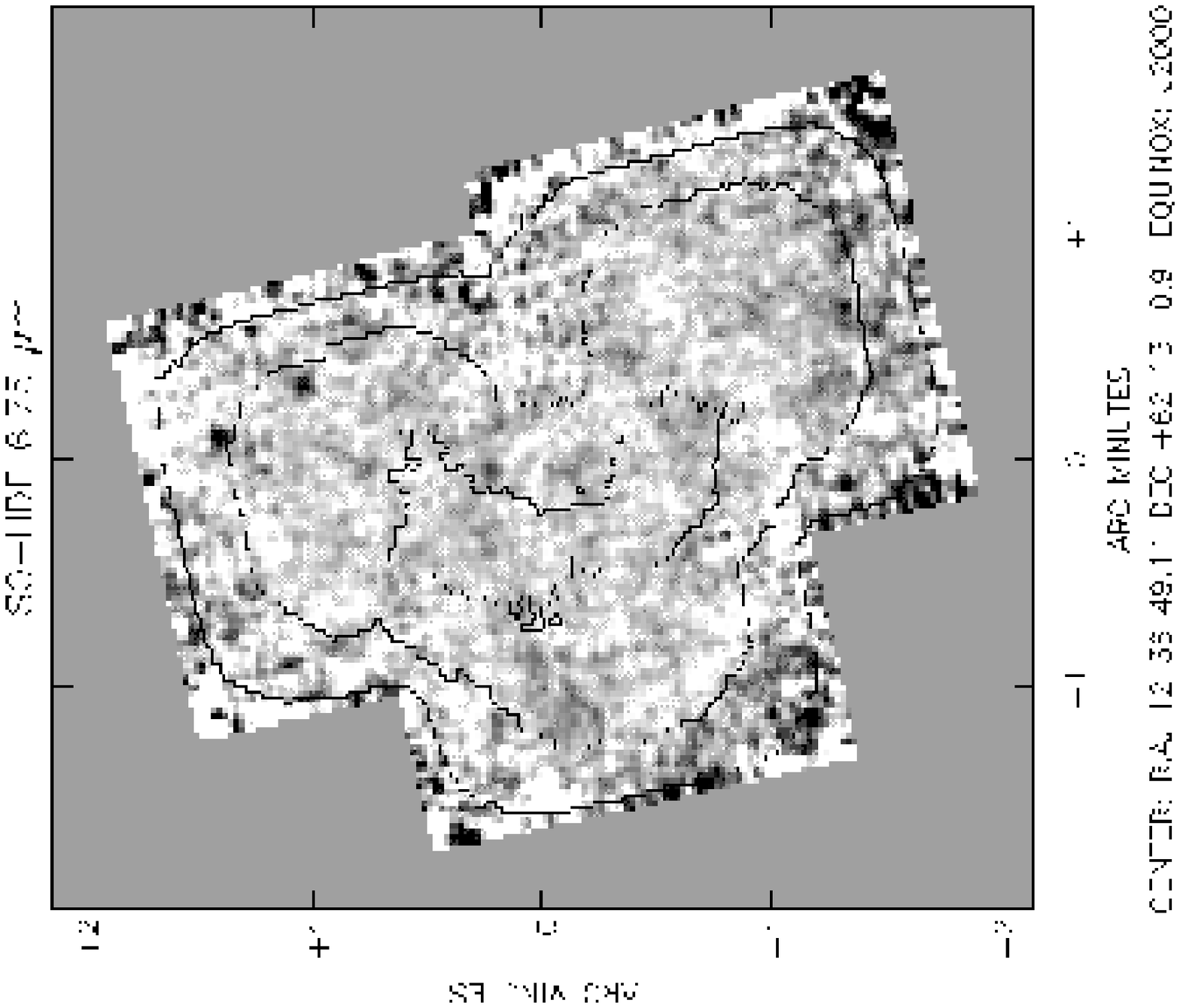,width=1.0\textwidth,angle=270}
\end{center}
\caption[]{Map of the ISO-HDF LW2 filter (6.75 $\mu$m). Resolution is of 1.5''/pixel. Contours are 5 $\tau_{w}$ detections levels at 35, 50 and 100$\mu$Jy}
\label{map_HDFlw2}
\end{figure*}

\begin{figure*}
\begin{center}
\psfig{file=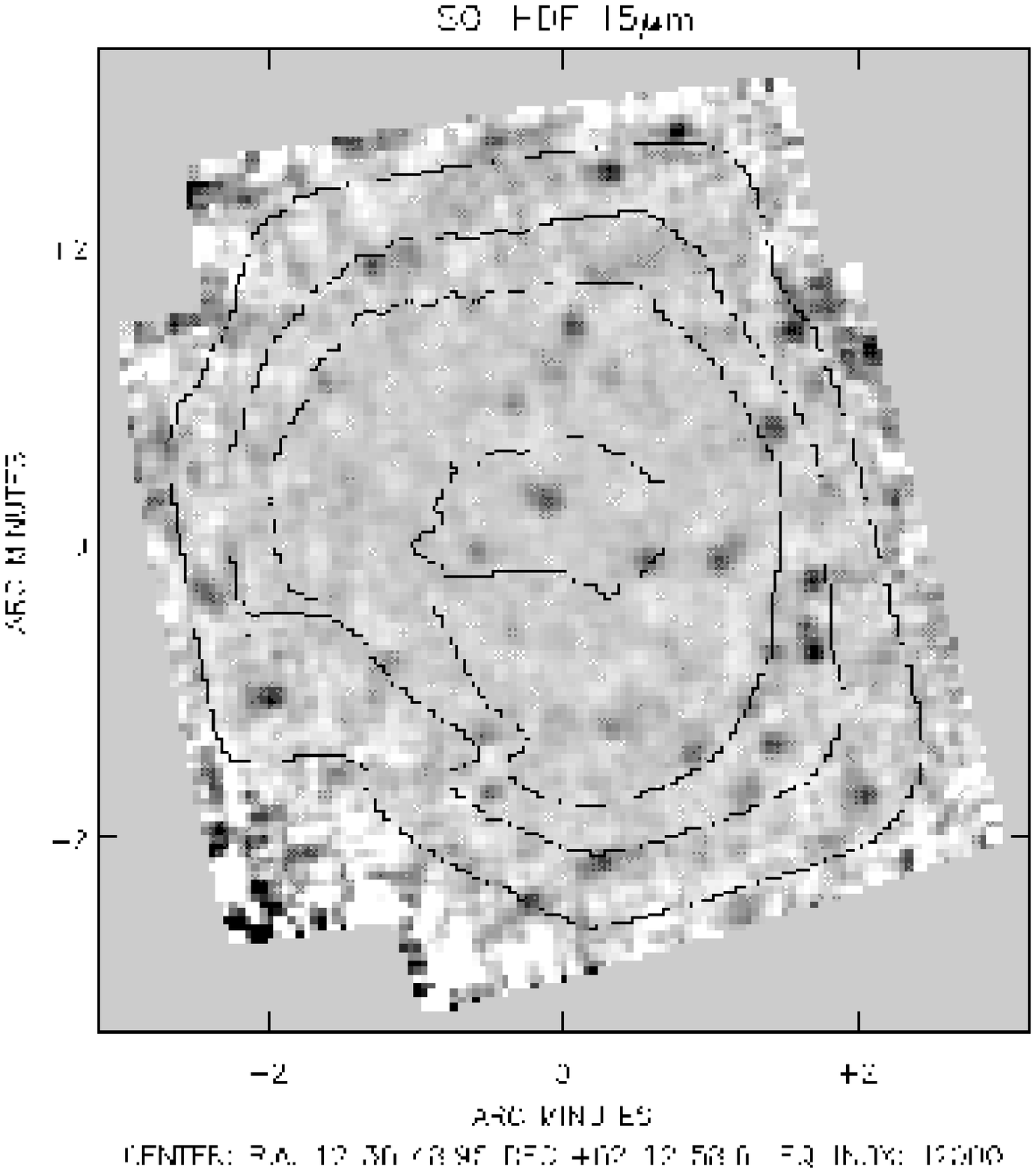,width=1.0\textwidth,angle=270}
\end{center}
\caption[]{Map of the ISO-HDF LW3 filter (15 $\mu$m). Resolution is of 3''/pixel. Contours are 5 $\tau_{w}$ detections levels at 35, 50, 100 and 200$\mu$Jy}
\label{map_HDFlw3}
\end{figure*}

\subsection{addition of sources}

Point sources of various fluxes are added using the PSF model of M.
P\'erault and K. Okumura, that match well the measurements (Okumura et
al.~\cite{Okumura}).  Each source flux is converted in ADU by using
the ISOCAM cookbook.  Object are randomly distributed on the raster
map, at a precision of a tenth of a pixel, then each position on the
detector array is recomputed, without taking the field distortion into
account.

Each raster image is then translated into a data cube by using the
transient model from Abergel et al.~(\cite{Abergel}), where the
detector response $D$ is a function of the incoming flux $I$ with~:
\begin{eqnarray*}
D(t) & = & r I(t) + (1-r)\int_{-\infty}^{t} I(t') \exp\left(-\frac{t-t'}{\alpha/I(t')}\right) dt'
\end{eqnarray*}
It is clear from this formula that the response depends on the 
intensity of the source and on the background, because the exponential 
time constant is a function of $I$. For this reason, we add to the 
simulated source images the median of the deglitched cube that is to 
be simulated before running the transient simulator. We subtract this 
image later, in order to have a simulated cube on a zero background, 
and add it to the raw data selected to perform the simulation.   

\subsection{ISO-HDF simulations}

We have searched the ISOCAM calibration database for 
observations matching the ISO-HDF configurations. For the LW3 
observation, we found a set of flat-field observations, called `FLAT16' 
that were done with the same filter, the same lens and same 
integration time. Seven bright sources present in the field were masked 
out. Fainter sources have been blurred since this observation is a raster with 
half an array of raster step (16 pixels). We checked the PRETI method 
by running it completely on this set. We set our deglitching 
parameters in order to avoid any detection in that field (since their 
should be no source), except the ones that could be related to a 
source present in the original dataset. We also checked that the 
`fader' and `dipper' rate is about the same in both observations.
This set of data is a perfect tool to build complete simulations. 
Since it is an empty field, we are able to study very carefully the 
level at which PRETI begins to give us false detections.

For the LW2 ISO-HDF, we could not find any satisfactory set of data~: 
only a few calibration have been performed with an integration time of 
10 s, and they are always short ones. We eliminated also data from our 
IDSPCO survey program that were matching all the configuration 
requirements, because we found a rate of long term effect cosmic rays 
about 20 \% higher~: this causes severe differences in the sensitivity 
that can be reached by an observation. Since our first reductions show 
us that the source density is not very high in the ISO-HDF LW2 
observation, we decided to use these data themselves for the 
simulations. The drawback is that this prevents us to fully study the 
confidence level and the false detections.

We performed two kinds of simulations. First, 10 sources of a given 
flux were randomly added, and the operation was repeated from 1 mJy 
to 100 $\mu$Jy by steps of 100 $\mu$Jy, and from 100 to 10 $\mu$Jy by 
steps of 10 $\mu$Jy. This allowed us to investigate our detection limits 
and to calibrate our source photometry. Finally, we performed 
simulations containing a wide range of fluxes, more or less matching 
number count models, in order to test the completeness of the method.

\subsection{Simulation results on the ISO-HDF}

Figure~\ref{fig_sim_counts} presents the results of our counts on simulated
data for LW2 (a) and LW3 (b). For the LW2 filter, all added sources in
the field are detected down to 60 $\mu$Jy for a 5 $\tau_{w}$
detection, showing that we are complete at this level. However, this
result does not take into account confusion. A faint source is not 
detected if is is too close to a bright one ( $\sim$ 9'' in LW3,
depending on the relative brightness of the two sources). This effect
has to be taken into account when deriving number counts. For the LW3
observation, the 99.9\% completion is obtained above 200 $\mu$Jy, for a
5$\tau_{w}$ detection and stays above 85\% at 100 $\mu$Jy. 7$\tau_{w}$
detection, that are more reliable in terms of avoiding false
detections, are only 99.9\% complete above 1 mJy, but stay of the order of
90 \% down to 220 $\mu$Jy. We could not measure the number of false
detections in the LW2 band since we used the data set itself to do the
simulations.  Since PRETI is run with the same arguments on both sets,
we assume that the difference should be small between the two filters
and tune our criteria according to LW3.

Figures~\ref{fig_sim_lw2} and \ref{fig_sim_lw3} show the calibrations
curves deduced from simulations, {\it i.e.} the relation between the
measured flux to the real flux that was simulated, for two measurement
methods : (1) an aperture photometry taking a radius of 3'' for LW2
and 6'' for LW3 (two pixels in each case), (2) the integrated flux of
the objects reconstructed by our wavelet based detection program
(Starck et al.~\cite{PRETI_moriond}). The two methods display very
different behaviors for the faint end of the plots.

Photometry on wavelet reconstructed objects (panels (c)) shows a
non-linearity effect for input fluxes below 100 $\mu$Jy in LW2 ({\it
resp.} 200 $\mu$Jy in LW3). Above these levels, the ratio of the
measured flux over the input flux is roughly constant ($\sim 0.45$ for
LW2). We interpret this factor in terms of the effect of transients
(factor of 0.6) , and by the fact that only the central core of the
PSF is reconstructed by the algorithm (factor of 0.8), thus giving a
theoretical correction of 0.48. Below these levels, the non-linearity
is explained by the fact that the wavelet analysis program
reconstructs objects with extended low levels wings. This effect might
be due to the fact that we use a b-spline wavelet transform of our map
that does not match closely enough the ISOCAM PSF. We note that the
non-linearity is more marked for LW3 than for LW2 where the PSF is less
extended, and is therefore closer to the wavelet shape. For the last
filter, the measured level stays constant at 30 $\mu$Jy while it rises
in the first one for decreasing fluxes. The reason for this difference
is not clear, but might be explained by the confusion of the the
observations in the LW3 map, or by residual glitches\footnote{the only
way to test this hypothesis is to achieve clean simulations of empty
fields !}. This non-linearity prevents to use the ``wavelet''
photometry for low levels objects, thus this method is not appropriate
for the HDF data set.

Aperture photometry (panels (d)) does not display such
non-linearity. In LW2 , the ratio of the measured flux over the input
flux decreases as the input flux decreases. The explanation for this
behaviour is the following~: when the flux decreases, the outer pixels
of the central core of the PSF are more and more dominated by the
noise, and the summed flux contains therefore negative terms. Indeed,
due to the baseline substraction, a few readouts are slightly negative
before and after the source, (the background level is
overestimated). Therefore, the average level on the final map around
the source is slightly negative. This effect is illustrated in
figure~\ref{fig_sim_lw2} (b) were a 3 pixel radius aperture photometry
has also been drawn (dash-dotted curve). For this method, the ratio at
high flux is larger than for a 2 pixel radius aperture (more of the
central core is taken into account), but decreases more abruptly
toward faint fluxes, because more noise and negative values are
integrated in the aperture. We note that at high flux, the ratio
between the measured flux in the 3 pixels and 2 pixels apertures is
$\sim 0.3/0.25 = 1.2$, in good agreement with the PSF model that
predicts 1.209. Again for this photometric method, the behaviour of the
LW3 filter is slightly different, because the slope of the calibration
curve is less steep toward faint fluxes. The change of slope occurs at
the same level (200 $\mu$Jy) where the ``wavelet'' method begin to be
non-linear. This indicates that the same process is at work for both
methods, but that aperture photometry is less affected. For these
reasons, we have used a two pixel aperture photometry for the HDF.

\section{Results on the ISO-HDF}

Figures~\ref{map_HDFlw2} and \ref{map_HDFlw3} present the reduced map
of the ISO-HDF LW2 and LW3. Many sources are clearly visible on the
latter, while the former appears almost empty. Various sensitivity
levels have been overlayed on the maps, showing that the noise is very
inhomogeneous. These sensitivity levels were computed by comparing the
noise level at each scale with the wavelet coefficients of a point
source image, the source being centered on a pixel. Thus, these levels
give only an indication. If a source fell between two pixels, its
wavelet coefficients at lower scales (high spatial frequencies) is
lower, while those at higher scales (low spatial frequencies) are
higher. Since the noise decreases when considering higher scales
(Starck et al.~\cite{PRETI_moriond}), sources can in fact be detected
at fluxes lower that the levels shown on figures
\ref{map_HDFlw2} and \ref{map_HDFlw3}.

\begin{figure}
\psfig{file=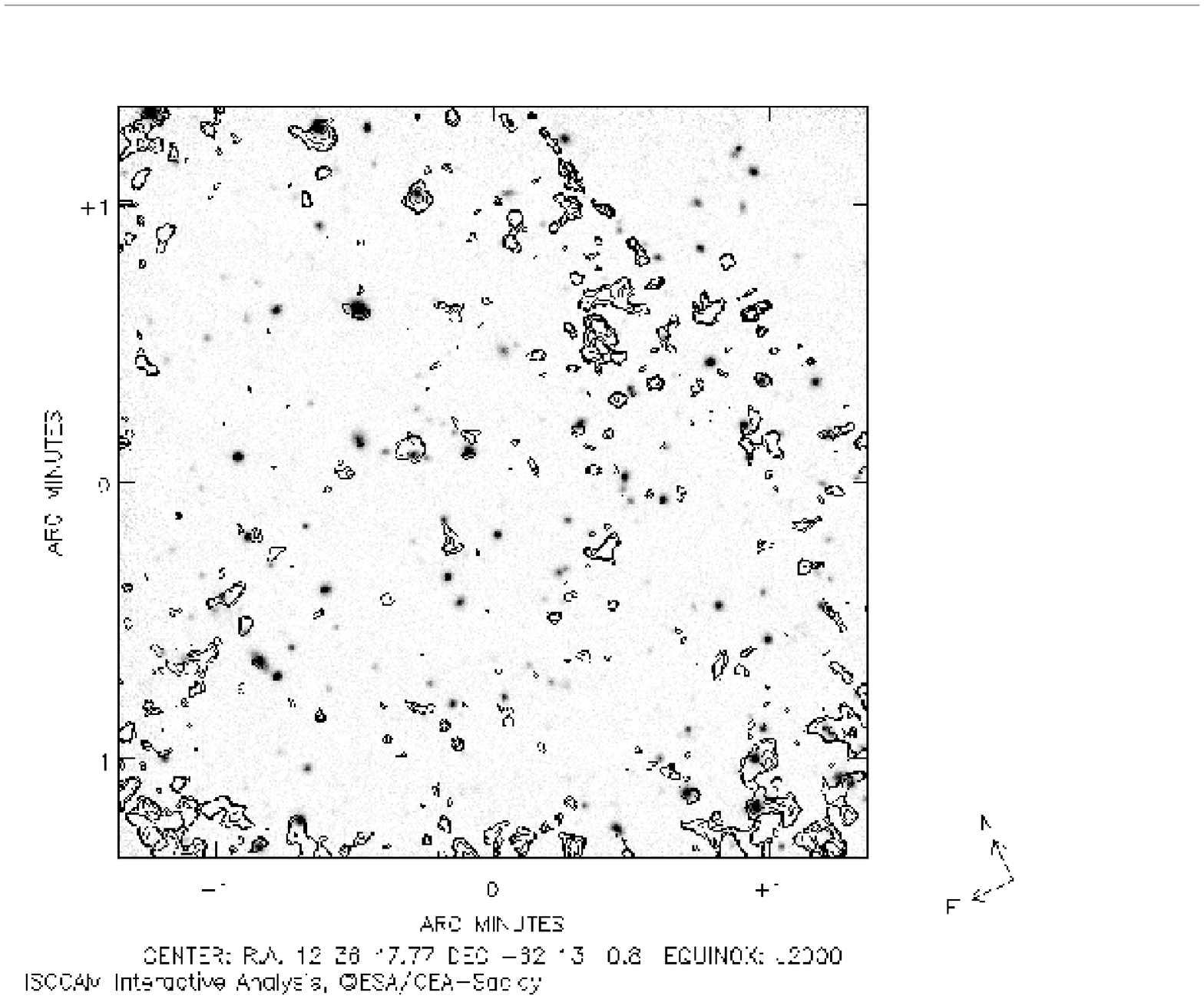,bbllx=60pt,bblly=110pt,bburx=559pt,bbury=650pt,width=0.5\textwidth,angle=270}
\caption[]{Comparison near-IR/LW2 filter. The contours of the central part of the LW2 map have been overlayed on the IRIM K map. Contours levels begin at  1.0 $\mu$Jy with steps of 2 $\mu$Jy. }
\label{irim_HDFlw2}
\end{figure}

\begin{figure}
\psfig{file=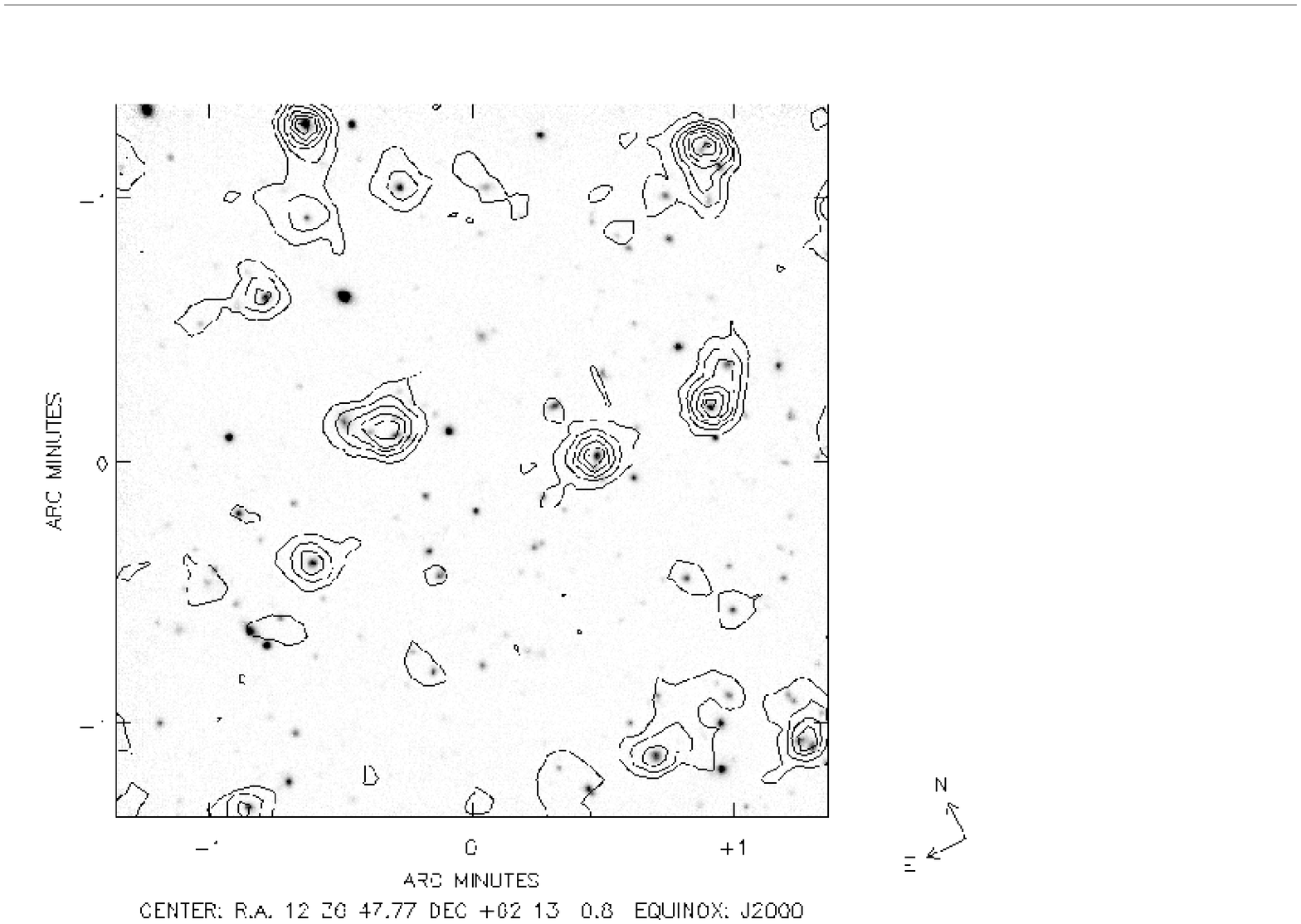,angle=90,bbllx=66pt,bblly=217pt,bburx=518pt,bbury=719pt,width=0.5\textwidth}
\caption[]{Comparison near-IR/LW3 filter. The contours of the central part of the LW3 map have been overlayed on the IRIM K map. Contour levels begin at 1.5 $\mu$Jy with steps of 2 $\mu$Jy. }
\label{irim_HDFlw3}
\end{figure}

\begin{table*}
\caption{PRETI main source list for the ISOHDF observations at $6.5\mu$m (LW2) and $15\mu$m (LW3). All sources are detected at $7\tau_{w}$ above readout and photon noise. (1)~:  Source identification `PM3' is for PRETI Main list LW3, `PM2' for PRETI Main list LW2 . (2),(3)~: source position for the J2000.0 equinox. (4)~: source flux in microJansky in the LW2 filter with error bar or upper limit ($5\tau_{w}$). A `-' indicates that the source is not in the LW2 field. (5)~: source flux in ADU/G/s in a 3'' aperture. (6)~: detection level ($n\tau_{w}$) in the LW2 image.(7)~: source flux in microJansky in the LW3 filter with error bar or upper limit ($5\tau_{w}$). (8)~: source flux in ADU/G/s in a 6'' aperture. (9)~: detection level ($n\tau_{w}$) in the LW3 image. (10)~: Identificator of the HST field in which the source falls (hd stands for HDF, others for flanking fields). (11)~: spectroscopic redshift with source indicated by a note. (c)~: data from Cohen et al.~(\cite{Cohen}), (p)~: data from Phillips et al.(\cite{Phillips}), (i)~: compiled in Cowie et al.~(\cite{Cowie}), (h) data from Hogg et al.~(\cite{Hogg98}). (12)~: morphological type of the object. '(i)' indicates a type compiled in Cowie et al.~(\cite{Cowie}),'(p)' type from Phillips et al.~(\cite{Phillips}), otherwise determinated by us. E : Elliptical or S0 galaxy. S : spiral. M : merger. P : peculiar/irregular, C : compact object from Phillips et al.~(\cite{Phillips}) sample, A : active nuclei galaxy, G : source falls on a group. U : unknown type. N : no optical counterpart. A '?' indicate a dubious type identification.}\label{liste_sures}    
\begin{tabular}{|c|c|c|c|c|c|c|c|c|c|c|c|}
\hline
Id$^{(1)}$& $\alpha^{(2)}$  & $\delta^{(3)}$ &\hspace{0.01cm} LW2$^{(4)}$\hspace{0.01cm}  & LW2$^{(5)}$  & N$\tau_{w}$ & \hspace{0.01cm}LW3$^{(7)}$\hspace{0.01cm}     &LW3$^{(8)}$ & N$\tau_{w}$&F&\hspace{0.1cm} z$^{(11)}$\hspace{0.1cm} & T$^{(12)}$\\
  &  (J2000)          &    (J2000)                 &$\mu$Jy & {\tiny ADU/G/s} &$^{(6)} $&$\mu$Jy & {\tiny ADU/G/s}   & $^{(9)}$   &   $^{(10)}$    &   &      \\
\hline
HDF\_PM3\_1& 12 36 31.4 & +62 11 15 & -      & - & - &$355^{+40}_{-60}$ & 0.211 & 7 & sw & -         & P  \\
HDF\_PM3\_2& 12 36 34.4 & +62 12 12 & -      & - & - &$448^{+68}_{-59}$ & 0.284 & 7 & sw & -         & M  \\
HDF\_PM3\_3& 12 36 34.4 & +62 12 42 & -      & - & - &$363^{+79}_{-38}$ & 0.227 & 7 & iw & 1.219 $^{(c)}$ & P\\
HDF\_PM3\_4& 12 36 34.8 & +62 12 24 & -      & - & - &$267^{+59}_{-62}$ & 0.155 & 7 & sw & 0.562 $^{(c)}$ & S\\
HDF\_PM3\_5& 12 36 35.6 & +62 14 24 & -      & - & - &$441^{+43}_{-82}$ & 0.264 & 7 & iw & -         & S   \\
HDF\_PM3\_6& 12 36 36.5 & +62 13 48 & -      & - & - &$353^{+40}_{-66}$ & 0.210 & 7 & iw & 0.960 $^{(p)}$ & CA$^{(p)}$ \\
HDF\_PM3\_7& 12 36 36.9 & +62 11 36 & $< 135$& - & - &$300^{+62}_{-67}$ & 0.115 & 7 & sw & 0.078 $^{(c)}$ & S   \\
HDF\_PM3\_8& 12 36 36.9 & +62 12 15 & $< 113$& - & - &$202^{+58}_{-50}$ & 0.116 & 7 & sw & -         & S   \\
HDF\_PM3\_9& 12 36 38.2 & +62 11 18 & -      & - & - &$212^{+58}_{-55}$ & 0.120 & 7 & sw & -         & G     \\
HDF\_PM3\_10& 12 36 38.2 & +62 11 51 & $< 61 $& - & - &$212^{+58}_{-55}$ & 0.119 & 7 & sw & -         & U   \\
HDF\_PM3\_11& 12 36 39.9 & +62 12 51 & $< 64 $& - & - &$302^{+67}_{-55}$ & 0.180 & 7 & iw & -         & E    \\
HDF\_PM2\_1& 12 36 40.4 & +62 11 41 &$127^{+99}_{-61}$ & 0.021 & 7 &  $<72$ & -   & - & hd & 0.585 $^{(i)}$ & E    \\
HDF\_PM3\_12& 12 36 41.2 & +62 11 33 & $< 66 $& - &- & $236^{+61}_{-58}$ & 0.133 & 7 & hd & 0.089 $^{(c)}$ & P  \\
HDF\_PM3\_13& 12 36 41.2 & +62 14 21 & -      & - &- & $ 76^{+71}_{-19}$ & 0.052 & 7 & iw & -         & P  \\
HDF\_PM3\_14& 12 36 41.6 & +62 12 06 & $< 37 $& - &- & $ 52^{+34}_{-09}$ & 0.042 & 7 & hd & 0.432 $^{(c)}$ & S$^{(i)}$ \\
HDF\_PM3\_15& 12 36 43.0 & +62 12 18 & $< 37 $& - &- & $ 49^{+36}_{-09}$ & 0.036 & 7 & hd & 0.454 $^{(c)}$ & S$^{(i)}$  \\
HDF\_PM2\_2& 12 36 43.4 & +62 11 42 &$185^{+99}_{-25}$&0.028& 7 &$<66$   &   -   & - & hd & 0.764 $^{(c)}$ & E$^{(i)}$ \\ 
HDF\_PM3\_16& 12 36 43.8 & +62 14 00 & $< 68 $& - & - &$ 50^{+33}_{-09}$ & 0.040 & 7 & iw & 0.201 $^{(c)}$ & S    \\
HDF\_PM3\_17& 12 36 44.2 & +62 12 51 & $< 50 $& - & - &$282^{+60}_{-64}$ & 0.163 & 7 & hd & 0.557 $^{(c)}$ & M$^{(i)}$ \\
HDF\_PM3\_18& 12 36 44.7 & +62 14 51 & $< 329$& - & - &$105^{+94}_{-21}$ & 0.076 & 7 & nw &  -        & N      \\
HDF\_PM3\_19& 12 36 46.4 & +62 11 42 & $88^{+45}_{-80}$&0.009& 5 & $170^{+59}_{-42}$ & 0.097 & 7 & hd & 1.016 $^{(c)}$ & S$^{(i)}$    \\
HDF\_PM3\_20& 12 36 46.4 & +62 14 06 &$191^{+41}_{-86}$&0.032& 7 &$107^{+95}_{-20}$ & 0.078 & 7 & hd & 0.960 $^{(c)}$ & S$^{(i)}$ \\
HDF\_PM3\_21& 12 36 46.4 & +62 15 30 & -      & - &- & $418^{+91}_{-94}$ & 0.250 & 7 & nw & 0.851 $^{(p)}$ & M    \\
HDF\_PM3\_22& 12 36 46.8 & +62 10 48 & -      & - &- & $327^{+39}_{-63}$ & 0.195 & 7 & se & -         & N      \\
HDF\_PM3\_23& 12 36 47.2 & +62 14 48 & $< 179$& - &- & $144^{+72}_{-47}$ & 0.084 & 7 & nw & -         & N      \\
HDF\_PM3\_24& 12 36 48.5 & +62 14 27 &$254^{+71}_{-73}$&0.051& 7 & $307^{+62}_{-67}$ & 0.177 & 7 & nw & -     & E    \\
HDF\_PM3\_25& 12 36 49.3 & +62 11 48 & $< 53 $& - & -& $ 75^{+71}_{-19}$ & 0.052 & 7 & hd & 0.961$^{(h)}$ & S    \\
HDF\_PM2\_3& 12 36 49.4 & +62 13 47 &$41^{+66}_{-30}$ &0.005& 7 &$< 52$ & -  & - & hd & 0.089 $^{(c)}$ & E$^{(i)}$    \\ 
HDF\_PM3\_26& 12 36 49.4 & +62 14 06 & $< 40 $& - & -&$150^{+74}_{-48}$ & 0.087 & 7 & hd & 0.752 $^{(c)}$ & S?$^{(i)}$ \\
HDF\_PM3\_27& 12 36 49.8 & +62 13 15 &$136^{+68}_{-57}$&0.019& 7 &$320^{+39}_{-62}$ & 0.191 & 7 & hd & -         & G  \\
HDF\_PM3\_28& 12 36 51.1 & +62 10 30 & -      & - & - &$341^{+40}_{-65}$ & 0.204 & 7 & se & 0.410 $^{(i)}$ & P    \\
HDF\_PM3\_29& 12 36 51.9 & +62 12 21 & $< 36 $& - &- & $ 48^{+32}_{-09}$ & 0.038 & 7 & hd & 0.299 $^{(c)}$ & S?$^{(i)}$\\
HDF\_PM3\_30& 12 36 51.9 & +62 13 54 & $< 39 $& - &- & $151^{+74}_{-68}$& 0.088 & 7 & hd & 0.557 $^{(c)}$ & M$^{(i)}$ \\
HDF\_PM3\_31& 12 36 53.2 & +62 11 15 & -      & - &- & $174^{+59}_{-43}$& 0.099 & 7 & se &   -       &P?G?\\
HDF\_PM3\_32& 12 36 53.7 & +62 11 39 & -      & - &- & $180^{+60}_{-43}$ & 0.103 & 7 & ie & -         & G  \\
HDF\_PM3\_33& 12 36 54.1 & +62 12 54 & $< 36 $& - &- & $179^{+60}_{-43}$ & 0.102 & 7 & hd & 0.642 $^{(c)}$ & S$^{(i)}$ \\
HDF\_PM3\_34& 12 36 54.1 & +62 13 57 & $< 42 $& - &- & $ 47^{+31}_{-09}$ & 0.038 & 7 & hd & 0.849 $^{(i)}$& S?$^{(i)}$ \\
HDF\_PM3\_35& 12 36 57.1 & +62 11 27 & -      & - &- & $ 75^{+71}_{-19}$ & 0.052 & 7 & ie & -         & N      \\
HDF\_PM3\_36& 12 36 57.5 & +62 13 00 & $< 38 $& - &- & $ 60^{+40}_{-23}$ & 0.043 & 7 & hd & 0.474 $^{(c)}$ & M?$^{(i)}$ \\
HDF\_PM3\_37& 12 36 58.0 & +62 14 54 & -      & - &- & $225^{+60}_{-56}$ & 0.127 & 7 & ne & -         & N      \\
HDF\_PM3\_38& 12 36 58.8 & +62 14 24 & $< 187$& - &- & $ 52^{+34}_{-09}$ & 0.041 & 7 & ne & -         & N      \\
HDF\_PM3\_39& 12 36 59.2 & +62 12 09 & $< 66 $& - &- & $157^{+75}_{-49}$ & 0.091 & 7 & ie & -         & P    \\
HDF\_PM3\_40& 12 37 00.1 & +62 14 51 & -      & - &- & $295^{+61}_{-66}$ & 0.170 & 7 & ne & 0.761 $^{(c)}$ & M    \\
\hline
\multicolumn{12}{|l|}{Continued on next page...}\\
\hline
\end{tabular}
\end{table*}

\begin{table*}[t]
\begin{tabular}{|c|c|c|c|c|c|c|c|c|c|c|c|}
\hline
\multicolumn{12}{|l|}{Continued from previous page...}\\
\hline
Id$^{(1)}$& $\alpha^{(2)}$  & $\delta^{(3)}$ &\hspace{0.01cm} LW2$^{(4)}$\hspace{0.01cm}  & LW2$^{(5)}$  & N$\tau_{w}$ & \hspace{0.01cm}LW3$^{(7)}$\hspace{0.01cm}     &LW3$^{(8)}$ & N$\tau_{w}$&F&\hspace{0.1cm} z$^{(11)}$\hspace{0.1cm} & T$^{(12)}$\\
  &  (J2000)          &    (J2000)                 &$\mu$Jy & {\tiny ADU/G/s} &$^{(6)}$ &$\mu$Jy & {\tiny ADU/G/s}   & $^{(9)}$   &    $^{(10)}$   &   &      \\
\hline
HDF\_PM3\_41& 12 37 01.8 & +62 13 24 & $< 97 $& - &- & $ 83^{+76}_{-19}$ & 0.058 & 7 & ie & 0.410 $^{(c)}$ & P    \\
HDF\_PM3\_42& 12 37 02.2 & +62 11 24 & -     & - &- & $162^{+76}_{-49}$ & 0.095 & 7 & ie & 0.136 $^{(p)}$ & S    \\
HDF\_PM3\_43& 12 37 03.1 & +62 12 18 & $< 296$& - &- & $ 49^{+74}_{-30}$ & 0.045 & 7 & ie & -         & N      \\
HDF\_PM3\_44& 12 37 03.1 & +62 14 03 & -     & - & - &$144^{+73}_{-47}$ & 0.084 & 7 & ne & 1.242 $^{(c)}$ & P    \\
HDF\_PM3\_45& 12 37 06.1 & +62 11 54 & -     & - & - &$431^{+43}_{-80}$ & 0.258 & 7 & ie & 0.904 $^{(p)}$ & S    \\
HDF\_PM3\_46& 12 37 06.5 & +62 13 33 & -     & - & - &$ 86^{+78}_{-20}$ & 0.060 & 7 & ie & 0.753 $^{(c)}$ & U   \\
\hline
\end{tabular}
\end{table*}

\begin{table*}[t]
\caption{PRETI supplementary source list for the ISOHDF observations at $6.5\mu$m (LW2) and $15\mu$m (LW3). All LW3 sources are at least detected at $3\tau_{w}$ above readout and photon noise, while all LW2 sources are detected atleast above $5\tau_{w}$ . (1)~:  Source identification `PS3' is for PRETI supplementary list LW3 and `PS2' for LW2. Sources with letter at the end, like `6a' are deblended neighbours of the source with same number in the main list. (2),(3)~: source position for the J2000.0 equinox. (4)~: source flux in microJansky in the LW2 filter with error bar or upper limit ($5\tau_{w}$). A `-' indicates that the source is not in the LW2 field. (5)~: source flux in ADU/G/s in a 3'' aperture. (6)~: detection level ($n\tau_{w}$) in the LW2 image.(7)~: source flux in microJansky in the LW3 filter with error bar or upper limit ($5\tau_{w}$). (8)~: source flux in ADU/G/s in a 6'' aperture. (9)~: detection level ($n\tau_{w}$) in the LW3 image. (10)~: Identificator of the HST field in which the source falls (hd stands for HDF, others for flanking fields,'--' means outside of all HST fields). (11)~: spectroscopic redshift with source incated by a note. (c)~: data from Cohen et al.~(\cite{Cohen}), (p)~: data from Phillips et al.~(\cite{Phillips}), (i)~: compiled in Cowie et al.~(\cite{Cowie}), (h) data from Hogg et al.~(\cite{Hogg98}). (12)~: morphological type of the object. '(i)' indicates a type compiled in Cowie et al.~(\cite{Cowie}),'(p)' type from Phillips et al.~(\cite{Phillips}), otherwise determinated by us. E : Elliptical or S0 galaxy. S : spiral. M : merger. P : peculiar/irregular, C : compact object from Phillips et al.~(\cite{Phillips}) sample, A : active nuclei galaxy, G : source falls on a group. U : unknown type. N : no optical counterpart. K : counterpart only in IRIM K image. st : star-like object . A '?' indicates a dubious type identification.}\label{liste_supp}    
\begin{tabular}{|c|c|c|c|c|c|c|c|c|c|c|c|}
\hline
Id$^{(1)}$& $\alpha^{(2)}$  & $\delta^{(3)}$ &\hspace{0.01cm} LW2$^{(4)}$\hspace{0.01cm}  & LW2$^{(5)}$  & N$\tau_{w}$ & \hspace{0.01cm}LW3$^{(7)}$\hspace{0.01cm}     &LW3$^{(8)}$ & N$\tau_{w}$&F&\hspace{0.1cm} z$^{(11)}$\hspace{0.1cm} & T$^{(12)}$\\
  &  (J2000)          &    (J2000)                 &$\mu$Jy & {\tiny ADU/G/s} & $^{(6)}$&$\mu$Jy & {\tiny ADU/G/s}   &  $^{(9)}$  &   $^{(10)}$    &   &      \\
\hline
HDF\_PS3\_1& 12 36 30.5 & +62 12 09 & -     & - & - &$145^{+73}_{-48}$ & 0.085 & 3 & sw & 0.456 $^{(c)}$ & E?      \\
HDF\_PS3\_2& 12 36 32.2 & +62 11 33 & -     & - & - & $ 44^{+28}_{-09}$ & 0.035 & 4 & sw & -         & N      \\
HDF\_PS3\_3& 12 36 33.5 & +62 13 21 & -     & - & - & $122^{+54}_{-40}$ & 0.069 & 4 & iw & 0.843  $^{(c)}$& P  \\ 
HDF\_PS3\_4& 12 36 33.9 & +62 14 03 & -     & - & - & $174^{+59}_{-43}$ & 0.099 & 3 & iw & -         & U   \\
HDF\_PS3\_5& 12 36 36.5 & +62 14 48 & -     & - & - & $ 44^{+30}_{-09}$ & 0.035 & 4 & iw & -         & N      \\
HDF\_PS3\_6a& 12 36 36.9 & +62 13 36 & -     & - & - & $184^{+61}_{-44}$ & 0.105 & - & iw & -         & P  \\
HDF\_PS3\_7& 12 36 38.2 & +62 14 33 & -     & - & - & $ 29^{+11}_{-13}$ & 0.026 & 3 & iw & -         & N       \\ 
HDF\_PS3\_8& 12 36 38.6 & +62 14 51 & -     & - & - & $ 52^{+34}_{-09}$ & 0.042 & - & nw & - 	     & N       \\
HDF\_PS3\_6b& 12 36 38.7 & +62 13 39 & -     & - & - & $ 52^{+34}_{- 9}$ & 0.042 & - & iw & -         & U       \\
HDF\_PS3\_11a& 12 36 39.1 & +62 12 57 & $< 90$& - & - & $211^{+58}_{-54}$ & 0.119 & - & iw & - 	     & S    \\
HDF\_PS3\_6c& 12 36 40.4 & +62 13 36 & -     & - & - & $ 36^{+11}_{-15}$ & 0.029 & - & iw & -         & U   \\
HDF\_PS3\_6d& 12 36 40.4 & +62 13 45 & -     & - & - & $ 24^{+10}_{-12}$ & 0.020 & 3 & iw & -         & N       \\
HDF\_PS3\_9& 12 36 40.8 & +62 10 48 & -     & - & - & $ 44^{+30}_{-09}$ & 0.035 & 3 & sw & -         & N       \\
HDF\_PS3\_6e& 12 36 41.9 & +62 13 33 & -     & - & - & $ 23^{+10}_{-12}$ & 0.019 & 4 & iw & -         & K       \\
HDF\_PS3\_10& 12 36 42.1 & +62 15 45 & -     & - & - & $459^{+46}_{-86}$ & 0.263 & 6 & -- &   -       & -       \\
HDF\_PS3\_12& 12 36 43.0 & +62 10 51 & -     & - & - & $ 18^{+ 9}_{-10}$ & 0.014 & - & sw & -         & N       \\
HDF\_PS3\_21a& 12 36 43.4 & +62 11 51 & $< 41$& - & - & $ 48^{+57}_{-10}$ & 0.044 & - & hd & 1.242$^{(i)}$ & U$^{(i)}$    \\
HDF\_PS3\_13& 12 36 43.4 & +62 14 30 & $<405$& - & - & $  6^{+ 8}_{- 7}$ & 0.005 & - & nw & -         & N      \\
HDF\_PS3\_21c& 12 36 44.7 & +62 11 51 & $< 41$& - & - & $111^{+52}_{-38}$ & 0.063 & - & hd & 2.803$^{(i)}$ & M$^{(i)}$    \\
HDF\_PS3\_17b& 12 36 44.7 & +62 13 03 & $< 60$& - & - & $ 33^{+11}_{-15}$ & 0.027 & - & hd & 0.485 $^{(p)}$ & S    \\
HDF\_PS3\_21b& 12 36 45.1 & +62 11 42 & $< 47$& - & - & $147^{+73}_{-48}$ & 0.086 & - & hd &   -       & C$^{(i)}$    \\
HDF\_PS2\_1& 12 36 45.4 & +62 12 39 &$ 77^{+67}_{-66}$ & 0.009 & 5 & $< 74$  &  -    & - & hd &   -       & S$^{(i)}$    \\
HDF\_PS3\_17a& 12 36 45.9 & +62 12 45 & $< 47$& - & - & $ 31^{+11}_{-14}$ & 0.025 & - & hd &   -       & E$^{(i)}$    \\
HDF\_PS2\_2& 12 36 46.2 & +62 11 46 &$ 88^{+45}_{-50}$ & 0.009 & 5 & $< 78$  &  -    & - & hd &0.504 $^{(c)}$ & E$^{(i)}$  \\ 
HDF\_PS3\_14& 12 36 47.7 & +62 10 21 & -     & - & - & $104^{+94}_{-21}$ & 0.075 & 3 & se & -         & N      \\ 
HDF\_PS2\_3& 12 36 47.7 & +62 13 12 &$ 36^{+65}_{-30}$ & 0.003 & 5 & $< 74$  &  -    & - & hd & 0.475 $^{(c)}$ & E   \\ 
HDF\_PS3\_15& 12 36 48.9 & +62 12 18 & -     & - & - & $ 32^{+11}_{-14}$ & 0.025 & 4 & hd & 0.953 $^{(p)}$ & M$^{(i)}$    \\ 
HDF\_PS3\_16& 12 36 50.2 & +62 12 39 & $< 34$& - & - & $ 22^{+10}_{-11}$ & 0.018 & 4 & hd & 0.474 $^{(c)}$ & M$^{(i)}$    \\
HDF\_PS3\_18& 12 36 51.9 & +62 10 24 & -     & - & - & $118^{+53}_{-39}$ & 0.067 & 5 & se & -         & P    \\ 
HDF\_PS3\_19& 12 36 51.9 & +62 11 48 & -     & - & - & $ 34^{+11}_{-15}$ & 0.027 & 4 & se & -         & N      \\ 
HDF\_PS3\_30a& 12 36 52.0 & +62 14 00 & $< 39$& - & - & $ 50^{+79}_{-20}$ & 0.040 & - & hd & 0.557$^{(i)}$ & S?$^{(i)}$   \\
HDF\_PS3\_20& 12 36 52.8 & +62 14 54 & -     & - & - & $ 42^{+29}_{- 9}$ & 0.033 & 3 & nw & 0.463 $^{(c)}$ & M    \\ 
HDF\_PS3\_22& 12 36 54.1 & +62 14 36 & $<136$& - & - & $ 51^{+75}_{-51}$ & 0.047 & 5 & ne & 0.577 $^{(c)}$ & U   \\ 
HDF\_PS3\_23& 12 36 54.9 & +62 11 27 & -     & - & - & $ 42^{+29}_{-09}$ & 0.032 & 3 & ie & -         & U   \\ 
HDF\_PS3\_24& 12 36 55.4 & +62 13 12 & $< 37$& - & - & $ 23^{+10}_{-11}$ & 0.018 & 3 & hd & 1.315$^{(h)}$ & E$^{(i)}$    \\ 
HDF\_PS3\_25& 12 36 55.5 & +62 15 15 & -     & - & - & $ 75^{+71}_{-19}$ & 0.052 & 3 & nw & -         & N      \\
HDF\_PS3\_26& 12 36 55.8 & +62 12 45 & $< 38$& - & - & $ 44^{+70}_{-09}$ & 0.035 & 5 & hd & 0.790 $^{(c)}$ & S?$^{(i)}$   \\ 
HDF\_PS3\_27& 12 36 56.2 & +62 11 33 & -     & - & - & $ 29^{+11}_{-13}$ & 0.023 & 3 & ie & -         & N   \\ 
HDF\_PS2\_4& 12 36 56.7 & +62 13 03 &$ 94^{+50}_{-69}$&0.012& 6 &  $< 54$   &   -   & - & hd & -         & E$^{(i)}$    \\      
\hline
\multicolumn{12}{|l|}{Continued on next page...}\\
\hline
\end{tabular}
\end{table*}

\begin{table*}[t]
\begin{tabular}{|c|c|c|c|c|c|c|c|c|c|c|c|}
\hline
\multicolumn{12}{|l|}{Continued from previous page...}\\
\hline
Id$^{(1)}$& $\alpha^{(2)}$  & $\delta^{(3)}$ &\hspace{0.01cm} LW2$^{(4)}$\hspace{0.01cm}  & LW2$^{(5)}$  & N$\tau_{w}$ & \hspace{0.01cm}LW3$^{(7)}$\hspace{0.01cm}     &LW3$^{(8)}$ & N$\tau_{w}$&F&\hspace{0.1cm} z$^{(11)}$\hspace{0.1cm} & T$^{(12)}$\\
  &  (J2000)          &    (J2000)                 &$\mu$Jy & {\tiny ADU/G/s} &$^{(6)}$ &$\mu$Jy & {\tiny ADU/G/s}   & $^{(9)}$ &  $^{(10)}$     &   &      \\
\hline
HDF\_PS3\_28& 12 36 56.7 & +62 13 30 & $< 50$& - & - & $ 26^{+10}_{-13}$ & 0.021 & 4 & hd & -         & N      \\ 
HDF\_PS3\_29& 12 36 58.8 & +62 11 18 & -     & - & - & $  9^{+ 8}_{- 7}$ & 0.007 & 4 & ie & -         & U   \\ 
HDF\_PS3\_31& 12 37 00.1 & +62 11 54 & $<150$& - & - & $ 48^{+56}_{-10}$ & 0.045 & 6 & ie & -         & G     \\
HDF\_PS3\_32& 12 37 01.8 & +62 11 45 & -     & - & - & $ 15^{+ 9}_{- 9}$ & 0.012 & 3 & ie & -         & S?   \\ 
HDF\_PS3\_33& 12 37 01.8 & +62 14 39 & -     & - & - & $ 21^{+10}_{-11}$ & 0.017 & 3 & ne & -         & N      \\
HDF\_PS3\_34& 12 37 02.7 & +62 12 48 & $<322$& - & - & $ 44^{+30}_{ -9}$ & 0.035 & 4 & ie & -         & E?   \\
HDF\_PS3\_35& 12 37 03.4 & +62 13 51 & -     & - & - & $ 29^{+11}_{-13}$ & 0.023 & 3 & ie & -         & N      \\
HDF\_PS3\_36& 12 37 04.4 & +62 14 48 & -     & - & - & $ 80^{+74}_{-19}$ & 0.056 & 3 & ie & -         & N      \\
HDF\_PS3\_37& 12 37 04.8 & +62 14 27 & -     & - & - & $ 72^{+65}_{-16}$ & 0.050 & 3 & ie & 0.561 $^{(c)}$ & S    \\
HDF\_PS3\_38& 12 37 08.2 & +62 12 48 & -     & - & - & $ 89^{+79}_{-20}$ & 0.061 & 6 & ie & 0.654 $^{(p)}$ & S?   \\
HDF\_PS3\_39& 12 37 08.2 & +62 12 54 & -     & - & - & $115^{+53}_{-39}$ & 0.066 & 6 & ie & 0.838 $^{(c)}$ & P    \\
HDF\_PS3\_40& 12 37 09.5 & +62 12 36 & -     & - & - & $306^{+62}_{-67}$ & 0.178 & 6 & ie & -         & st   \\
\hline
\end{tabular}
\end{table*}

\begin{figure*}
\psfig{file=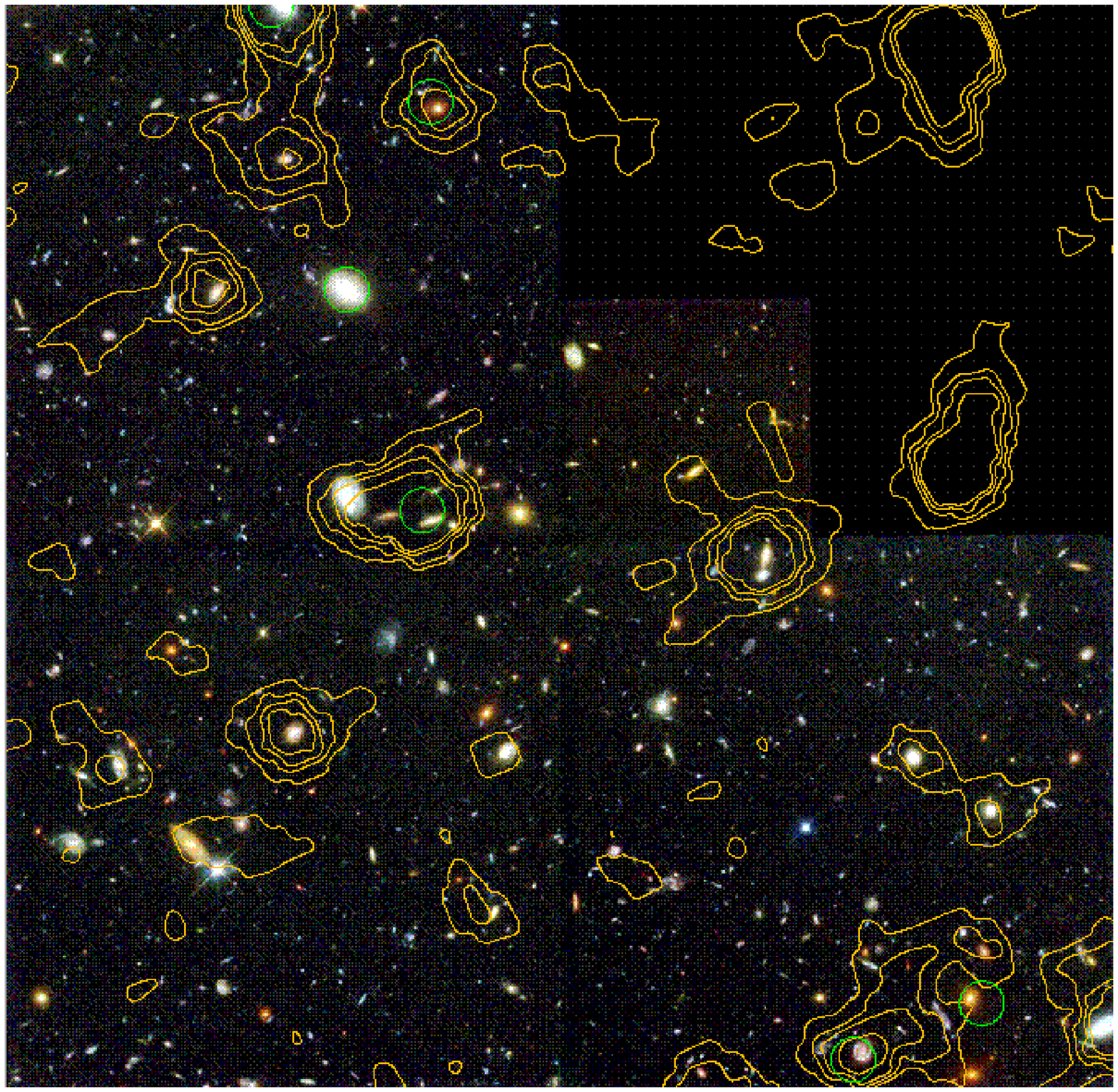,width=1.0\textwidth,angle=270}
\caption[]{Overlay of the ISO-HDF observation on the HDF color composite image produced by Williams {\it et al. } (\cite{Williams} )from the F450W, F606W and F814W images obtained with HST. Orange contours : contours of the 15 $\mu$m (LW3) ISOCAM observation. Contours levels begin at  1.0 $\mu$Jy with steps of 2 $\mu$Jy. Green circles : ISOCAM 6.75 $\mu$m (LW2) detections. }
\label{HST_ISO}
\end{figure*}

\subsection{Source catalog}
Table \ref{liste_sures} gives the catalog of detected sources at 6.7
$\mu$m (LW2 filter) and 15 $\mu$m (LW3) above a secure $7\tau_{w}$ of
photon and readout noise. The first column is our ISO-HDF
identificator, the second and third column are the right ascension and
declination of the object. The pointing accuracy is of 1 pixel on the
final map, {\it i.e.} 3'' for the LW3 image and 1.5'' for the LW2
image . Columns 4 and 5 give the objects fluxes in the LW2 band in
$\mu$Jy and ADU/G/s, that is in analog to digital unit per second
divided by the instrument gain, and columns 7 and 8 give the same in
the LW3 band. Columns 6 and 9 give the detection level in the LW2
({\it resp.} LW3) band. Column 10 identifies the field where the
object is observed (HDF or flanking field). Column 11 gives the
spectrometric redshift when available. Finally, column 12 give the
morphological type of the object.

\begin{figure}
\centering
\psfig{file=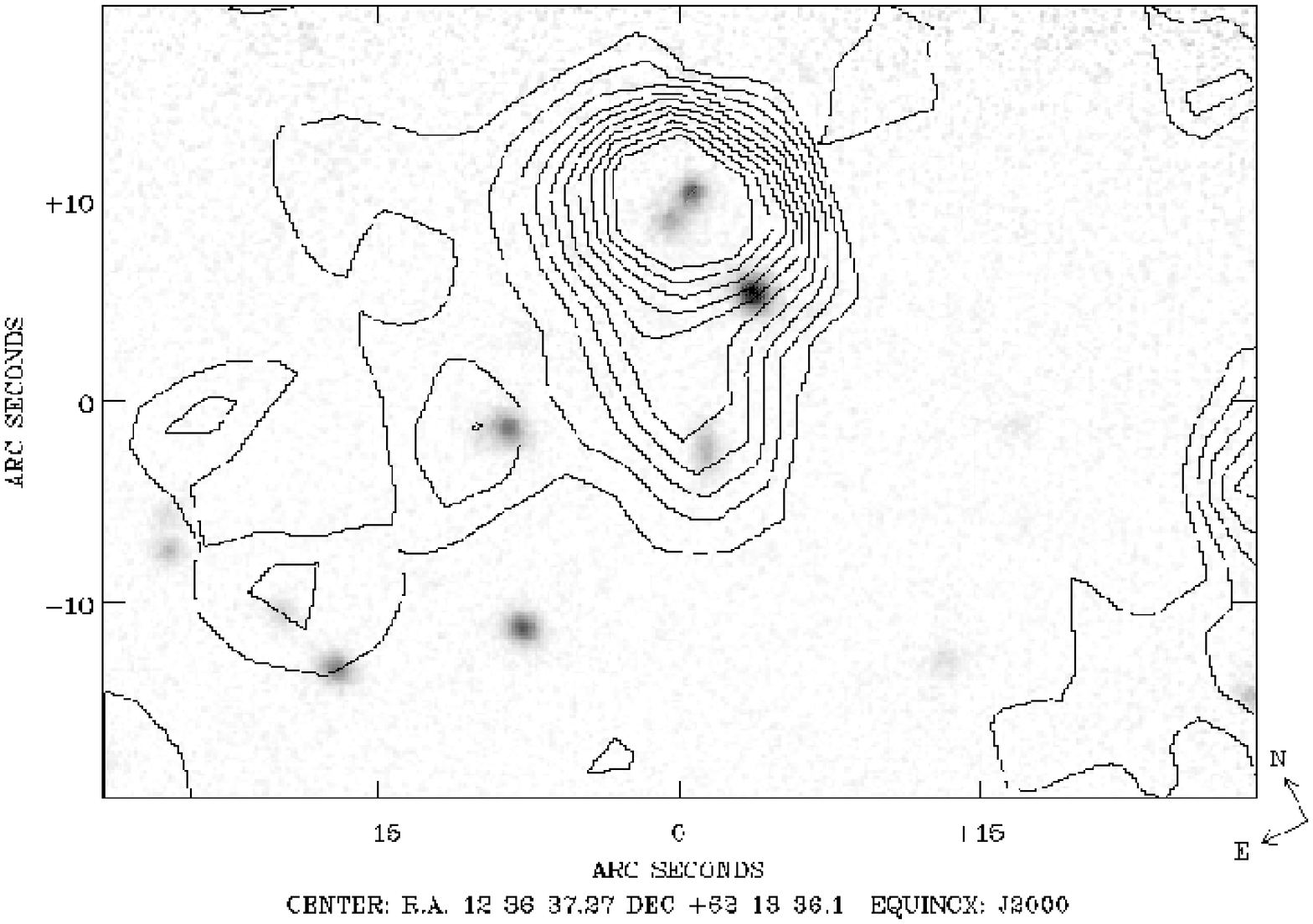,angle=90,width=0.5\textwidth}
\caption[]{Comparison near-IR/LW3 filter for source HDF\_PM\_6. The contours of the details of the LW3 map around source C6 have been overlayed on the IRIM K map. Contour levels begin at 1.5 $\mu$Jy with steps of 2 $\mu$Jy. }
\label{objet6}
\end{figure}

The redshift and morphological type of each ISOCAM source depend on
the identification of its optical counterpart. Hopefully, our
astrometric correction is good enough to allow us a good superposition
with optical images, at a scale of 1 pixel. This is clearly visible on
figures \ref{irim_HDFlw2} and \ref{irim_HDFlw3} where we have
overlayed our ISOCAM maps on the KPNO IRIM deep K exposure
\footnote{Based on observations made at the Kitt Peak National
Observatory, National Optical Astronomy Observatories, which is
operated by the Association of Universities for Research in Astronomy,
Inc. (AURA) under cooperative agreement with the National Science
Foundation.} (Dickinson at al.~\cite{Dickinson}), as well as on figure
\ref{HST_ISO} where we have overlayed the LW3 ISOCAM map and the LW2
detections on the color composite image of the HDF observed by HST
from williams {\it et al. } (\cite{Williams}). In most cases, the ISOCAM
source matches an IR or optical source, and the identification is
easy. However, due to the large PSF (15'' at 15$\mu$m) of ISOCAM,
sometimes more than one object could be associated with our
detection. Figure~\ref{objet6} displays an example of such a case :
the main peak of source HDF\_PM3\_6 correspond to two objects. In this
particular case, we have taken the brightest object in K as
counterpart. When K data are not available, we have used the `G' (for
group) symbol in our table, because we could not determine the right
counterpart.

\begin{figure}
\centering
\psfig{file=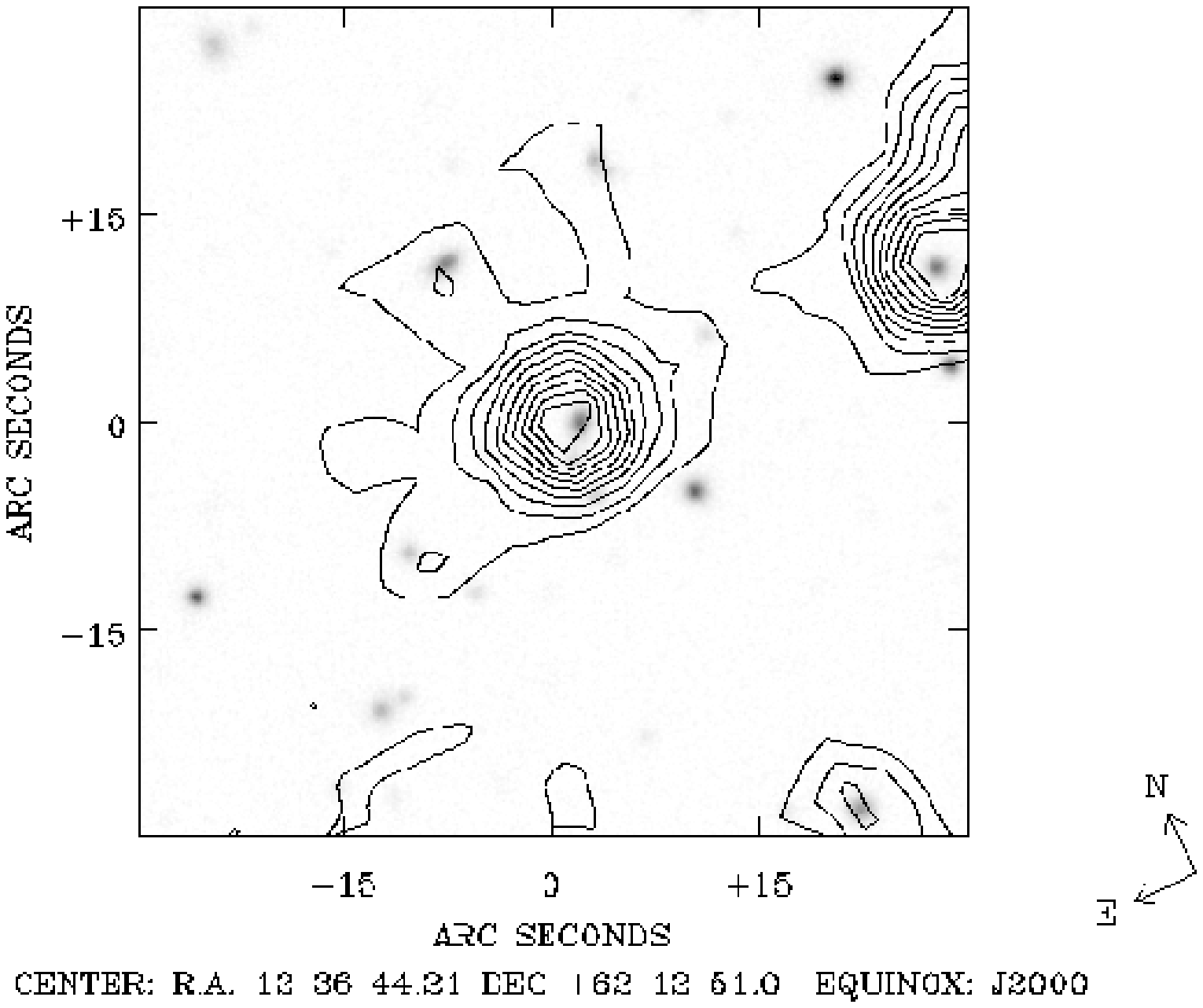,bbllx=434pt,bblly=307pt,bburx=68pt,bbury=738pt,angle=90,width=0.5\textwidth}
\caption[]{Comparison near-IR/LW3 filter for source HDF\_PM\_17. The contours of the details of the LW3 map around source 17 have been overlayed on the IRIM K map. Contour levels begin at 1.5 $\mu$Jy with steps of 2 $\mu$Jy. }
\label{objet17}
\end{figure}

\begin{figure}
\centering
\psfig{file=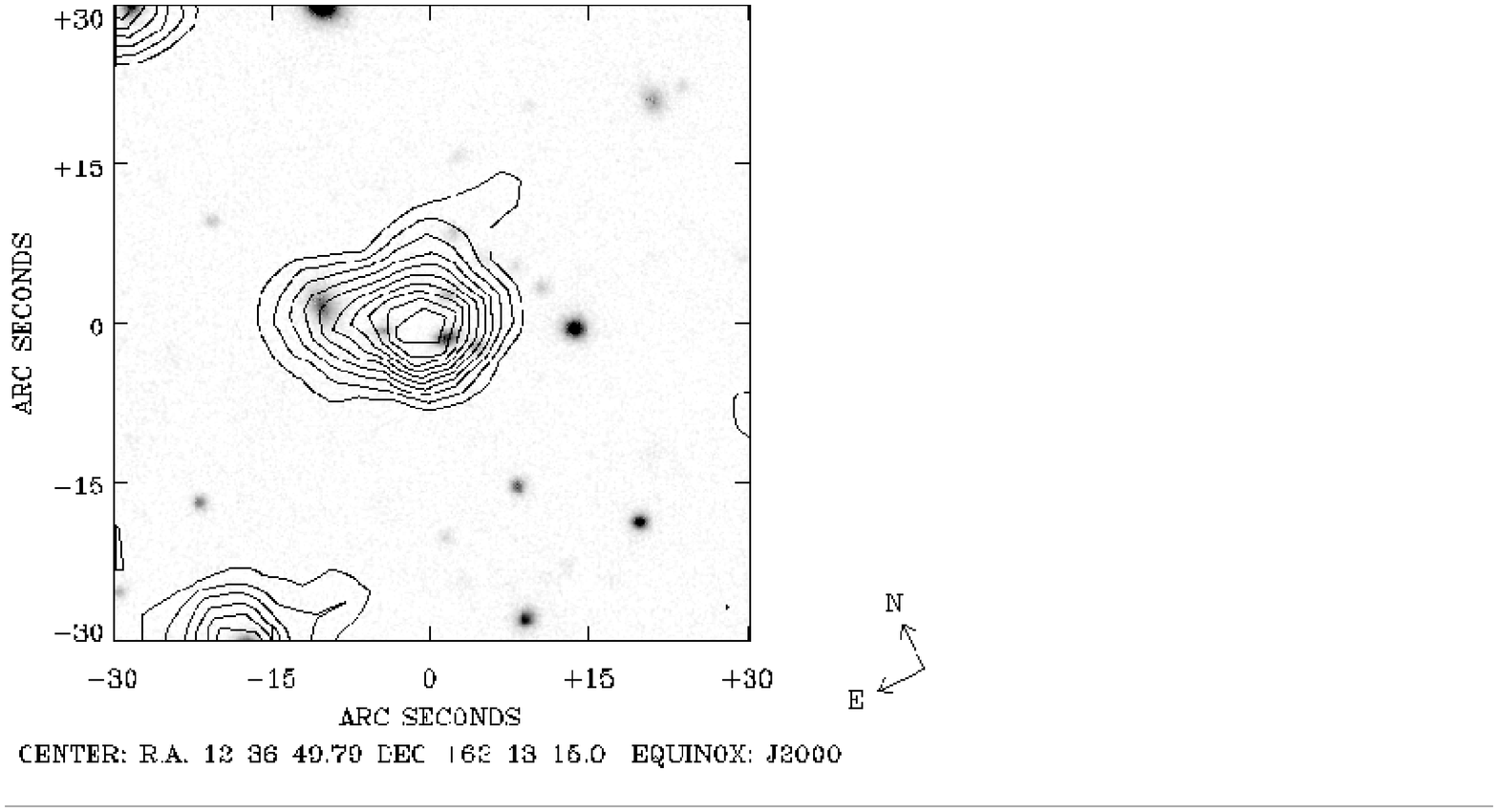,bbllx=434pt,bblly=307pt,bburx=68pt,bbury=738pt,angle=90,width=0.5\textwidth}
\caption[]{Comparison near-IR/LW3 filter for source 27. The contours of the details of the LW3 map around source HDF\_PM\_27 have been overlayed on the IRIM K map. Contour levels begin at 1.5 $\mu$Jy with steps of 2 $\mu$Jy. }
\label{objet27}
\end{figure}

Looking closely at figure \ref{objet6}, one can see that the object
HDF\_PM3\_6, detected as a whole, displays substructures. Our
detection program does not separate the various components of the
source because it is not robust against blending~: to avoid detection
of cosmic ray residuals, we choose to perform the detection starting
from the second scale of the wavelet transform of our map, {\it i.e.}
on scales of the order of 3 pixels that prevent us to separate very
closeby sources, especially if they are much weaker than the main
detection. However, in some cases, this separation can be done by
hand, when blending is weak enough to allow for the separation of the
source (Rayleigh criteria). This is the case for source HDF\_PM3\_6,
for which 4 extensions are visible on figure \ref{objet6}, South and
East of the main peak. These extensions are in our supplementary list
(see table \ref{liste_supp}) with the names HDF\_PS3\_6a,
HDF\_PS3\_6b, {\it etc...}. Figure~\ref{objet17} displays another case
where the separation is possible, for source HDF\_PM3\_17
. Figure~\ref{objet27} displays a closeup on source HDF\_PM3\_27~: the
source is extended, with multiple counterparts, but we failed to
separate the various contributions. We note however that this source
is the radio source 3639+1313 in the catalog of Fomalont at
al. (\cite{Fomalont}), whose contours in the radio map (Richards et
al.~\cite{Richards}) are also clearly extended.

We produced a supplementary list of sources with these extensions of
reliable sources, together with results of PRETI at lower detection
levels (6 and 5 $\tau_{w}$), that is listed in table
\ref{liste_supp}. However, this catalog may contain spurious
detections, as it is certainly the case for sources HDF\_PS3\_13 and
HDF\_PS3\_29, with fluxes below 10 $\mu$Jy. In the future, we expect
to obtain data of a long staring observation on an empty field. Using
this observation as a basis for simulations, we hope to be able to
obtain higher reliability on the sources at these low level of flux.

\subsection{Catalog properties.}

Our main source list consist in 49 objects detected above a $7\tau_{w}$
detection threshold in both filters. 42 sources are detected at 15
$\mu$m only, 4 at 6.5 and 15 $\mu$m, and 3 at 6.5 $\mu$m only. Since
the LW3 field is larger than the LW2 field, this last number means
that 21 of the LW3 sources are not detected in LW2.  The supplementary
list adds 51 sources to this catalog for a total of 100 objects. 47
are observed in the LW3 filter only, and 4 in the LW2 filter only. None are
detected in both filters, 15 of the LW3 sources that could be visible
in the LW2 field remain undetected. For these missed detection, we
have computed  $5\tau_{w}$ upper limit as follows : a test image
containing only the source is computed by projecting a point source as
observed at each raster pointing in a copy of the final raster
map. This ensures that the modification to the PSF produced by
microscan are taken into account. The b-spline wavelet transform of
this image is then calculated and compared to 5 times the noise map at
each scale. The minimum flux ensuring the detection is then converted
in microJansky, using our calibration from simulations.

All sources from the main list within the HDF have optical
counterparts, thanks to the depth of the HST map, but 6 sources within
the flanking fields do not have visible counterparts in the much less
deep F814W images. However, these 6 sources are very reliable and were
also found by other teams (see below). The number of sources without
counterpart rises to 11 in the supplementary list, of which 1 lays in
the HDF. Source HDF\_PS3\_6e does not have any counterpart in the `iw'
HST image, but has a counterpart in the IRIM K image. This example
shows that the reliability of the source should not be established on
the existence of counterparts in optical images only, especially in
the flanking fields. It is remarkable that all but one (3652+1444) of
the radio sources observed at 8.4 GHz by Fomalont et
al. (\cite{Fomalont}) in the field of view, are detected, mostly in the
LW3 filter (Table~\ref{compare_other_lw3}), and one in the LW2 filter
only (Table~\ref{compare_other_lw2}).

Other teams have used different algorithms to produce source lists. We
have compared our results to those of the Imperial College group
(hereafter IC) (Goldschmidt et al.~\cite{Goldschmidt}) and to those of
IAS (D\'esert et al.~\cite{IAS}). The common source lists are given in Table
\ref{compare_other_lw3} for the LW3 filter, and in Table 
\ref{compare_other_lw2} for the LW2 one , as well as the names of the radio sources observed at
8.4 GHz by Fomalont et al. (\cite{Fomalont}) matching ISOCAM sources.

\begin{figure}
\psfig{file=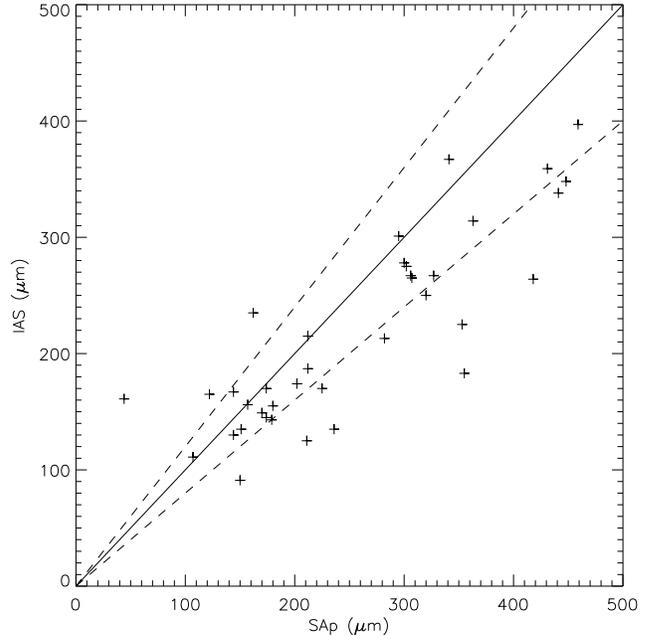,width=0.5\textwidth}
\caption[]{Comparison of the photometry of common sources in the ISO-HDF at 15 $\mu$m for PRETI and IAS. Solid line is 1 to 1 relation, dashed lines are a 20 \% error with respect to the 1 to 1 relation.}\label{figcompsapias}
\label{counpfluxiaslw3}
\end{figure}

PRETI and the ``triple beam switch method'' of IAS (D\'esert et al.~\cite{IAS})
give very similar results (see Table \ref{compare_other_lw2} and
\ref{compare_other_lw3}), especially in the LW3 band where out of the
41 sources detected by them, we find 37. We do not detect
HDF\_ALL\_LW3\_8,14,29,40 that are at 3.4$\sigma$, 3.3$\sigma$,
3.9$\sigma$ and 3.1$\sigma$ respectively, that is among the faintest
of their list. Both number counts agrees above 200 $\mu$Jy where PRETI
reaches 99.9\% completeness~: the thwo methods detect 24
sources. Figure~\ref{figcompsapias} displays a comparison between the
fluxes derived by the two methods for the common source list. The
fluxes derived by the ``triple beam switch method'' are lower than our
fluxes by a systematic factor of 0.82. We note that both methods relie
on the ISOCAM cookbook values for the correction of ADUs into
milli-Janskies. For one part, we can explain the difference by an
implicit color correction that we have introduced in our simulations~:
our PSFs are simulated assuming a rising spectrum with spectral index
of 3, while standard ISOCAM PSFs, used by D\'esert et al.~\cite{IAS},
are measured on stars, thus assuming a spectral index of (-2). When
comparing our PSFs to standard ones, a mean difference of 10 \% is
found. Moreover, source fluxes are measured by us with aperture
photometry while the ``triple beam switch method'' uses fits of
gaussian of fixed width. Since at 6'', the ISOCAM PSF is only roughly
approximated by a gaussian, this can account for the remaining
discrepancy. At these faint level of fluxes, such differences are not
very important, considering that the source lists of the two
catalogues matche very well. However, PRETI allows to reach deeper
sensitivity, as we consider sources at 40 $\mu$Jy as reliable.  For
the LW2 filter, the ``triple beam switch method'' detects 6 sources, 3
of which are also detected with PRETI. Of the 3 undetected objects, 2
lay on the edges of the map where we do not allow for detection. The
last one is measured by D\'esert et al. at 58 $\mu$Jy where we are not complete.

\begin{figure}
\psfig{file=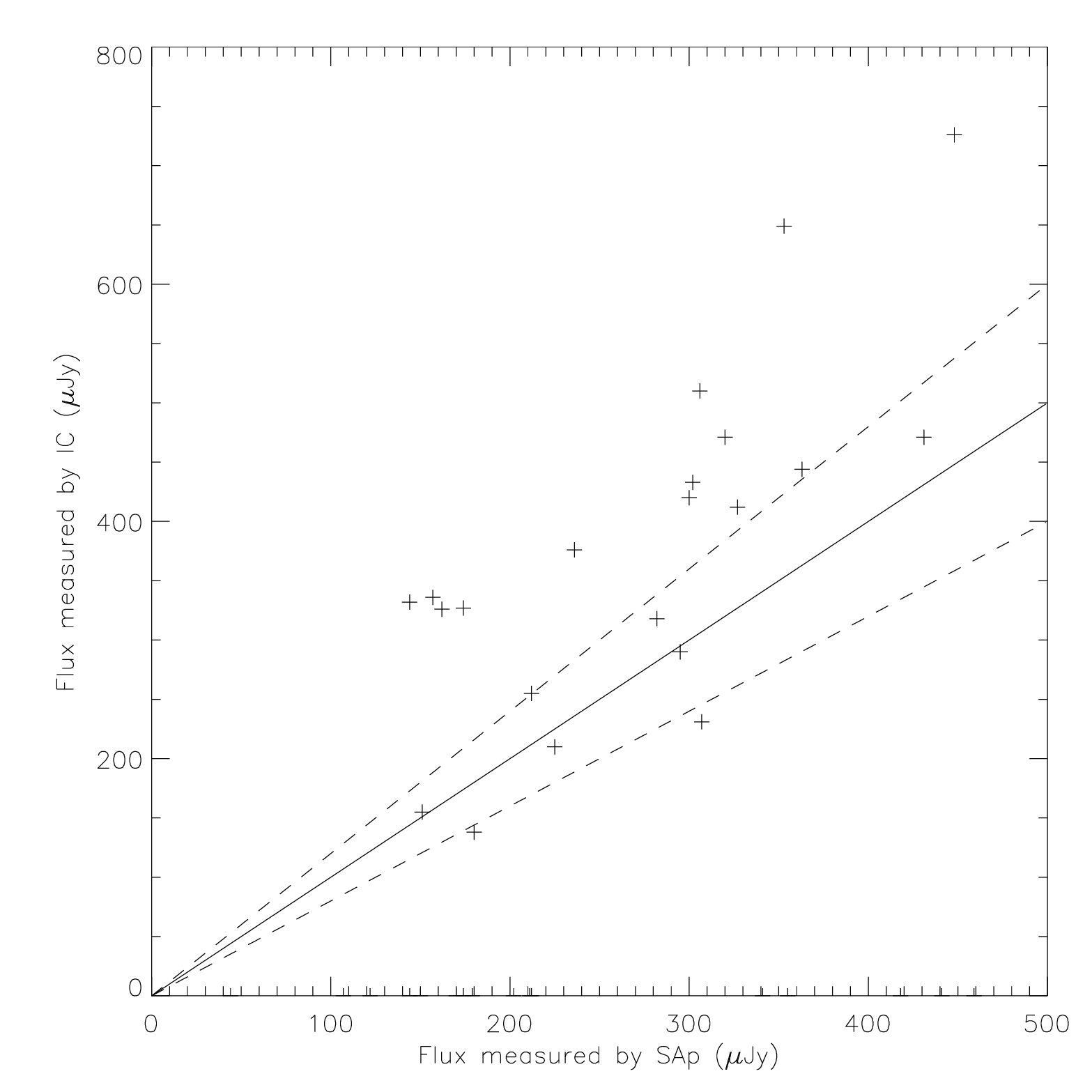,width=0.5\textwidth}
\caption[]{Comparison of the photometry of common sources in the ISO-HDF at 15 $\mu$m for PRETI and IC. Solid line is 1 to 1 relation, dashed lines are a 20 \% error with respect to the 1 to 1 relation}
\label{compfluxiclw3}
\end{figure}

The IC group have made the original data reduction of the ISO-HDF,
when some of the latest refinements of ISOCAM calibration such as
field distortion were not available. We find 21 of the 22 sources
detected by them in their LW3 image (Goldschmidt et
al.~\cite{Goldschmidt}). We could not find the source J123659.4+621337
in our map. Source photometry differs from our catalog~: on the
average, their flux estimates are higher than ours by a factor of 1.5,
as shown in figure
\ref{compfluxiclw3}. They performed simulations, and concluded that
they should be 70\% complete at a level of 225 $\mu$Jy. According to
the 1.5 factor of difference in photometry, they should have detected
at least 26 of our 37 sources above 150 $\mu$Jy, where they only find
19. This indicates that they are only 50\% complete at this level,
while PRETI reaches 95\% at a level of 150 $\mu$Jy, according to our
simulations, if source confusion is ignored. Comparing the LW2
detections, we only find 5 of the 27 sources in their list (3 from
their main list of 6 objects and 2 from their supplementary list of 21
objects), as shown in table
\ref{compare_other_lw2}. Moreover, we measure much fainter fluxes for
these objects. Since this observation is made with 10s integration
time, so that pixels affected by cosmic ray glitches are more
numerous, two explanations can account for this~: either we set too
heavy a deglitching criterion and we have cleaned real sources out, or
they did not clean their data enough against cosmic rays. Our analysis
lead us to favor the second explanation, because we do not find that
we are missing sources in the simulations. Here also the results of
D\'esert at al. (\cite{IAS}) are in good agreement with us.

\begin{table*}
\caption{Comparison of the LW2 catalog obtained here with PRETI, with the catalogs by IC (Goldschmidt et al.~\cite{Goldschmidt}) and IAS (D\'esert et al.~\cite{IAS}), as well as radio sources detected at 8.4 GHz (Fomalont et al.~\cite{Fomalont})}\label{compare_other_lw2}
\begin{tabular}{|c|c|c|c|c|c|c|c|}
\hline
ID & Flux	&  IC 	& Flux	& IAS & Flux & 8.4 GHz & Flux \\
\hline
HDF\_PM2\_1& $127^{+99}_{-61}$ & J123641.6+621142 & 51.6 & - & - & - & - \\
HDF\_PM2\_2& $185^{+99}_{-25}$ &- & - & - & - & 3644+1133 & 458 \\ 
HDF\_PM3\_20&  $191^{+41}_{-86}$& J123646.4+621406 & 52.1 & HDF-2\_LW2\_1 & 69.& 3646+1404 & 152 \\ 
HDF\_PM3\_24& $254^{+71}_{-73}$& J123648.2+621427 & 65.7 & HDF-2\_LW2\_2 & 96.& - & - \\ 
HDF\_PM2\_3& $41^{+66}_{-30}$ & - & - & HDF-2\_LW2\_3 & 48. & - & - \\ 
HDF\_PM3\_27& $136^{+68}_{-57}$& J123649.7+621315 & 48.1 & - & - & 3649+1313 & 22 \\ 
HDF\_PS2\_4& $94^{+50}_{-69}$& J123656.6+621307 & 31.3 & - & - & - & - \\
\hline
\end{tabular}
\end{table*}

\begin{table*}
\caption{Comparison of the LW3 catalog obtained here with PRETI, with the catalogs by IC (Goldschmidt et al.~\cite{Goldschmidt}) and IAS (D\'esert et al.~\cite{IAS}), as well as radio sources detected at 8.4 GHz (Fomalont et al.~\cite{Fomalont})}\label{compare_other_lw3}
\begin{tabular}{|c|c|c|c|c|c|c|c|}
\hline
ID & Flux       &  IC   & Flux  & IAS & Flux & 8.4 GHz & Flux \\
\hline
HDF\_PM3\_1 & $355^{+40}_{-60}$ &      -            &  -   & HDF\_ALL\_LW3\_1  & 183    & - & - \\
HDF\_PM3\_2 & $448^{+68}_{-59}$ & J123633.9+621217 & 726 & HDF\_ALL\_LW3\_3  & 348      & 3634+1212 & 40\\
HDF\_PS3\_3 & $122^{+54}_{-40}$ &      -           &  -  & HDF\_ALL\_LW3\_2  & 165      & - & - \\
HDF\_PS3\_4 & $174^{+59}_{-43}$ &      -           &  -  & HDF\_ALL\_LW3\_5  & 145      & - & - \\
HDF\_PM3\_3 & $363^{+79}_{-38}$ & J123634.3+621238 & 444 & HDF\_ALL\_LW3\_4  & 314      & 3634+1240 & 40\\
HDF\_PM3\_5 & $441^{+43}_{-82}$ &     -            &  -  & HDF\_ALL\_LW3\_6  & 338      & - & - \\
HDF\_PM3\_6 & $353^{+40}_{-66}$ & J123636.5+621348 & 649 & HDF\_ALL\_LW3\_10 & 225      & - & - \\
HDF\_PS3\_6e& $ 23^{+10}_{-12}$ &     -            &  -  &  -  &        -       & 3642+1331 & 80\\
HDF\_PM3\_7 & $300^{+62}_{-67}$ & J123635.9+621134 & 420 & HDF\_ALL\_LW3\_7  & 278   & - & - \\
HDF\_PM3\_8 & $202^{+58}_{-50}$ &     -            &  - & HDF\_ALL\_LW3\_9  & 174       & - & - \\
HDF\_PM3\_9 & $212^{+58}_{-55}$ & J123637.5+621109 & 255 & HDF\_ALL\_LW3\_12 & 215      & - & - \\
HDF\_PM3\_10 & $212^{+58}_{-55}$ &     -            &  -  & HDF\_ALL\_LW3\_11 & 187     & - & - \\
HDF\_PM3\_11 & $302^{+67}_{-55}$ & J123639.3+621250 & 433 & HDF\_ALL\_LW3\_15 & 275     & - & - \\
HDF\_PS3\_11a& $211^{+58}_{-54}$ &      -           &  -  & HDF\_ALL\_LW3\_13 & 125     & - & - \\
HDF\_PM3\_12 & $236^{+61}_{-58}$ & J123641.1+621129 & 376 & HDF\_ALL\_LW3\_17 & 135 & - & - \\
HDF\_PS3\_9& $ 44^{+30}_{-09}$ &      -           &  -  & HDF\_ALL\_LW3\_16 & 161       & - & - \\
HDF\_PS3\_10 & $459^{+46}_{-86}$ &      -           &  -  & HDF\_ALL\_LW3\_18 & 397     & - & - \\
HDF\_PM3\_17 & $282^{+60}_{-64}$ & J123643.7+621255 & 318 & HDF\_ALL\_LW3\_19 & 213     & 3644+1249 & 13\\
HDF\_PM3\_19& $170^{+59}_{-42}$ &      -           &  -  & HDF\_ALL\_LW3\_20 & 149      & - & - \\
HDF\_PM3\_20 & $107^{+95}_{-20}$ &      -           &  -  & HDF\_ALL\_LW3\_21 & 111    & 3646+1404 & 152\\
HDF\_PM3\_21 & $418^{+91}_{-94}$ &      -           &  -  & HDF\_ALL\_LW3\_22 & 264     & - & - \\
HDF\_PM3\_22& $327^{+39}_{-63}$ & J123646.9+621045 & 412 & HDF\_ALL\_LW3\_23 & 267      & - & - \\
HDF\_PM3\_23& $144^{+72}_{-47}$ &      -           &  -  & HDF\_ALL\_LW3\_24 & 130      & 3646+1448 & 36\\
HDF\_PM3\_24& $307^{+62}_{-67}$ & J123648.1+621432 & 231 & HDF\_ALL\_LW3\_25 & 265   & - & - \\
HDF\_PM3\_26& $150^{+74}_{-48}$ &      -           &  -  & HDF\_ALL\_LW3\_26 &  91      & - & - \\
HDF\_PM3\_27& $320^{+39}_{-62}$ & J123649.8+621319 & 471 & HDF\_ALL\_LW3\_27 & 250      & 3649+1313 & 22 \\
HDF\_PM3\_28& $341^{+40}_{-65}$ &      -           &  -  & HDF\_ALL\_LW3\_28 & 367      & - & - \\
HDF\_PM3\_29& $ 48^{+32}_{-09}$ &      -           &  -  &  -  &         -      & 3651+1221 & 18\\
HDF\_PM3\_30 & $151^{+74}_{-68}$ & J123651.5+621357S& 155 & HDF\_ALL\_LW3\_30 & 135     & - & - \\
HDF\_PM3\_31 & $174^{+59}_{-43}$ & J123653.0+621116 & 327 & HDF\_ALL\_LW3\_31 & 170     & - & - \\
HDF\_PM3\_32& $180^{+60}_{-43}$ & J123653.6+621140 & 138 & HDF\_ALL\_LW3\_32 & 155      & 3653+1139 & 15\\
HDF\_PM3\_33& $179^{+60}_{-43}$ &       -          &  -  & HDF\_ALL\_LW3\_33 & 143              & - & - \\
HDF\_PS3\_24& $ 23^{+10}_{-11}$ &       -          &  -  &  -  &          -     & 3655+1311 & 12\\
HDF\_PS3\_37& $225^{+60}_{-56}$ & J123658.1+621458S& 210 & HDF\_ALL\_LW3\_34 & 170      & - & - \\
HDF\_PS3\_39& $157^{+75}_{-49}$ & J123658.7+621212 & 336 & HDF\_ALL\_LW3\_35 & 156      & - & - \\
HDF\_PS3\_40& $295^{+61}_{-66}$ & J123700.2+621455 & 290 & HDF\_ALL\_LW3\_36 & 301      & - & - \\
HDF\_PS3\_32& $ 15^{+ 9}_{- 9}$ &       -          &  -  & -   &        -        & 3701+1147 & 19\\
HDF\_PM3\_42& $162^{+76}_{-49}$ & J123702.0+621127 & 326 & HDF\_ALL\_LW3\_37 & 235      & - & - \\
HDF\_PM3\_44& $144^{+73}_{-47}$ & J123702.5+621406 & 332 & HDF\_ALL\_LW3\_38 & 167      & - & - \\
HDF\_PM3\_45& $431^{+43}_{-80}$ & J123705.7+621157 & 471 & HDF\_ALL\_LW3\_39 & 359              & - & - \\
HDF\_PS3\_40& $306^{+62}_{-67}$ & J123709.8+621239 & 510 & HDF\_ALL\_LW3\_41 & 267      & - & - \\
HDF\_PS3\_38& $ 89^{+79}_{-20}$ &     -            &  -  &  -  &        -        & 3708+1245 & 15 \\
\hline
\end{tabular}
\end{table*}

\subsection{Number counts}

\begin{figure}
\psfig{file=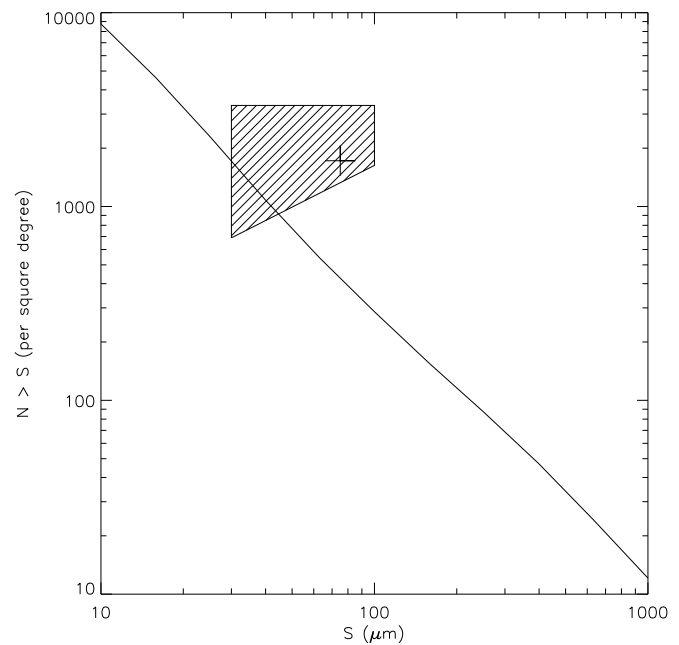,width=0.5\textwidth}
\caption[]{Number counts in the ISO-HDF at 6.75 $\mu$m, based on the 7$\tau_{w}$ sample. \\ 
Solid : no evolution model from Franceschini et al. (\cite{Franceschini97}) \\
Hatched area : $1\sigma$ error area for the integrated counts point (cross) from the 7$\tau_{w}$ sample.}
\label{fig_counts_lw2}
\end{figure}

The small number of LW2 sources does not allow to compute reliable
number counts, as the poissonian fluctuations introduce an uncertainty
of plus or minus 40~\% 1 $\sigma$ when using the reliable 7$\tau_{w}$
source list. Nevertheless, we have, at the level of 100 $\mu$Jy, 5
reliable detections in an area of $10.45'^{2}$, that is : $N(>100\mu
Jy) = 1.7 \times 10^{3} \pm 8 \times 10^{2}$ sources per square
degree, a value compatible with the prediction of a pure passive
luminous evolution model of number counts (Franceschini et
al.~\cite{Franceschini97}),as shown on Fig.~\ref{fig_counts_lw2}.

The more numerous detections in the LW3 band allow
us to compute some points of the Log(N)-Log(S) curve, and to try
do determine its slope. We determined two curves, using the
7$\tau_{w}$ and 5$\tau_{w}$ detection thresholds, for which a few more
spurious detections are expected.

In order to compute reliable number counts, we have taken into account
the following parameters :
\begin{enumerate}
\item the area available for detection at a given flux with a given
detection threshold (N$\tau_{w}$) has been computed, using the same
technique as for determining the upper limits of undetected
sources. However, an additional hypothesis was made~: each
source is considered centered on a pixel of the map.
\item the blending of faint sources by bright ones is treated in a very simple way : the area occupied by the sources of all the brighter bins of flux is subtracted from the area available for the detection of sources at a given bin. For each source, we have taken out a circle of radius 9'', that is approximately the distance between sources separated by PRETI.
\item we take the completeness of the detection at a given level into account by using the results of our simulations (see Fig.~\ref{fig_sim_counts}).
\item when dealing with the 5$\tau_{w}$ detections, we have not taken source HDF\_PS3\_40 into account, because of its high probability of being a star.
\end{enumerate}

\begin{figure}
\psfig{file=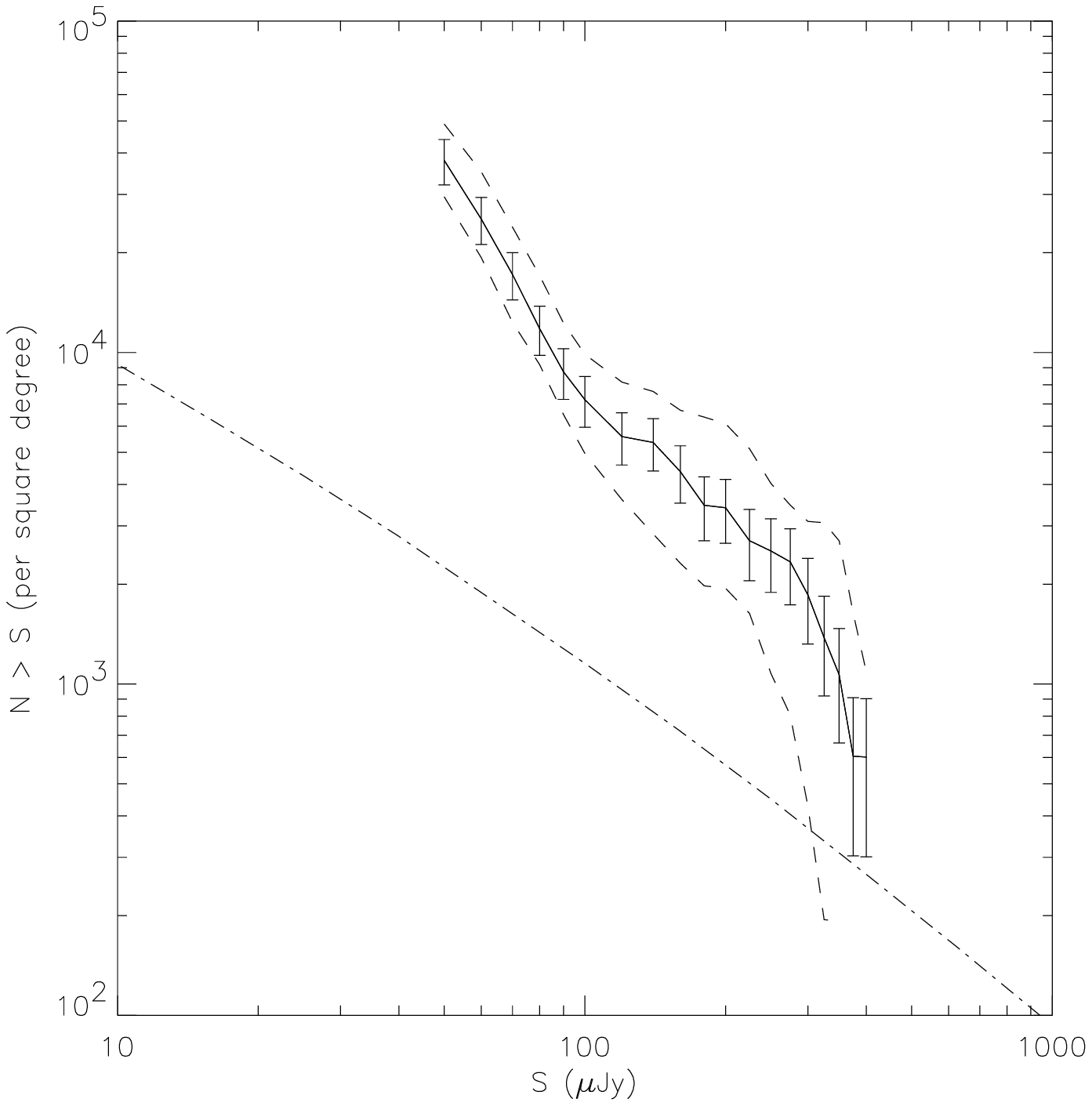,width=0.5\textwidth}
\caption[]{Number counts in the ISO-HDF at 15 $\mu$m, based on the 7$\tau_{w}$ sample. \\ 
Solid : 7$\tau_{w}$ with poissonian error bars. \\
dashed : `upper' and `lower' limits of counts.\\
dash-dotted : no evolution model }
\label{count7sig}
\end{figure}

\begin{figure}
\psfig{file=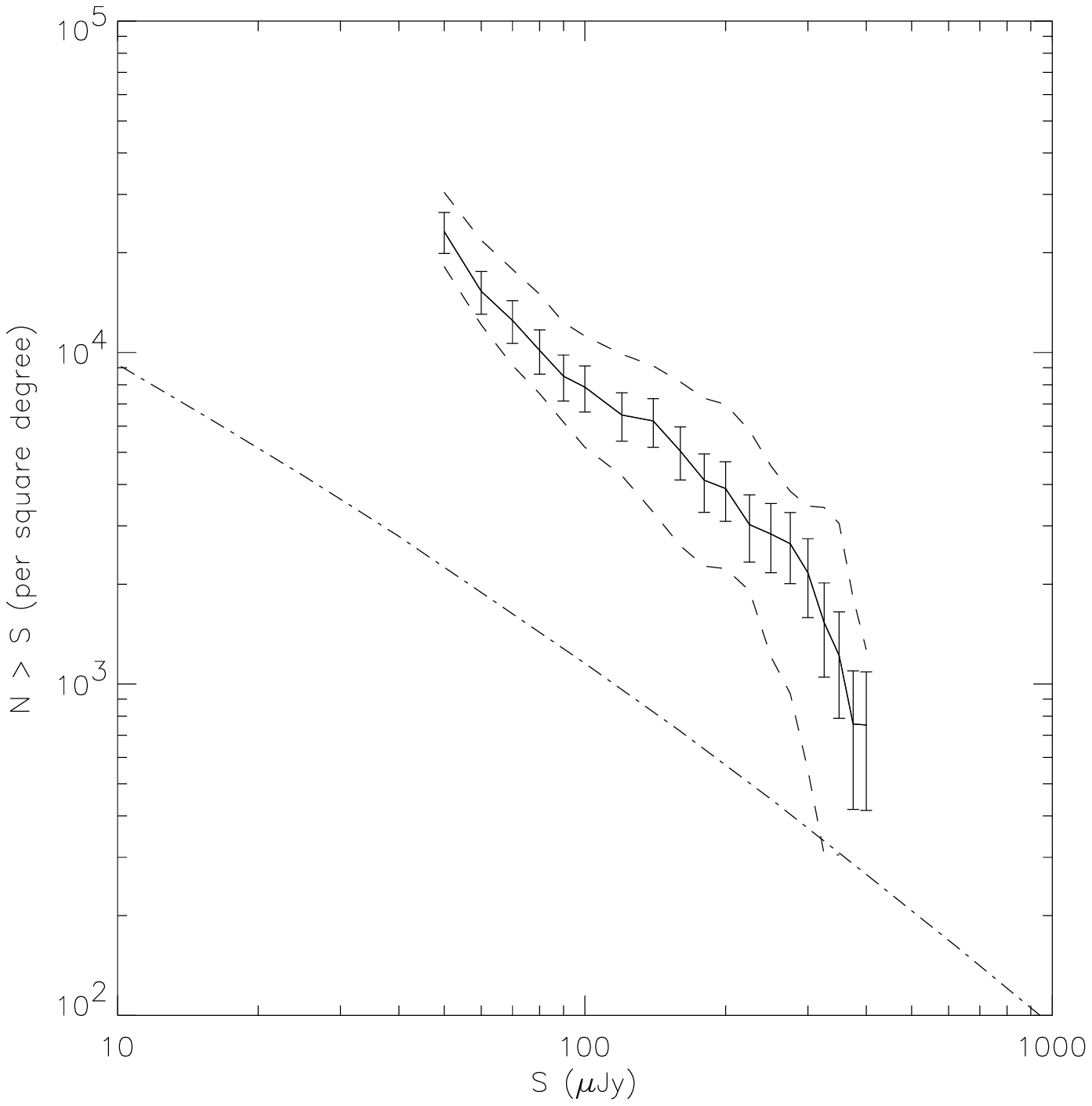,width=0.5\textwidth}
\caption[]{Number counts in the ISO-HDF at 15 $\mu$m, based on the 5$\tau_{w}$ sample plus unblended sources. \\ 
Solid : 5$\tau_{w}$ with poissonian error bars. \\
dashed : `upper' and `lower' limits of counts.\\
dash-dotted : no evolution model }
\label{count5sig}
\end{figure}

This method leads us to derive the number counts that are plotted on
Figs. \ref{count7sig} and \ref{count5sig}, with error bars
corresponding to 1$\sigma$ poissonian fluctuations (as the square root
of the number of events). Moreover, we have also plotted on these
figure two extreme limits :
\begin{enumerate}
\item The `upper' limit, where all sources are given their measured flux plus their positive error bar, to which we add the poissonian fluctuations.
\item The `lower' lower, where all sources are given their measured flux minus their negative error bar, to which we subtract the poissonian fluctuations.
\end{enumerate}
Finally, we have also plotted on these figures the prediction of a
model with no evolution (Franceschini et al.~\cite{Franceschini97}).
 
In both 7$\tau_{w}$ and 5$\tau_{w}$ sample, the straightforward
result is that the counts derived from ISO-HDF in the LW3 band show a
strong excess~: nearly a factor of ten at 200 $\mu$Jy, with respect to
the prediction of the no evolution model. This excess is higher than
the one obtained by Oliver et al. (\cite{Oliver}) from the first
reduction of the ISO-HDF data. This might be due to the
underestimation of the completeness of the sample.

In order to test our simple method to account for blending, we
have plotted in figure \ref{count5sig} the number counts obtained with
the 5$\tau_{w}$ sample with the method described when blending is not
taken into account, but adding to our sample the unblended sources in
our supplementary list. On this last plot, the slope of counts is
constant down to 60 $\mu$Jy, and slightly steeper than predicted in
the framework of the no evolution model.

\section{Discussion}

The number counts derived from the ISO-HDF observations implying strong
evolution at 15 $\mu$m and no evolution at 6.75$\mu$m, may appear
surprising at first sight. This apparent discrepancy is in fact easy
to interpret, considering that we are not observing the same physical
processes at these two wavelengths.

As shown in Fig.~\ref{sketch_ellipitical}, the LW2 filter
tends to select the old stellar population of elliptical galaxies;
taking advantage of the K correction, the detection of such systems
are favored at z $>$ 0.5. For gaseous systems, dominated by dust and
UIB emission, as the redshift rises, the LW2 filter crosses a gap
between the UIB features above 6.2 $\mu$m and stellar emission below
4$\mu$m in the rest frame, leading to a small number of detections,
even though the sensitivity is higher than with the LW3 filter.

The LW3 filter picks up hot dust in nearby systems as
well as redshifted UIB emission at z as high a 1.5. Studies of nearby
galaxies show that this emission is enhanced in the regions
surrounding a starburst, so that the detection of starbursts, often in
mergers, is favoured. This is borne out in the results. Indeed, a close
examination of the morphology of ISO sources shows that early type
galaxies are numerous among the few LW2 detections (6 out 11
detections), while spirals, irregulars, or mergers dominate in the LW3
sample.

\begin{figure}
\psfig{file=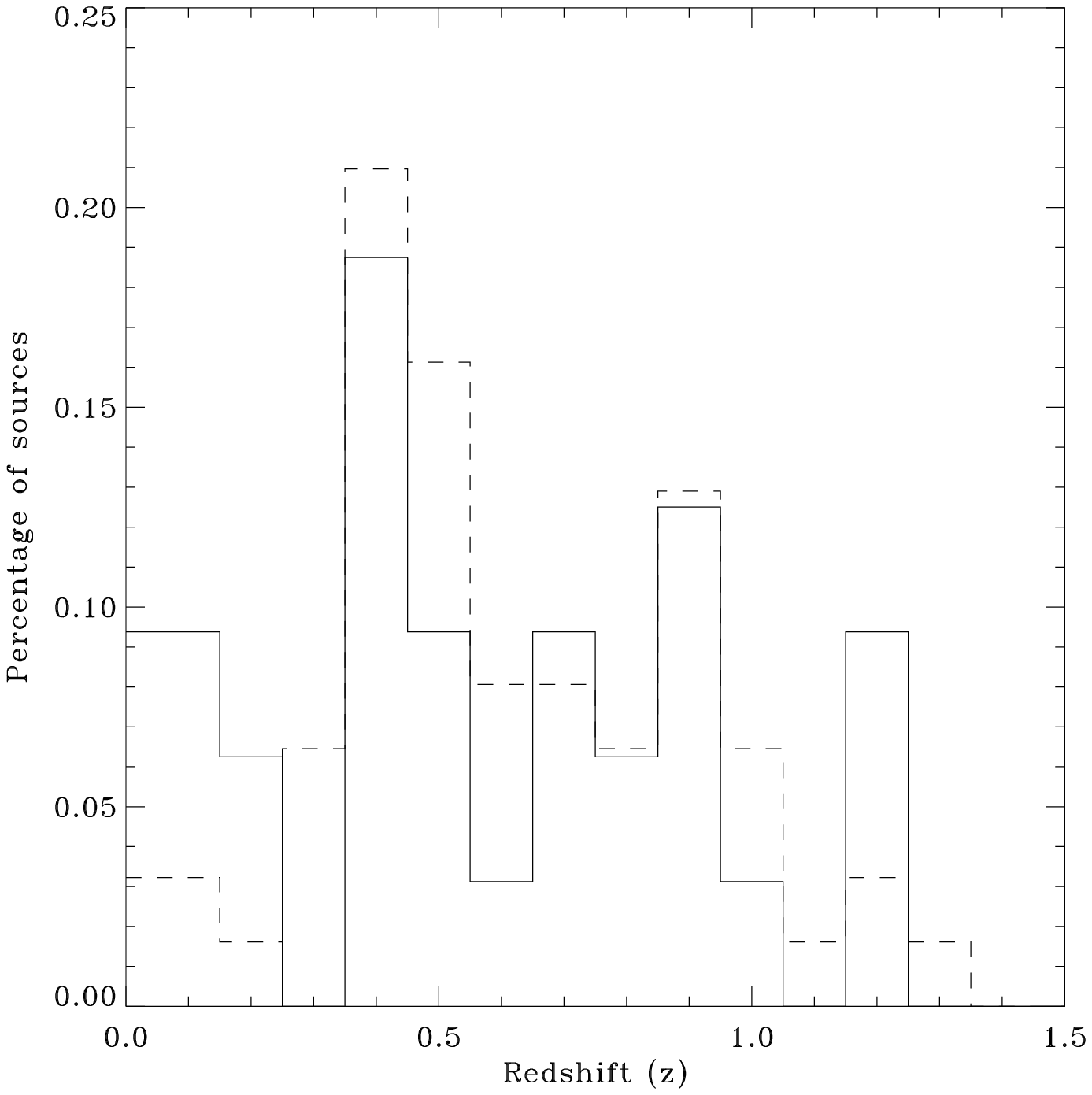,width=0.5\textwidth}
\caption[]{Histogram of redshifts distribution in two samples \\ 
Solid : ISOCAM LW3 sources . \\
dashed : Cowie et al.~(\cite{Cowie}) sample on HDF with  $B< 25$ and $K < 22$.}
\label{histoz}
\end{figure}

The z distribution of our sample of LW3 sources (with signal in
wavespace 7 times above the threshold) is displayed in
Fig.~\ref{histoz}, and compared to that of sources with B $<$ 25 and
K $<$ 22., taken from the list by Cowie et al. (\cite{Cowie}). The
redshifts in our sample range from z = 0.078 to z=1.242, with a median
value of z=0.585 ; most of ISOCAM LW3 sources have z $>$ 0.4 . The
distribution in redshift of our two samples is very similar, implying
that the LW3 filter does not select a particular range in z. 

A forthcoming paper (Elbaz et al., {\it in prep.}) will present a
detailed study of each ISO-HDF sources. This study will allow to
determine if the evolution observed at 15 $\mu$m is a general trend in
galaxies or is due to certain types in the galaxy population, and put
limits on the evolution of the early type systems at z $<$1.5.

\section{Conclusion}

The PRETI method for faint source detection with ISOCAM has been
successfully applied to the observation of the HDF and flanking
fields. Faint fluxes are reached with a good reliability. Moreover,
our simulations allow us to test the completeness and photometric
accuracy of our results. On the ISO-HDF, the method gives a 99.9\%
completeness at 200 $\mu$Jy in LW3 when detecting above five times the
noise level in wavelet space. The same completeness is obtained at 65
$\mu$Jy in LW2. Using PRETI, we detected 49 sources in our main source
list, plus 51 in a less secure supplementary list. Most of the sources
are detected at 15 $\mu$m, 11 are detected at 6.7 $\mu$m only.  Our
results for the brightest sources are in good agreement with those of
D\'esert et al. (\cite{IAS}) in both filters, as well as with those of
Rowan-Robinson et al. (\cite{RowanRobinson}) at 15 $\mu$m, but not at
6.75$\mu$m. The number counts derived for the LW3 filter show a clear
excess of sources with respect to the predictions of a no evolution
model (Franceschini et al.~\cite{Franceschini97}). A forthcoming paper (Elbaz et al. \cite{Elbaz}) will refine these results with a source by source study and
discuss extinction and star formation rates.

\acknowledgements
We wish to thank Alberto Franceschini for interesting discussions and
comments along the preparation of this paper, as well as for having
provided his models of number counts. We thank also
Fran\c{c}ois-Xavier D\'esert for useful discussions. We made use of
the CAM Interactive Analysis Software (CIA), a joint development by
the ESA Astrophysics Division and the ISOCAM Consortium. The ISOCAM
Consortium is led by the ISOCAM PI, C. J. Cesarsky, Direction des
Sciences de la Matiere, C.E.A., France.


\end{document}